\documentclass[letterpaper,journal,onecolumn]{IEEEtran}

\ifCLASSINFOpdf
	\usepackage[pdftex]{graphicx}
	\graphicspath{{./figures/},{./data/}}
	\DeclareGraphicsExtensions{.pdf,.jpeg,.jpg,.png}
\else
	\usepackage[dvips]{graphicx}
	\graphicspath{{./figures/},{./data/}}
	\DeclareGraphicsExtensions{.eps}
\fi

\usepackage[update,prepend]{epstopdf}

\usepackage[cmex10]{amsmath}
\usepackage{amssymb}

\usepackage{xfrac}

\usepackage{url}

\ifCLASSOPTIONcompsoc
	\usepackage[caption=false,font=normalsize,labelfont=sf,textfont=sf]{subfig}
\else
	\usepackage[caption=false,font=footnotesize]{subfig}
\fi

\title{Towards Predictions of the Image Quality of Experience for Augmented Reality Scenarios}
\author{Brian~Bauman~and~Patrick~Seeling,~\IEEEmembership{Senior Member, IEEE}%
\thanks{B. Bauman and P. Seeling are with the Department of Computer Science, Central Michigan University, Mount Pleasant, MI 48859, USA.}%
\thanks{Please direct correspondence to P. Seeling, pseeling@ieee.org.}%
}

\begin{document}
\maketitle

\begin{abstract}
	\boldmath
Augmented Reality (AR) devices are commonly head-worn to overlay context-dependent information into the field of view of the device operators.  One particular scenario is the overlay of still images, either in a traditional fashion, or as spherical, i.e., immersive, content.  For both media types, we evaluate the interplay of user ratings as Quality of Experience (QoE) with ($i$) the non-referential BRISQUE objective image quality metric and ($ii$) human subject dry electrode EEG signals gathered with a commercial device.  
Additionally, we employ basic machine learning approaches to assess the possibility of QoE predictions based on rudimentary subject data.

Corroborating prior research for the overall scenario, we find strong correlations for both approaches with user ratings as Mean Opinion Scores, which we consider as QoE metric.  In prediction scenarios based on data subsets, we find good performance for the objective metric as well as the EEG-based approach.  While the objective metric can yield high QoE prediction accuracies overall, it is limited i its application for individual subjects.  The subject-based EEG approach, on the other hand, enables good predictability of the QoE for both media types, but with better performance for regular content.  Our results can be employed in practical scenarios by content and network service providers to optimize the user experience in augmented reality scenarios.

\end{abstract}


\section{Introduction}
\label{s:intro}

Multimedia accounts for a significant fraction of data transmitted over networks to be consumed on wired and wirelessly connected devices.  The range of devices capable and desired for media consumption has similarly increased over time as well.  Traditional media playout devices, such as computer screens and television sets, are complemented by mobile devices, such as tablet computers and smart phones, and wearable computing devices, such as virtual reality (VR) and augmented reality (AR) devices.
Commonly, a trade-off exists between the quality offered and the required amounts of (compressed) media data. This relationship is typically one of diminishing returns, whereby significantly more bits are required to increase media qualities at the high end. Content and network providers, in turn, have a clear incentive to optimize the delivery of media to connected users.
This relationship has furthermore given rise to a broad range of research efforts, targeting the assessment of quality in objective and subjective forms.  Originally network-centric, the Quality of Service (QoS) is employed to determine several objectively measurable metrics and deduct how well (typically multimedia) services are offered to end users.  From a general viewpoint, network-dependent metrics (such as throughput, delay variations, or buffer fill levels) and media-dependent, but objectively determined quality metrics (such as traditional objective audio quality~\cite{ReFiRe0502:Voice-quality}, image quality~\cite{ShRoLo14:No-reference-image,HeGaHo14:Objective-image,LiKu11:Perceptual-visual}, or video quality metrics~\cite{ChWuZh15:QoStoQoE,ChSuRe1106:Objective-Video}) can all be considered objective QoS metrics.

With human consumption being main target of delivered media content (as in contrast to e.g., automatic video surveillance or annotation systems), perceptual considerations have made great inroads into the evaluation of service qualities.  Overarching, the Quality of Service is generally in the process of being replaced with the Quality of Experience (QoE), formally defined first in~\cite{BrBeDeDoEg13}.  The QoE is typically expressed as a Likert-type categorization of acceptance of a service, based on intrinsic human cognitive and emotional states mixed with the delivery of services (for which one could refer back to QoS metrics).  Several efforts were made in determining the interplay of QoS and QoE, see, e.g., \cite{Fiedler:IQX,Reichl:WFL}.  One motivation for evaluation of this interdependency is that sufficient experimentation is commonly required to establish a ground truth across multiple human subjects.  This approach results in several challenges stemming from the psycho-physiological nature of human subject experimentation, including Institutional Review Board approvals and repeatability.  Several approaches were made in the past to determine standardized human subject trial configurations, such as typical ITU-T standardization descriptions, see, e.g., \cite{ITUT:BT500,ITUT:P910}.  However, the changing landscape of devices employable for media consumption requires that new modalities for environmental impact parameters become part of the experimentation protocol, see, e.g., \cite{Seel1506:Augmented}.  

The shift to the experiential focus in media quality evaluations has resulted in several research efforts directed at capturing the perceived impacts of quality-size trade-offs in new metrics, such as those described in~\cite{PaXuYa16:Exploiting-neural, LiKu11:Perceptual-visual,YoReHa10:Perceptual-based-quality}.  The cognitive and emotional aspects that are inherent to the shift from QoS to QoE is accompanied by an increase of research in brain-computer interfaces (BCI) in order to support direct quantification~\cite{BoMuWi1610:Brain-Computer-Interfacing}.  In past endeavors, such as \cite{DaCrKr16:The-effect-of-perceptual,ScBoTr1205:Toward-a-Direct,LiMa11:Assessing-the-quality}, event-related potentials were successfully employed for quality evaluations.  The common potential utilized in these evaluations occurs 300 to 500 ms after an external stimulus, (i.e., the media display as considered here).
The contribution of our work following these prior efforts is the evaluation of EEG signals within this duration for prediction of the QoE. 
As outlined in \cite{AcBoPoCuMu1504:EEG-based}, the analysis for this short period has potential drawbacks.  In turn, the authors employ a steady state visual evoked potential (SSVEP) approach to the determination of the QoE.  Our work for the prediction follows a similar trajectory, whereby we normalize the overall session EEG measurements and subsequently utilize these values in our prediction approach.  The equivalent could be considered a user-specific overall EEG level availability. 

In typical BCI research, commonly a large amount of wet electrodes with sophisticated setup procedures are employed to gather fine-grained measurements, including in the domain of describes QoE evaluations.  Such setup, however is not feasible for practical implementations.  Consumer-grade hardware has emerged that can capture at a reduced level of complexity, commonly employing dry electrodes at a reduced set of contact points for ease of use.  Employing these already available devices, the main contributions of our work are the evaluation of different approaches to predict the QoE for regular and spherical (i.,e., immersive) images in augmented reality settings.  Employing non-referential metrics and EEG measurements, we find that highly accurate predictions of the QoE are enabled by our approach with little overheads.  The ease of use paired with availability is testamentary to the applicability of our approach in real world implementations.

The remainder of this article is structured as follows. 
We describe our general approach, including metrics employed, in Section~\ref{s:approach}.
Subsequently, evaluation results for QoE and QoS with objective image quality are provided for spherical images in Section~\ref{s:eval}.  We perform a first prediction evaluation for this type of content from the objective BRISQUE metric in Section~\ref{s:predB}.
We widen the evaluation to include BCI information from EEG scans employed to perform predictions of user experiences in the AR domain in Section~\ref{s:predEEG}, followed by a discussion of implications in Section~\ref{s:disc} and a conclusion in Section~\ref{s:conc}.

\section{Approach}
\label{s:approach}
\begin{figure*}
	\centering
	\subfloat[Bamboo]{\includegraphics[width=0.475\linewidth]{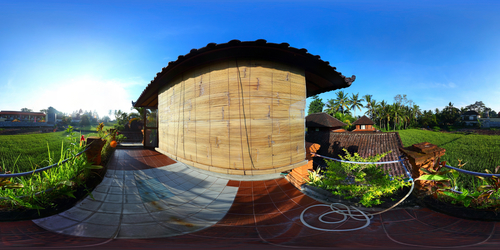}}
	\qquad
	\subfloat[Beach House]{\includegraphics[width=0.475\linewidth]{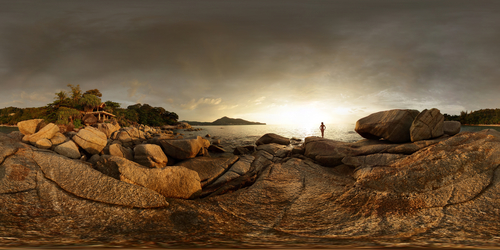}}    
	\\
	\subfloat[Garden]{\includegraphics[width=0.475\linewidth]{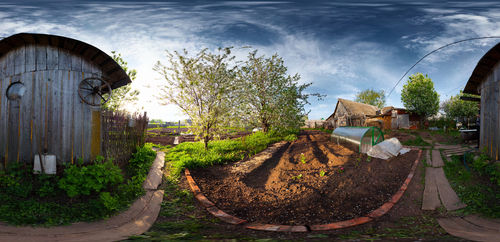}}
	\qquad
	\subfloat[Golf]{\includegraphics[width=0.475\linewidth]{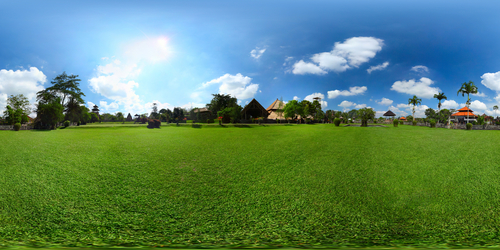}}    
	\\
	\subfloat[Mosque]{\includegraphics[width=0.475\linewidth]{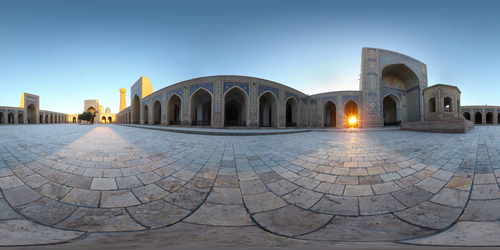}}
	\qquad
	\subfloat[Ocean]{\includegraphics[width=0.475\linewidth]{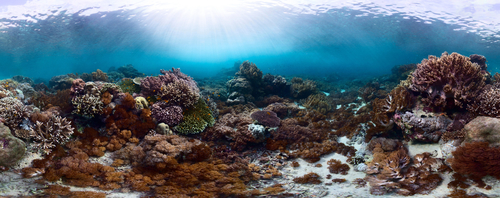}}    
	\caption{Spherical images employed in the human subject experiments, all from Adobe Stock photos.}
	\label{fig:images}
\end{figure*}

Overall, our approach follows the setup we described in~\cite{Seel1506:Augmented} and employed in subsequent evaluations~\cite{BaSe1701:Towards-Still,Se1601:Visual}.  Specifically, we utilize an individual media rating session for each human subject~\footnote{Central Michigan University Institutional Review Board \#568993.} during which several images are displayed and subsequently rated as follows.
Users are instructed in the general usage of the viewing and monitoring devices and overall procedure, including consent, before the experimental phase.
We employ professionally generated spherical images, as illustrated in Figure~\ref{fig:images}, derived from the Adobe Stock photo database.
These images, denoted as $i$, are displayed on a customized viewer application that displays them on the head-worn mobile augmented reality viewer, taking device limitations into account. Different levels of JPEG compression were introduced to the original source images to generate varying levels of quality degradations, which are abstracted in our overall scenario as QoS or impairment levels $l$.  We focus on the non-referential BRISQUE image quality metric~\cite{MiMoBo1212:No-Reference}. In our prior research, we found high correlations between this particular metric and its application for AR scenarios, see~\cite{BaSe1701:Towards-Still} for more details.  An inherent benefit of utilizing a non-referential quality metric is that the original uncompressed source media is not required for comparisons and, subsequently, metric values could be determined at any point from content over network service providers to end user devices.  An overview of the resulting BRISQUE values $B_{i,l}$ is provided in Table~\ref{tab:brisque}.
\begin{table}
	\centering
	\caption{BRISQUE metric values determined for the employed spherical test images.\label{tab:brisque}}
	\begin{tabular}{|c|r|r|r|r|r|r|}
		\hline
		Imp. $l$ &  Gard. &  Bamb. & Mosque &  Ocean &   Golf & Beachh. \\ \hline\hline
		Org., 0     & 27.602 & 11.325 &  8.643 & 10.107 & 11.632 &   3.402 \\ \hline
		1        & 34.082 & 31.327 & 15.609 &  15.020 & 31.747 &  23.422 \\ \hline
		2        & 33.384 & 41.784 & 27.429 & 22.109 & 32.648 &  25.361 \\ \hline
		3        & 35.943 & 51.389 & 46.592 & 33.902 & 34.323 &  32.967 \\ \hline
		4        & 49.366 & 57.931 & 61.083 & 43.216 &  46.32 &   52.790 \\ \hline
		5        & 80.646 & 90.495 & 86.582 & 65.843 & 86.237 &  85.074 \\ \hline
	\end{tabular}
\end{table}
We notice that most images exhibit original BRISQUE metric values around or below ten, with \textit{Garden} being the only exception with a BRISQUE value of $B_{\mathrm{\it Garden},0}=27.602$. Subsequent impairments increase these values in varying degrees, with the high impairment level commonly exhibiting BRISQUE values above 80, with the \textit{Ocean} image being an exception here with a BRISQUE value of $B_{\mathrm{\it Ocean},0}=65.843$.  

These images are displayed on the head-worn AR viewer (Epson Moverio BT-200 without shades) and shown to subjects for a time period of 15 seconds, following the ACR-HR approach (i.e., we include the original image in the evaluation).
In addition to the head-mounted AR viewer, subjects also wore a commercial-grade EEG headband, which captured EEG data for TP9, Fp1, Fp2, and TP10 positions $p$. For these positions, common brain wave band data is gathered at 10Hz in absolute values.  Specifically, the common EEG data we utilize include
($i$) low $\iota^p$ at 2.5-6.1 Hz, 
($ii$) delta $\delta^p$ at 1-4 Hz,
($iii$) theta $\theta^p$ at 4-8 Hz,		
($iv$) alpha $\alpha^p$ at 7.5-13 Hz,			
($v$) beta $\beta^p$ at 13-30 Hz, and		
($vi$) gamma $\gamma^p$ at 30-44 Hz.
Given our overall experimental setup, we sample the EEG signals from image display up to the potential component P3 between 300 and 500 ms after stimulus, motivated by results in prior research endeavors.  With $t_{i,l}$ denoting the time an image with impairment level $l$ is displayed, we utilize EEG readings until $t_{i,l}+500$ ms.  We employ the average channel levels at individual positions from the image display until 500 ms in our subsequent evaluations. 
Let $\epsilon_t$ denote an arbitrary EEG band measured at time $t$ and position $p$ to simplify ideas.  Following our notation, we employ
\begin{equation}\label{e:p500}
\epsilon_{i,l} = \frac{ \sum_{t=t_{i,l}}^{t_{i,l}+500 \text{ms}} \epsilon_t }{ \sum_{t=t_{i,l}}^{t_{i,l}+500 \text{ms}} [\epsilon_t \in \mathbb{R}] }
\end{equation}
in our evaluations of EEG signals (noting that $[\cdot]$ represents the Iverson Bracket).
We additionally note that the start of the image display time and P3 threshold cutoff around 500 ms are determined within the real-world experimental configuration employed (i.e., with accuracy challenges) and, thus are representative of realistically attainable results. 
Let $t=0$ denote the start time of an experimental session recording and $t=T$ denote the last recording time. During a particular session, we recorded a total of $J=\sum_{t=0}^{T} [\epsilon_t \in \mathbb{R}]$ EEG readings. 
For normalization of EEG measurements, we employ the z-Score for $\epsilon$ as follows.  Let $\bar{\epsilon} = \frac{1}{J} \sum_t \epsilon_t$ denote the average individual position EEG channel value over the experimental session. Similarly, let $\sigma(\epsilon)$ denote the corresponding standard deviation.  The z-Score for each individual measurement is subsequently defined as 
\begin{equation}
\epsilon^z_t = \frac{\epsilon_t - \bar{\epsilon}}{\sigma(\epsilon)}.
\end{equation}
In turn, the z-scored version of Equation~\ref{e:p500} is given as 
\begin{equation}\label{e:p500z}
\epsilon^z_{i,l} = \frac{ \sum_{t=t_{i,l}}^{t_{i,l}+500 \text{ms}} \epsilon^z_t }{ \sum_{t=t_{i,l}}^{t_{i,l}+500 \text{ms}} [\epsilon_t \in \mathbb{R}] }.
\end{equation}

Subsequently, the subjects or users $u$, $u=1, \ldots, U$, were asked to rate the quality on a 5-point Likert scale.
We denote the thus derived user ratings for image $i$ at impairment level $l$ as $q^u_{i,l}$, with $q^u_{i,l}\in{1,2,3,4,5}$.  The resulting mean opinion score (MOS) as QoE for a given image and impairment level is given as 
\begin{equation}
Q_{i,l}=\frac{1}{U} \sum_u q^u_{i,l}, ~l=1,\ldots,5.
\end{equation}
We denote the MOS as AR-MOS for regular content display in the field of view and as SAR-MOS for spherical, i.e., immersive, content.
We additionally include the differential MOS (DMOS), determined as crushed differential viewer score according to ITU-T P.910~\cite{ITUT:P910}.
Specifically, for an individual user rating we determine 
\begin{equation}
q'^u_{i,l}=
\begin{cases}
q^u_{i,l}-q^u_{i,0}+5,& \text{if } q^u_{i,l}-q^u_{i,0}+5 \le 5\\
\frac{7 (q^u_{i,l}-q^u_{i,0}+5)}{q^u_{i,l}-q^u_{i,0}+7}, & \text{otherwise}
\end{cases}
\end{equation}
for $l=1,\ldots,5$.
Similar to the MOS, we denote the DMOS $Q'_{i,l}$ as AR-DMOS and SAR-DMOS for regular and immersive media, respectively.

\section{General Evaluation of Spherical Image Quality Experiences}
\label{s:eval}

\begin{figure}
	\centering
	\includegraphics[width=0.75\linewidth]{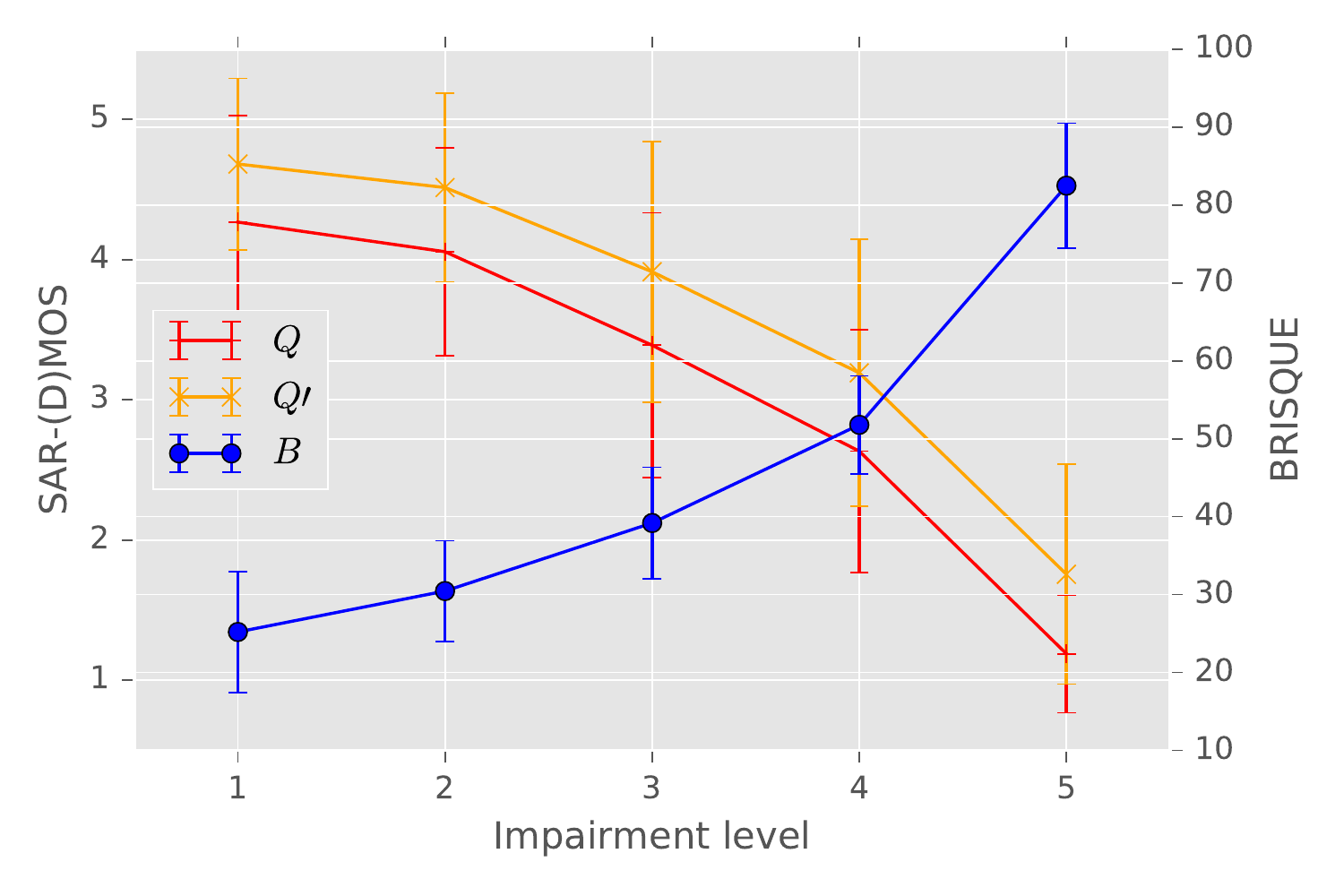}
	\caption{Overview of subject-rated QoE (SAR-MOS $Q$ and SAR-DMOS $Q'$) and non-referential BRISQUE metric $B$ values across all images evaluated.}
	\label{fig:SAR-all}
\end{figure}

Initially, we perform an overall summary of metrics and SAR-MOS and SAR-DMOS values across all images $I$.  In turn, the BRISQUE average is given as $\overline{B}_l=\frac{1}{I} \sum_i B_{i,l}$ and the corresponding MOS values are calculated similarly.
We illustrate the overall SAR-(D)MOS and the BRISQUE objective image quality metric values and standard deviations in Figure~\ref{fig:SAR-all}.
We note that the overall SAR-MOS values follow the set impairment level trend, albeit at a rate that resembles a logarithmic decay function.  For the SAR-DMOS, we notice a general shift to slightly higher values.  These observations are overall in line with prior findings for regular images in this overall experimental configuration.  We refer the interested reader to~\cite{BaSe1701:Towards-Still,Se1601:Visual,Seel1506:Augmented} for these prior evaluations.  For the BRISQUE values, we observe an opposite behavior, which directly results from BRISQUE being an impairment metric.  We additionally note that the medium SAR-DMOS is slightly shifted towards higher impairment levels, whereas the SAR-MOS is closer to the medium impairment level.

We provide the Pearson Correlation Coefficients~\cite{} for subject ratings with the set level and the BRISQUE metric in Table~\ref{tab:pearsonsar}.
\begin{table}
	\centering
	\caption{Pearson correlation coefficients between individual subject ratings ($q^u_{i,l}$ and $q'^u_{i,l}$) and set level $l$ as well as BRISQUE metric values $B_{i,l}$ for spherical test images.\label{tab:pearsonsar}}

\begin{tabular}{|l||r|r||r|r|}
	\hline
	Image       & \multicolumn{2}{|c||}{Reg. $q^u_{i,l}$} & \multicolumn{2}{|c|}{Diff. $q'^u_{i,l}$} \\
	            &   Level $l$ &           BRISQUE  $B_{i,l}$ &   Level $l$ &           BRISQUE  $B_{i,l}$\\ \hline\hline
	All         & -0.7925 &           -0.7887 & -0.7625 &           -0.7712 \\ \hline\hline
	Garden      & -0.9935 &           -0.9978 & -0.9903 &           -0.9965 \\ \hline
	Bamboo      & -0.9946 &           -0.9967 & -0.9775 &           -0.9820 \\ \hline
	Mosque      & -0.9930 &           -0.9948 & -0.9776 &           -0.9811 \\ \hline
	Ocean       & -0.9943 &           -0.9958 & -0.9807 &           -0.9837 \\ \hline
	Golf        & -0.9943 &           -0.9954 & -0.9817 &           -0.9840 \\ \hline
	Beach house & -0.9940 &           -0.9951 & -0.9829 &           -0.9850 \\ \hline
\end{tabular}
\end{table}
We initially note that across all images rated by subjects, the correlation coefficient indicates a strong negative linear relationship with the set level for the direct user ratings not considering the original media.  The reason for the negative relationship is that both levels and quality metric are based on impairments, not quality.  Furthermore, a strong linear relationship is also indicated for the BRISQUE metric values across all images and levels.  This further corroborates the previously made graphical observations for both and motivates a further utilization of the BRISQUE metric here.  For the differential viewing scores, we notice a slight decline in the correlation between the impairment level as well as the objective quality metric.

Investigating the results for the individual spherical images more closely, we note that the highest linear relationship for level and BRISQUE metric values is attained for the \textit{Garden} image, whereas the lowest one is attained for the \textit{Mosque} image.
The latter exhibits a more narrow range of colors and textures, which are complemented by large similar floor areas as well as a fairly uniform background sky.  This results in detail loss that is well distributed as compression and impairment levels increase.  In turn, discerning the potential level could be seen as a more difficult task and result in slightly less of a clear valuation by subjects.  
With the various color texture and cloudy skies of the \textit{Garden} image, complemented by very earthy color tones throughout, higher compression levels are not impacting individual content items directly.  This may be responsible for the better distinction of the levels, as subjects could infer from the compression artifacts and decolorizing of the image content.   
We note that in all cases, however, the levels of the correlation coefficients indicate a very strong linear relationship between level and BRISQUE values and resulting subject ratings, respectively.

\begin{figure}
	\centering
	\includegraphics[width=0.75\linewidth]{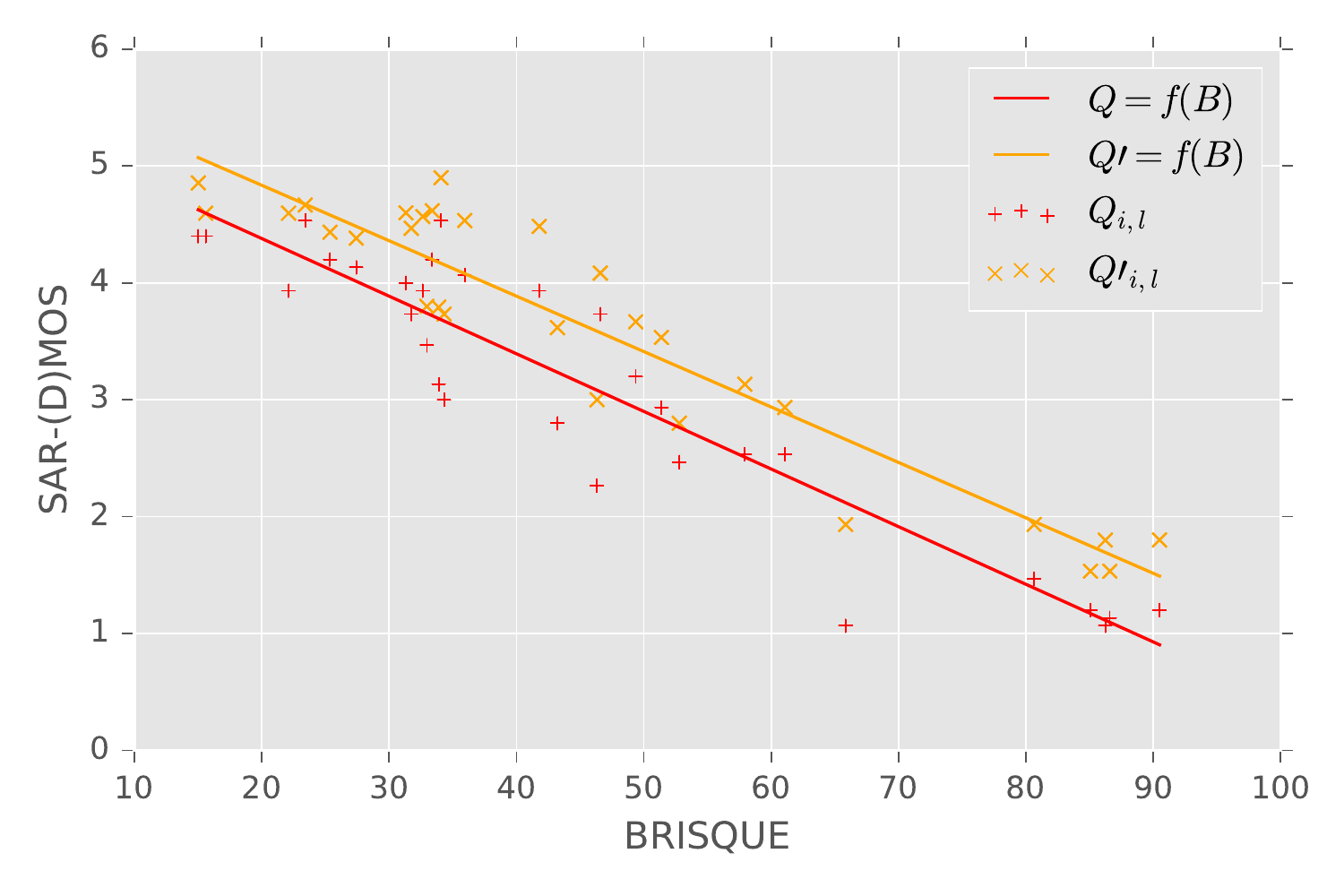}
	\caption{Linear regression for the SAR-MOS $Q_{i,l}$ and SAR-DMOS $Q'_{i,l}$ based on BRISQUE metric values $B_{i,l}$ across all images and levels.}
	\label{fig:SAR-all-linreg}
\end{figure}

To evaluate the relationship between the objective non-referential BRISQUE metric and the attained SAR-(D)MOS, we illustrate the individual $Q_{i,l}, Q'_{i,l}$ ratings as a function of the individual image BRISQUE metrics $B_{i,l}$ in Figure~\ref{fig:SAR-all-linreg}.
We observe an overall somewhat banded increase in the SAR-MOS values together with an increase of the objective BRISQUE values.  Simple linear regression, illustrated as well, results in an $R^2=0.861$ for $Q = 5.3661 -0.0493 \cdot B$.  The coefficient of determination $R^2$ is generally employed to evaluate the closeness of a fit, whereby $R^2=1$ indicates a perfect fit~\cite{}.  Similarly, we note that the inclusion of the hidden reference image results in an almost parallel shift of the resulting SAR-DMOS as a function of BRISQUE.  For the differential case, we obtain $R^2=0.886$ with $Q' = 5.7843 -0.0474 \cdot B$.  Overall, these findings indicate a general linear relationship between the BRISQUE metric and resulting subject ratings, albeit with significant deviations.

\begin{figure}
	\centering
	\includegraphics[width=0.95\linewidth]{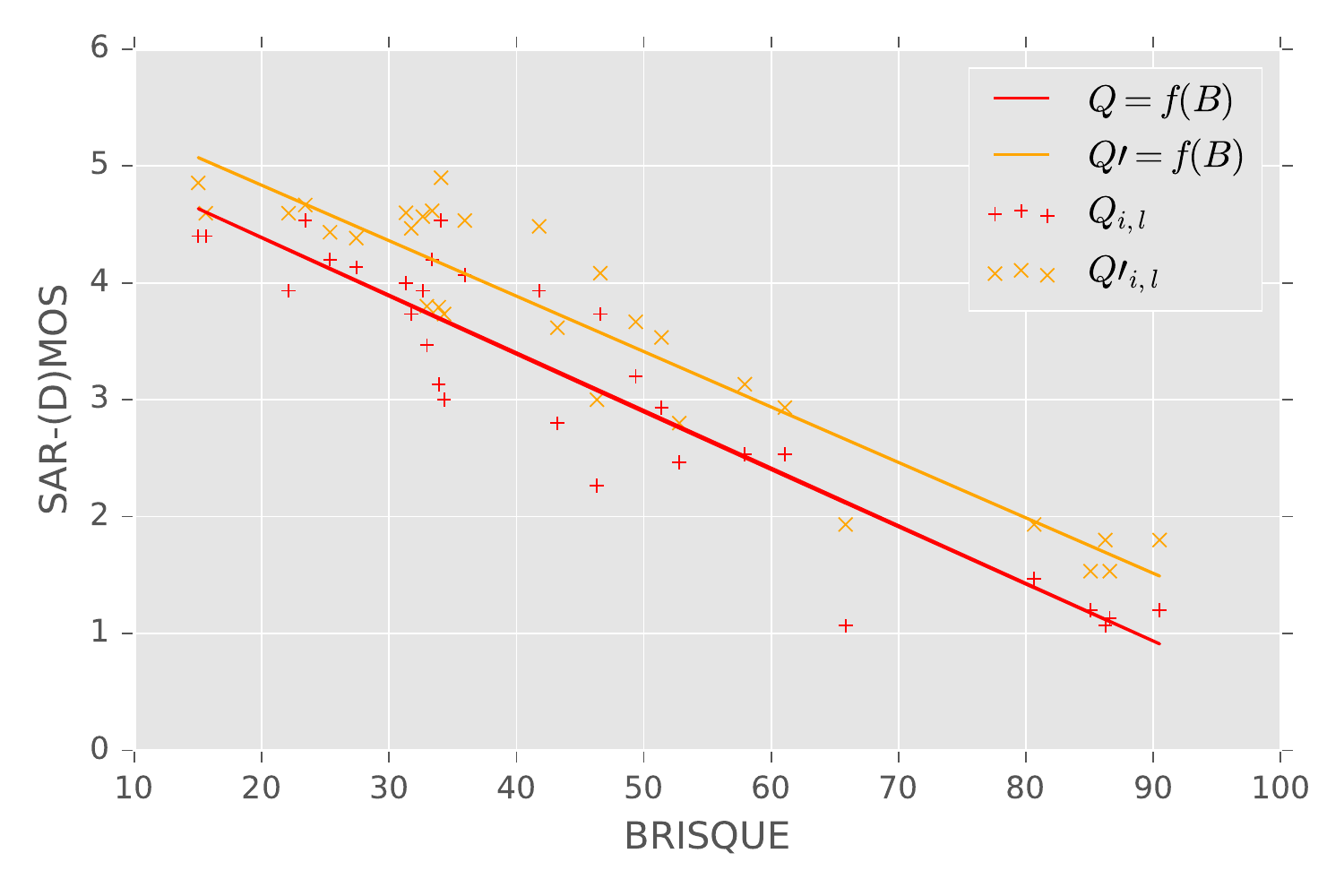}
	\caption{IQX-based interdependency of SAR-MOS and SAR-DMOS as function of the BRISQUE metric employed as QoS input factor.}
	\label{fig:SAR-all-iqx}
\end{figure}

We corroborate this relationship by applying the Exponential Interdependency between QoS and QoE (IQX) theory as described in~\cite{Fiedler:IQX}.  Specifically, we employ the generic formula of $QoE = \alpha \cdot e^{-\beta \cdot QoS} + \gamma$.  In our scenario, we let QoE denote the SAR-(D)MOS and the QoS factor denote the BRISQUE metric's values.
Non-linear regression for determination of the coefficients results in 
\begin{eqnarray}
Q = 130.021 \cdot e^{-3.872 \cdot 10^{-4} \cdot B} -124.632\\
Q' = 8489.479 \cdot e^{-5.589 \cdot 10^{-6} \cdot B} -8483.694.
\end{eqnarray}
We illustrate the resulting fit in Figure~\ref{fig:SAR-all-iqx}.
As indicated by the high $R^2$ values for the linear regression results, the IQX in this scenario has a rather linear relationship to the underlying QoS metric.  This, in turn, causes the resulting non-linear regression to equally result in a highly linear outcome.  The resulting $R^2=0.861$ and $R^2=0.886$ for SAR-MOS and SAR-DMOS, respectively, indicate that the IQX and linear approaches both result in equally suited fits due to the nature of the underlying QoS metric.

%
%
\section{Prediction from Non-Referential Objective Metric}
\label{s:predB}
Motivated by the prior observations for the BRISQUE image quality metric, in this section we evaluate its suitability for prediction of subjectively experienced image qualities.
Specifically, we evaluate linear as well as logistic regressions at different degrees, whereby the latter approach is motivated by the categorical Likert-scale.  Due to the almost linear shift between SAR-MOS nd SAR-DMOS, we focus our evaluation on the former.

\subsection{Mean Opinion Score}
We employ regression on random 80\% parts of the BRISQUE metric $B_{i,l}$ and subject rating data $q^u_{i,l}$ and utilize the resulting fit for prediction of the remaining pairs. 
As a pre-processing step, the existing data is converted to polynomials of varying degrees $d$, i.e., we employ a transformation of $q^u_{i,l}=\sum_{k=0}^{d}a_k B_{i,l}^k$ in the prediction part.
For each individual regression fit, we obtain $d+1$ of resultant coefficients $a_k$ for the employed linear regression approach.  For logistic regression, we derive a $5 \cdot (d+2)$ total coefficients due to the five classes from the underlying Likert-type scale.  In addition to the linear regression, which results in $q^u_{i,l}$ not necessarily as integer on the desired Likert-type scale, we additionally modify the regression prediction results to full integers, bound to the Likert scale range under consideration.

\begin{figure}
	\centering
	\subfloat[Linear: $\hat{R}^2$=0.622, MSE=0.691, MAE=0.662, MedAE=0.626]{\includegraphics[height=0.29\textheight]{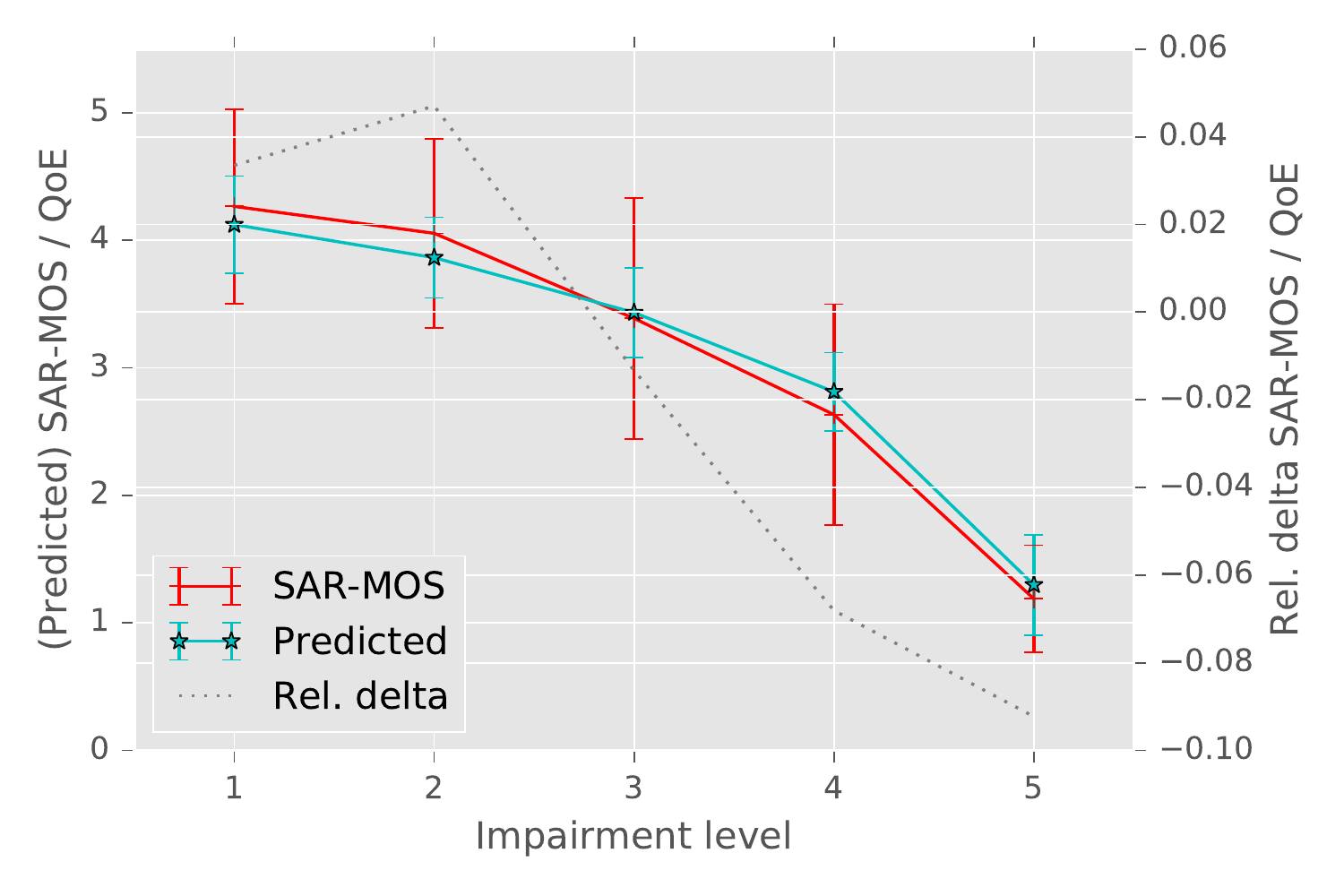}}
	\\
	\subfloat[Bound linear: $\hat{R}^2$=0.586, MSE=0.758, MAE=0.616, MedAE=1.0]{\includegraphics[height=0.29\textheight]{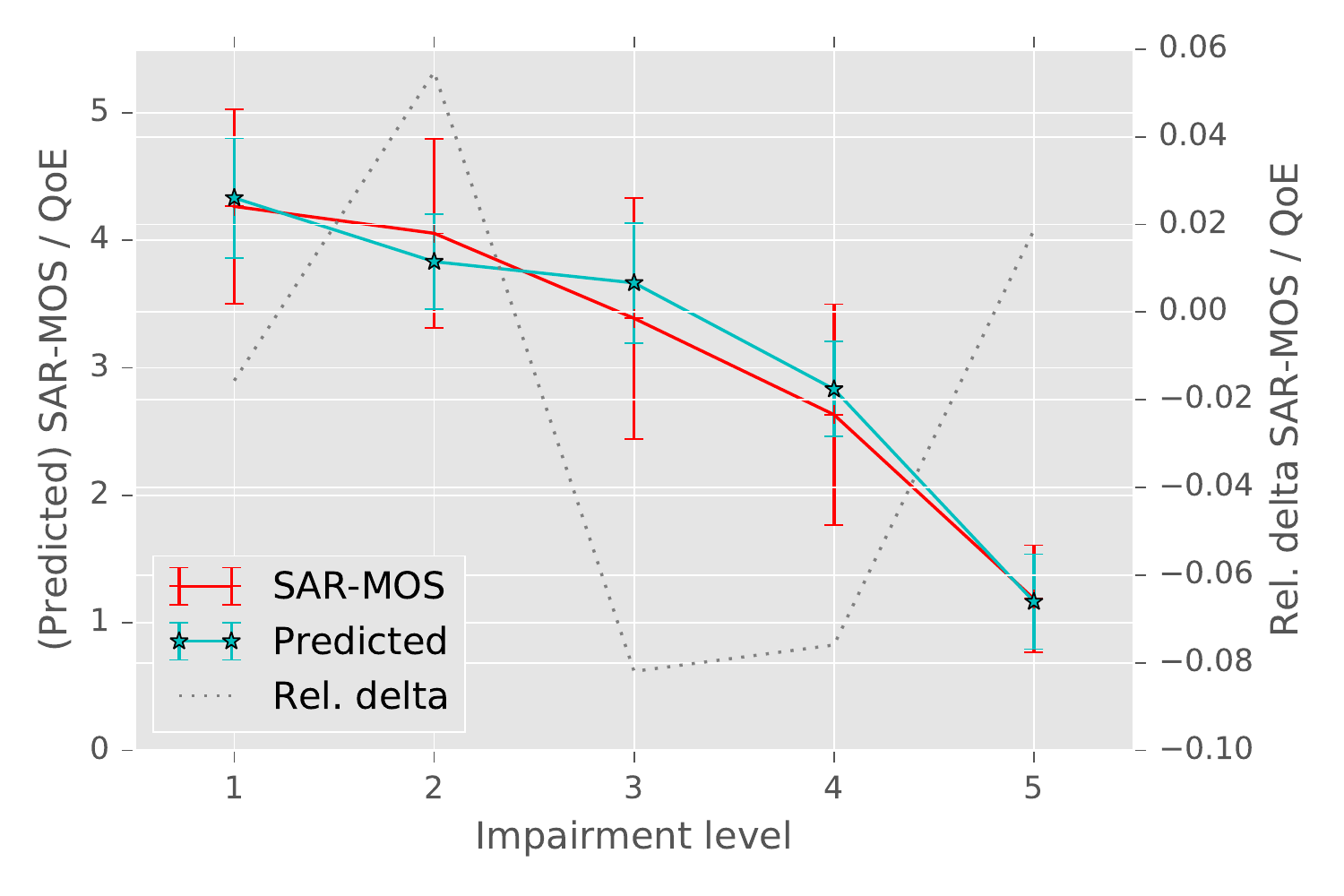}}
	\\
	\subfloat[Logistic: $\hat{R}^2$=0.477, MSE=0.956, MAE=0.667, MedAE=1.0]{\includegraphics[height=0.29\textheight]{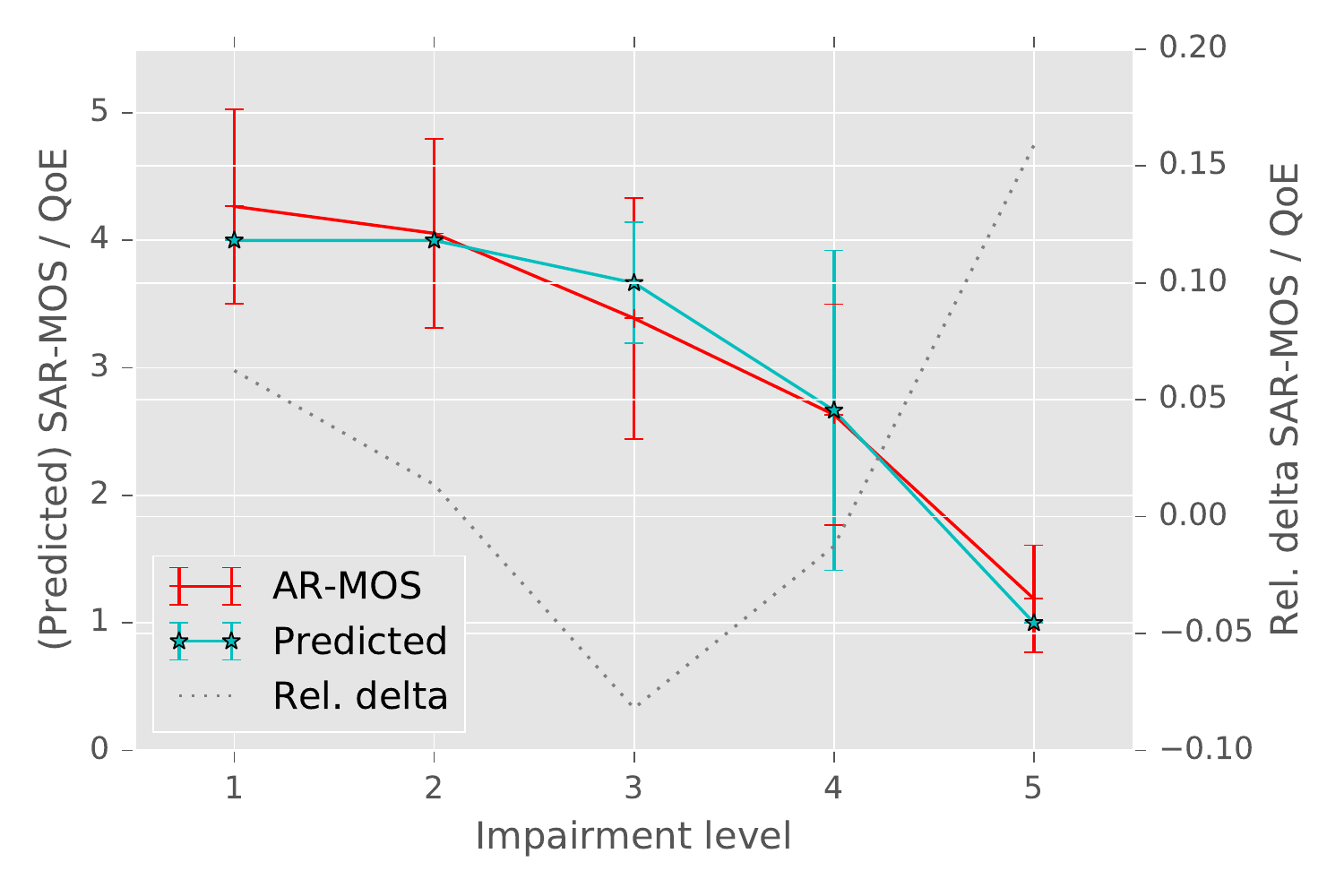}}
	\caption{Linear, bound linear, and logistic SAR-MOS prediction approaches for first degree $d=1$ underlying QoS metric.}
	\label{fig:SAR-all-logregd1}
\end{figure}

We repeat this prediction process $n, 500 \le n \le 50000$ times, interrupting when a 95\% confidence interval width~\cite{} below 5 percent of the estimated $\dot{R}^2$ value is reached.  The determined coefficients for these runs are subsequently averaged, resulting in $\dot{a_k} = \frac{1}{n}\sum_n a_k(n)$.
We subsequently employ these averaged model parameters to perform prediction and determine the $\ddot{R}^2$ for the individual fit of averaged values in a similar fashion. Deriving a narrow level of overall confidence, we again repeat this process with $50 \le n \le 250$, again employing a 95\% confidence interval width as criterion for stopping these second-level prediction runs.  The result is a final estimated $\hat{R}^2$ for the thus determined coefficient estimates $\hat{a_k}$.  We illustrate the resulting average prediction outcomes in Figure~\ref{fig:SAR-all-logregd1} for a first degree $d=1$.
We initially note that the thus predicted QoE curve average closely follows the one obtained through subject ratings for all approaches.  Furthermore, we observe that the variability of results, illustrated as standard variation, is commonly smaller for the predictions than the actual SAR-MOS.  The additionally presented relative prediction error is also fairly close to an average of below 10\%, indicating an overall decent prediction performance. This is corroborated by an estimated determination coefficient of $\hat{R}^2=0.622$ for the linear prediction. The bound linear prediction approach results in a slightly lower  $\hat{R}^2=0.586$, trailed by  $\hat{R}^2=0.477$ for the logistic regression approach.  Evaluating the errors through prediction, we additionally consider the Mean Squared Error (MSE), Mean Absolute Error (MAE), and Median Absolute Error (MedAE).  The MedAE is particularly interesting due to its robustness concerning outliers.
In comparison, we observe that the MSE is inversely following the $\hat{R}^2$ values observed.  The mean absolute errors are generally comparable, whereas the median absolute errors are one for the integer-based predictions, while the linear prediction results in a lower error.  Jointly, we note a general indication that the each approach generally captures the overall SAR-MOS, with on average being off by one level. 

\begin{figure}
	\centering
	\subfloat[Linear: $\hat{R}^2$=0.622, MSE=0.691, MAE=0.662, MedAE=0.636]{\includegraphics[height=0.29\textheight]{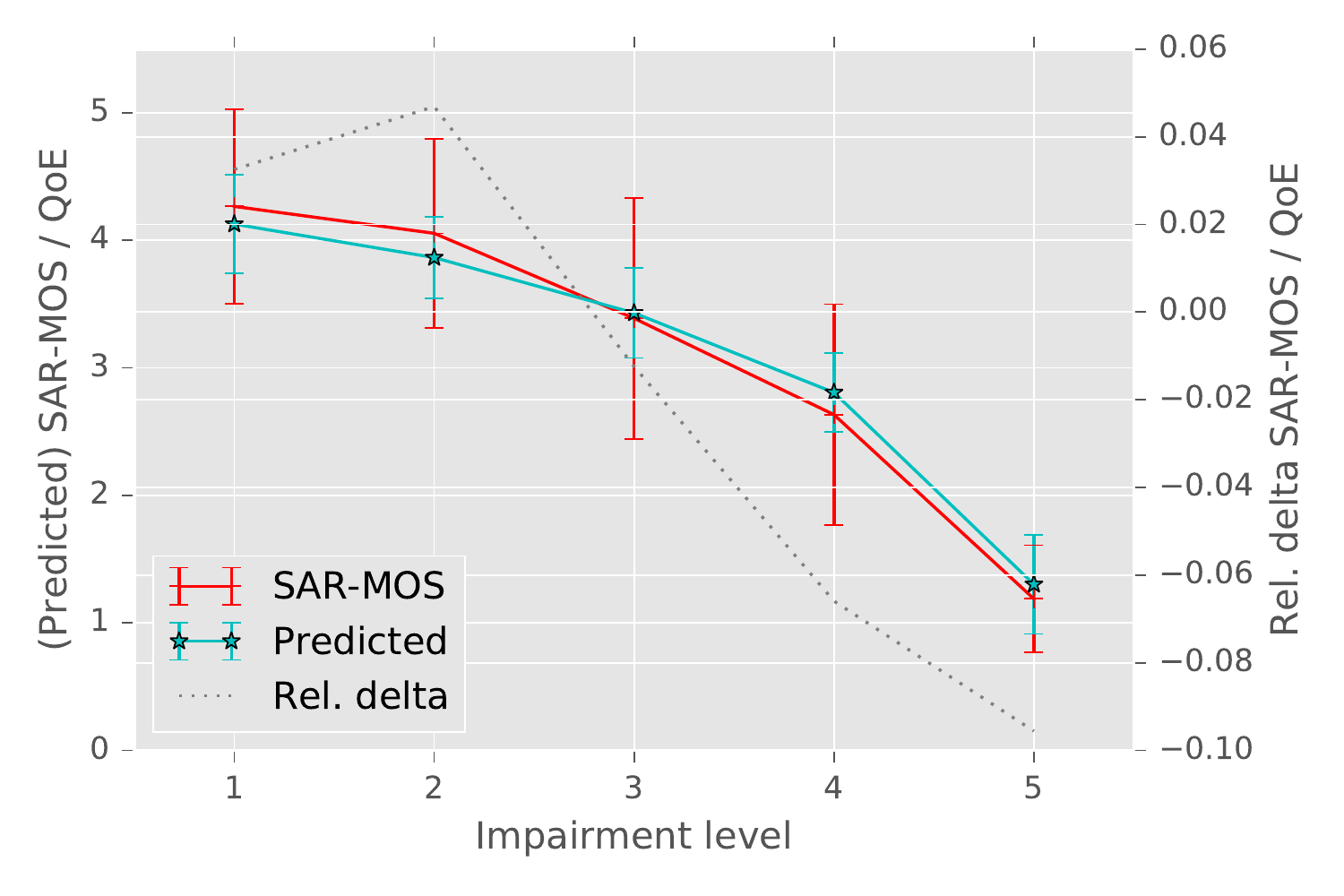}}
	\\
	\subfloat[Bound linear: $\hat{R}^2$=0.586, MSE=0.758, MAE=0.616, MedAE=1.0]{\includegraphics[height=0.29\textheight]{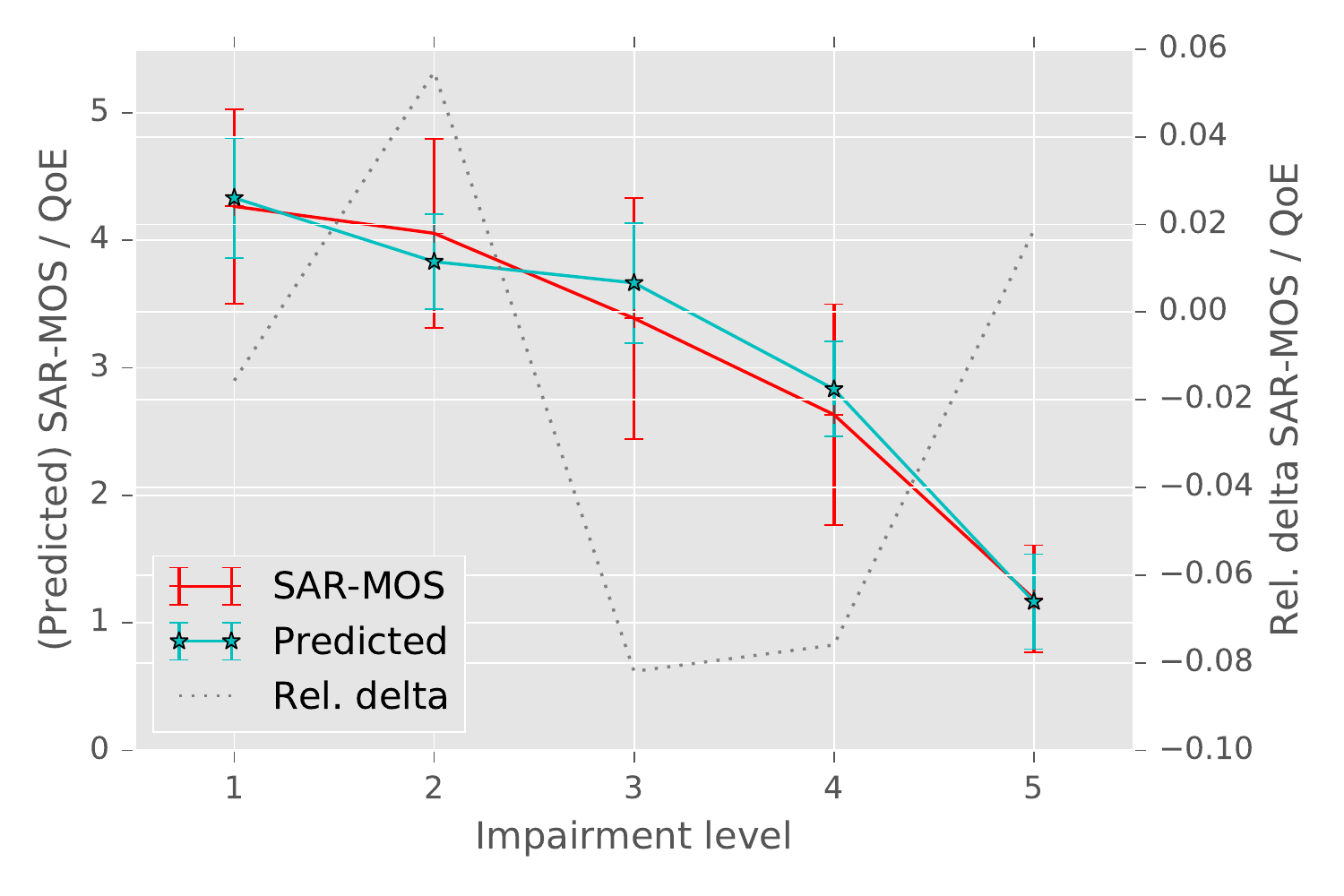}}
	\\	
	\subfloat[Logistic: $\hat{R}^2$=0.592, MSE=0.747, MAE=0.613, MedAE=1.0]{\includegraphics[height=0.29\textheight]{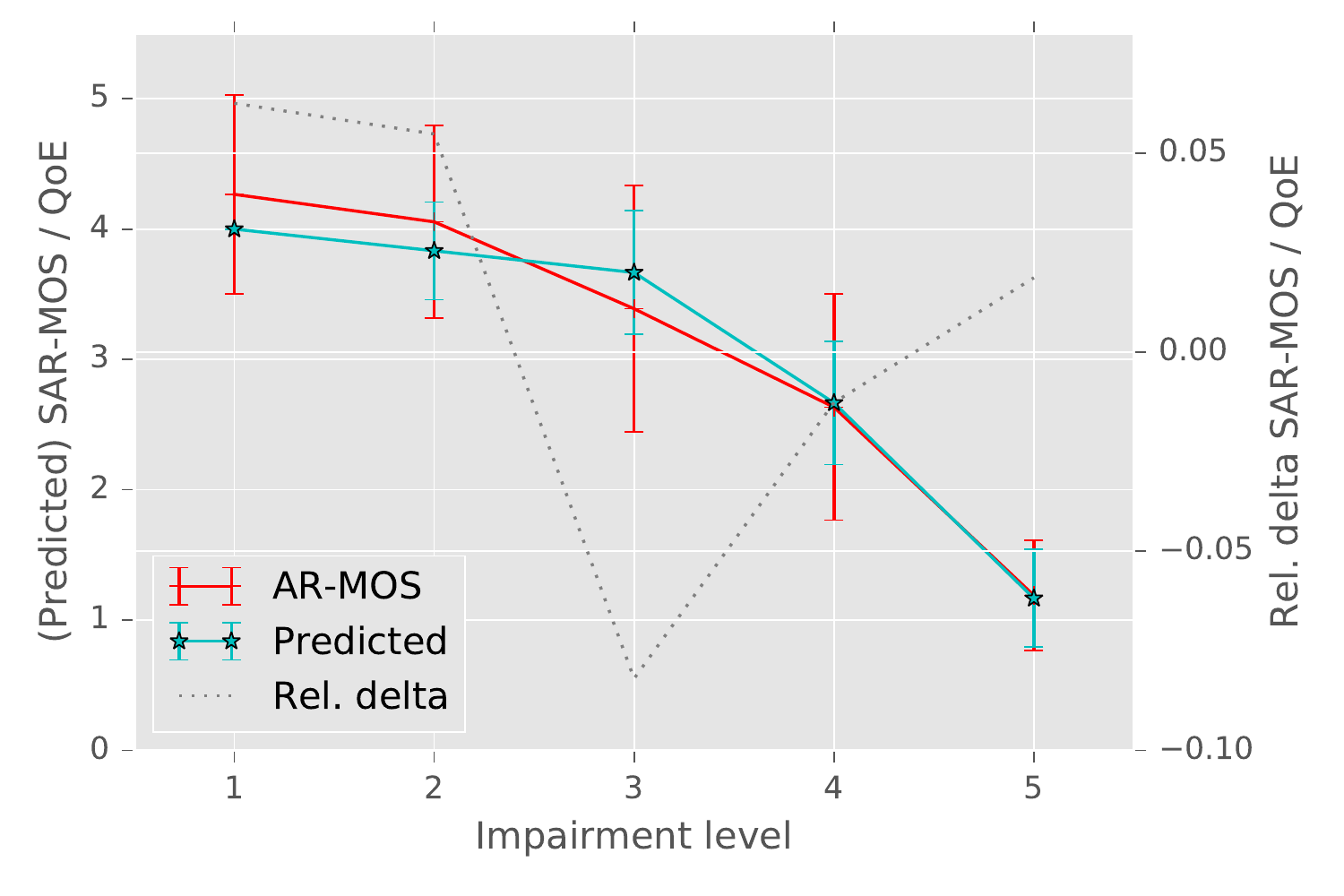}}
	\caption{Linear, bound linear, and logistic SAR-MOS prediction approaches for second degree $d=2$ underlying QoS metric.}
	\label{fig:SAR-all-logregd2}
\end{figure}

\begin{figure}
	\centering
	\subfloat[Linear: $\hat{R}^2$=0.632, MSE=0.674, MAE=0.649, MedAE=0.647]{\includegraphics[height=0.29\textheight]{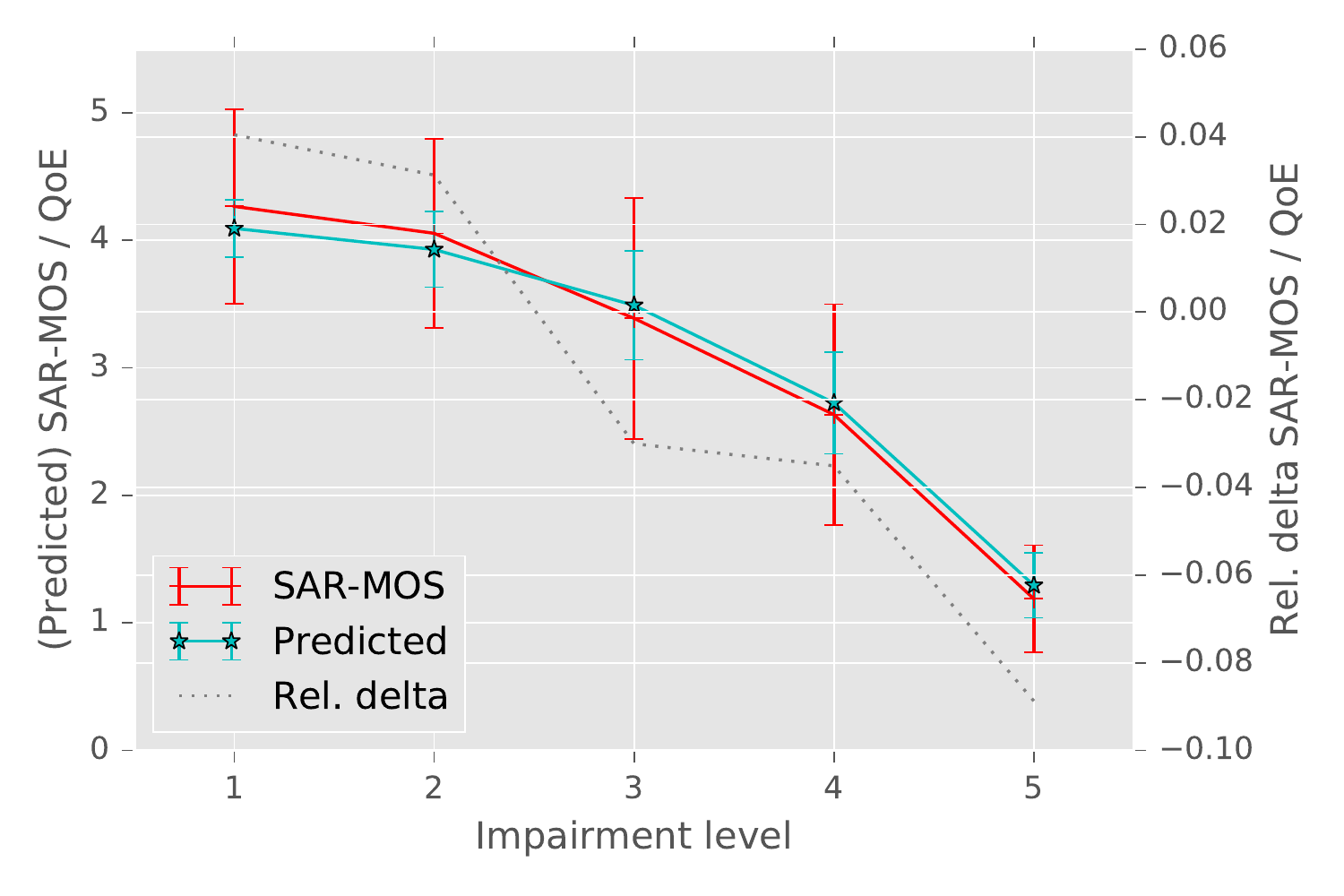}}
	\\
	\subfloat[Bound linear: $\hat{R}^2$=0.592, MSE=0.747, MAE=0.613, MedAE=1.0 ]{\includegraphics[height=0.29\textheight]{all_pred_lin_bound_BRISQUE_degree_1}}
	\\
	\subfloat[Logistic: $\hat{R}^2$=0.583, MSE=0.762, MAE=0.589, MedAE=1.0]{\includegraphics[height=0.29\textheight]{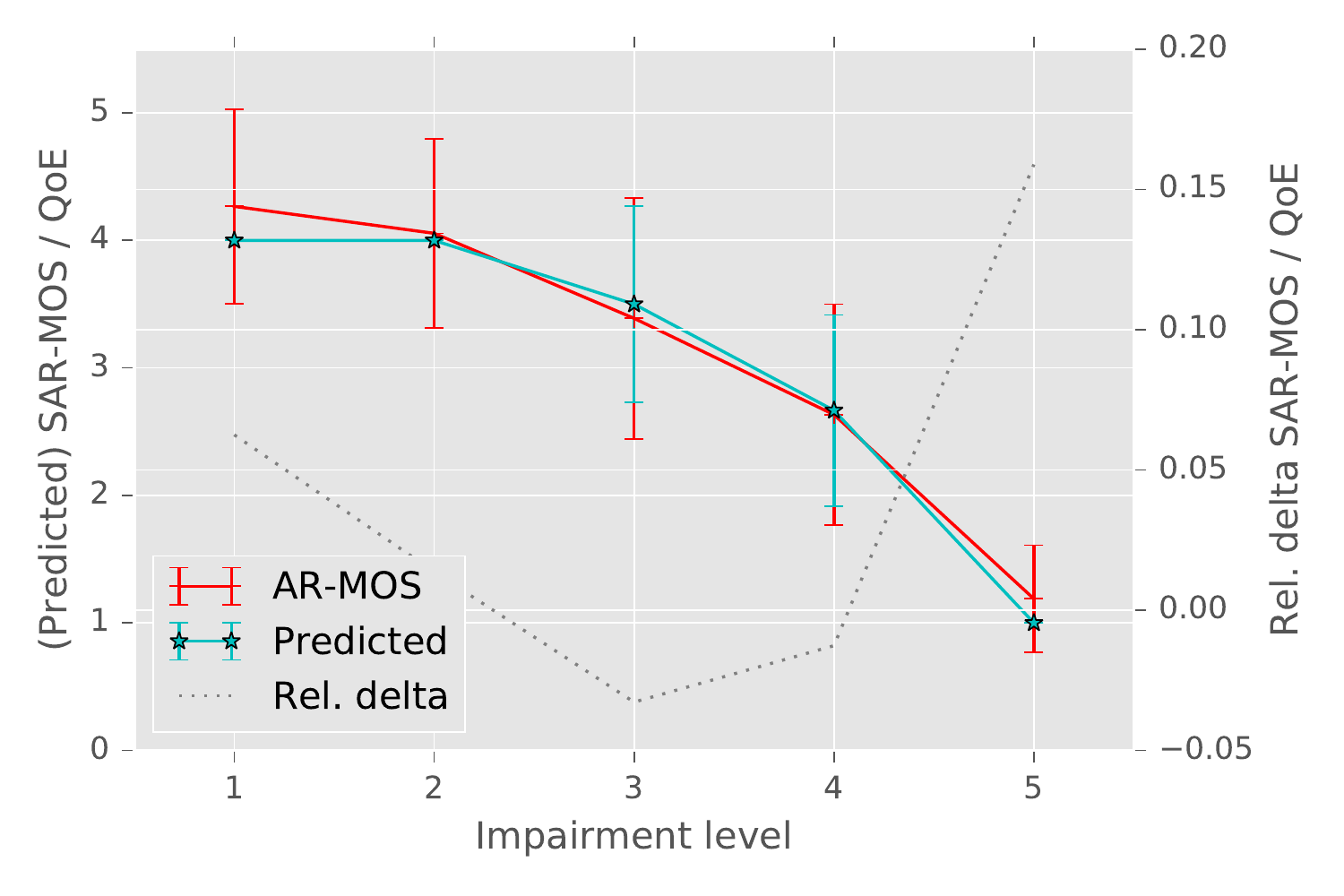}}
	\caption{Linear, bound linear, and logistic SAR-MOS prediction approaches for third degree $d=3$ underlying QoS metric.}
	\label{fig:SAR-all-logregd3}
\end{figure}

To evaluate whether a higher degree for the regression and prediction would yield better performance, we evaluated the second degree polynomial extension, with results illustrated in Figure~\ref{fig:SAR-all-logregd2}.
While overall comparable in following the subject-rated MOS and variabilities to the first degree scenario, we here note that the determination has increased for the logistic regression to $\hat{R}^2=0.592$, indicating an overall slight improvement.  Similarly, the reduced MSE and MAE values indicate a better fit.  The other two approaches do not exhibit any significant differences with the increased degree.
Lastly, we evaluate the case of $d=3$, with outcomes illustrated in Figure~\ref{fig:SAR-all-logregd3}.
Interestingly, we do not observe a significant increase in the performances for the different approaches.  In contrast, the performance of the logistic regression approach for the prediction exhibits a slightly worse result, a reversal effect.
Overall, we note that the increase to higher degree polynomials for the prediction approach does not result in significant performance increases beyond the second degree, for which the logistic approach showcases an increase in the performance.

\subsection{Prediction by Image}
With the common content dependency of the SAR-MOS and interdependency with the BRISQUE metric's values, we now investigate the individual images in a more detailed approach.  Specifically, we focus on the linear and logistic approaches at varying degrees $d$.  We provide the commonly utilized performance metrics in Table~\ref{tab:sar-images}.
\begin{table}
	\centering
	\caption{Overview of employed images and prediction results employing the non-referential BRISQUE image quality metric. Provided values include the $\hat{R}^2$ score, the mean squared error (MSE), the mean absolute error (MAE), and the median absolute error (MedAE). \label{tab:sar-images}}
	\begin{tabular}{|cl|r|r|r|r|r|r|r|r|}
		\hline
		$d$ & Image       & \multicolumn{4}{|c|}{Linear}  & \multicolumn{4}{|c|}{Logistic} \\
		  &             &    R2 &   MSE &   MAE & MedAE &    R2 &   MSE &   MAE &     MedAE \\ \hline
		1 & \textit{Garden}      & 0.698 & 0.518 & 0.619 & 0.579 & 0.480 & 0.893 & 0.627 &         0 \\
		  & \textit{Bamboo}      & 0.598 & 0.684 & 0.603 & 0.437 & 0.208 & 1.347 & 0.787 &         1 \\
		  & \textit{Mosque}      & 0.686 & 0.633 & 0.634 & 0.453 & 0.445 & 1.120 & 0.693 &         1 \\
		  & \textit{Ocean}       & 0.709 & 0.538 & 0.573 & 0.419 & 0.264 & 1.360 & 0.800 &         1 \\
		  & \textit{Golf}        & 0.671 & 0.464 & 0.547 & 0.516 & 0.500 & 0.707 & 0.573 &         1 \\
		  & \textit{Beach house} & 0.719 & 0.558 & 0.599 & 0.713 & 0.408 & 1.173 & 0.747 &         1 \\ \hline
		2 & \textit{Garden}      & 0.701 & 0.513 & 0.610 & 0.533 & 0.480 & 0.893 & 0.627 &         0 \\
		  & \textit{Bamboo}      & 0.600 & 0.680 & 0.610 & 0.443 & 0.459 & 0.920 & 0.653 &         1 \\
		  & \textit{Mosque}      & 0.712 & 0.581 & 0.587 & 0.509 & 0.657 & 0.693 & 0.533 &         0 \\
		  & \textit{Ocean}       & 0.710 & 0.535 & 0.566 & 0.356 & 0.546 & 0.840 & 0.600 &         0 \\
		  & \textit{Golf}        & 0.737 & 0.372 & 0.477 & 0.394 & 0.689 & 0.440 & 0.413 &         0 \\
		  & \textit{Beach house} & 0.739 & 0.518 & 0.603 & 0.600 & 0.603 & 0.787 & 0.573 &         0 \\ \hline
		3 & \textit{Garden}      & 0.701 & 0.513 & 0.609 & 0.534 & 0.480 & 0.893 & 0.627 &         0 \\
		  & \textit{Bamboo}      & 0.618 & 0.649 & 0.579 & 0.446 & 0.451 & 0.933 & 0.640 &         1 \\
		  & \textit{Mosque}      & 0.716 & 0.573 & 0.602 & 0.465 & 0.676 & 0.653 & 0.547 &         0 \\
		  & \textit{Ocean}       & 0.713 & 0.530 & 0.567 & 0.426 & 0.546 & 0.840 & 0.600 &         0 \\
		  & \textit{Golf}        & 0.754 & 0.347 & 0.432 & 0.263 & 0.689 & 0.440 & 0.413 &         0 \\
		  & \textit{Beach house} & 0.744 & 0.507 & 0.598 & 0.499 & 0.603 & 0.787 & 0.573 &         0 \\ \hline
	\end{tabular}

\end{table}
For the first degree, we notice that the linear prediction results in $\hat{R}^2$ scores above $0.59$ for all images, $\hat{R}^2=0.68$ on average.  This indicates an overall decent prediction fit.  This is corroborated by the error performance metrics, which each, on average, are below $0.6$.  The only outlier is the \textit{Bamboo} image, which exhibits the lowest $\hat{R}^2$ score paired with the highest MSE value.  For the logistic approach, we notice a lower score for each individual image, with an average of $\hat{R}^2=0.384$. Again, the \textit{Bamboo} image represents the lowest performing prediction outcome. The MSE on average is 1.1, indicating significant deviations between the predicted values and the SAR-MOS.  For this category-based approach, the MedAE additionally showcases that for most images, the prediction is off by one on average.  Overall, the $\hat{R}^2$ performance is 94\% better with a 47\% lower MSE for the linear approach versus the logistic one. 

Increasing the degree for the prediction does not result in significant increases for the linear approach, but the logistic prediction increases $\hat{R}^2$ performance by 60\% paired with a reduction of the MSE by 30\%.  In turn, the performance difference between both approaches is reduced to 24\% on average.  Most visible is the average error based on the median for the logistic approach, which for most images is reduced to 0, indicating a significantly better fit than for $d=1$.
Increasing the degree further to $d=3$ no longer yields significant improvements in the performance of each approach individually.  Some small improvements can be observed for individual images and performance indicators, but not on a significant level overall.

\subsection{Subjective Prediction}
Completing the investigation of employing the BRISQUE metric for QoE predictions, we now focus on the individual users, with performance values provided in Table~\ref{tab:sar-subjects}.
\begin{table}
	\centering
	\caption{Overview of individual subject-based prediction results employing the non-referential BRISQUE image quality metric. Provided values include the $R^2$ score, the mean squared error (MSE), the mean absolute error (MAE), and the median absolute error (MedAE) for degree $d=2$.\label{tab:sar-subjects}}
\begin{tabular}{|c|r|r|r|r|r|r|r|r|}
	\hline
	Subject & \multicolumn{4}{|c|}{Linear}  & \multicolumn{4}{|c|}{Logistic} \\
	        &    $\hat{R}^2$ &   MSE &   MAE & MedAE &    $\hat{R}^2$ &   MSE &   MAE &     MedAE \\ \hline
	  16    & 0.718 & 0.412 & 0.555 & 0.514 & 0.521 & 0.700 & 0.567 &       0.5 \\ \hline
	  17    & 0.779 & 0.452 & 0.573 & 0.573 & 0.723 & 0.567 & 0.500 &         0 \\ \hline
	  18    & 0.621 & 0.681 & 0.664 & 0.631 & 0.593 & 0.733 & 0.533 &         0 \\ \hline
	  19    & 0.827 & 0.333 & 0.425 & 0.281 & 0.810 & 0.367 & 0.300 &         0 \\ \hline
	  20    & 0.689 & 0.496 & 0.555 & 0.381 & 0.666 & 0.533 & 0.400 &         0 \\ \hline
	  21    & 0.672 & 0.671 & 0.624 & 0.531 & 0.593 & 0.833 & 0.567 &         0 \\ \hline
	  22    & 0.617 & 0.732 & 0.668 & 0.509 & 0.460 & 1.033 & 0.567 &         0 \\ \hline
	  23    & 0.630 & 0.676 & 0.649 & 0.443 & 0.526 & 0.867 & 0.600 &         0 \\ \hline
	  24    & 0.706 & 0.560 & 0.620 & 0.471 & 0.528 & 0.900 & 0.633 &       0.5 \\ \hline
	  25    & 0.771 & 0.374 & 0.466 & 0.345 & 0.673 & 0.533 & 0.467 &         0 \\ \hline
	  26    & 0.632 & 0.577 & 0.560 & 0.362 & 0.404 & 0.933 & 0.600 &         0 \\ \hline
	  27    & 0.548 & 0.530 & 0.552 & 0.450 & 0.432 & 0.667 & 0.467 &         0 \\ \hline
	  28    & 0.722 & 0.492 & 0.513 & 0.371 & 0.622 & 0.667 & 0.533 &         0 \\ \hline
	  29    & 0.602 & 0.689 & 0.657 & 0.515 & 0.595 & 0.700 & 0.500 &         0 \\ \hline
	  30    & 0.542 & 0.884 & 0.731 & 0.723 & 0.378 & 1.200 & 0.800 &         1 \\ \hline
\end{tabular}
\end{table}
We initially note that the approximation of user-dependent ratings based on linearly predicted BRISQUE values results in a fairly broad range of $\hat{R}^2$ values, from $\hat{R}^2=0.542$ for subject 30 to $\hat{R}^2=0.827$ for subject 19.  The lower model scores are, in turn, reflected in the higher MSE and MAE values that result from the less accurate capturing of variability.  The same relationship is obtained for the logistic approach, albeit at a lower level of $\hat{R}^2$ values ranging from $\hat{R}^2=0.378$ to $\hat{R}^2=0.81$ for the same test subjects.  On average, the logistic approach results in approximately 15\% lower $\hat{R}^2$ scores and 32\% higher MSE values.

\begin{figure*}
	\centering
	\subfloat[Subject 30, linear: $\hat{R}^2=0.542$]{\includegraphics[width=0.475\linewidth]{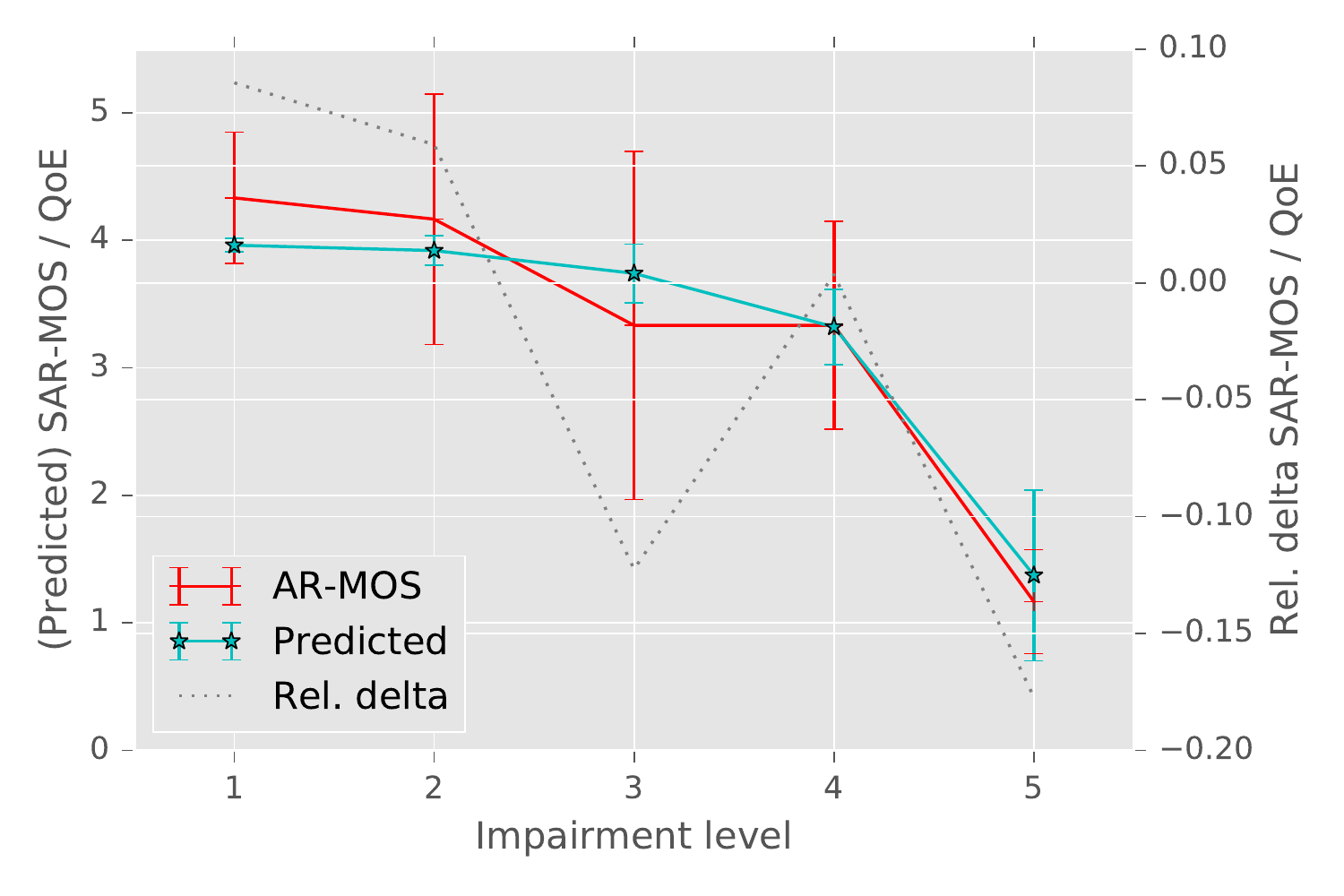}}
	\qquad
	\subfloat[Subject 19, linear: $\hat{R}^2=0.827$]{\includegraphics[width=0.475\linewidth]{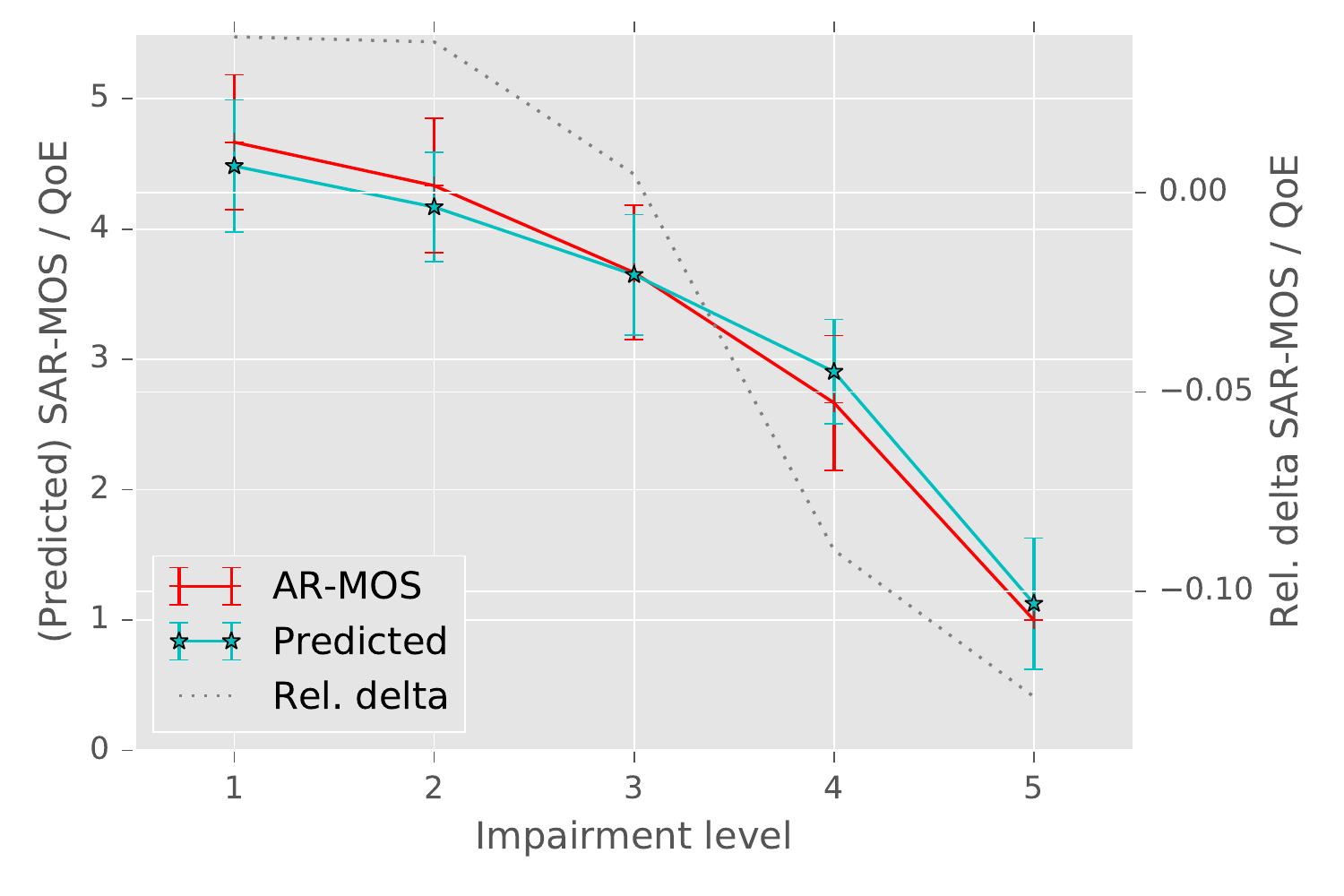}}
	\\
	\subfloat[Subject 30, logistic: $\hat{R}^2=0.378$]{\includegraphics[width=0.475\linewidth]{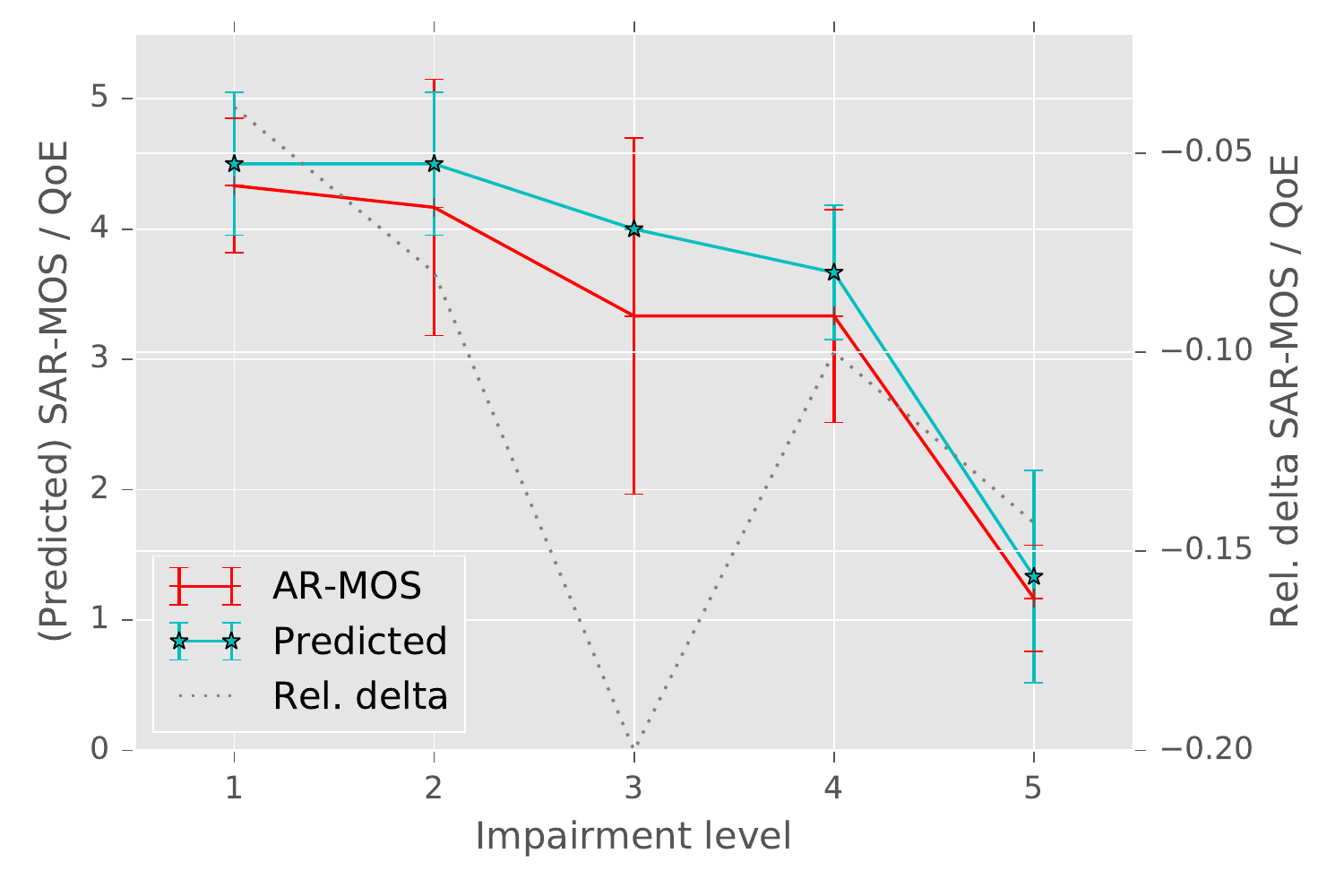}}
	\qquad
	\subfloat[Subject 19, logistic: $\hat{R}^2=0.810$]{\includegraphics[width=0.475\linewidth]{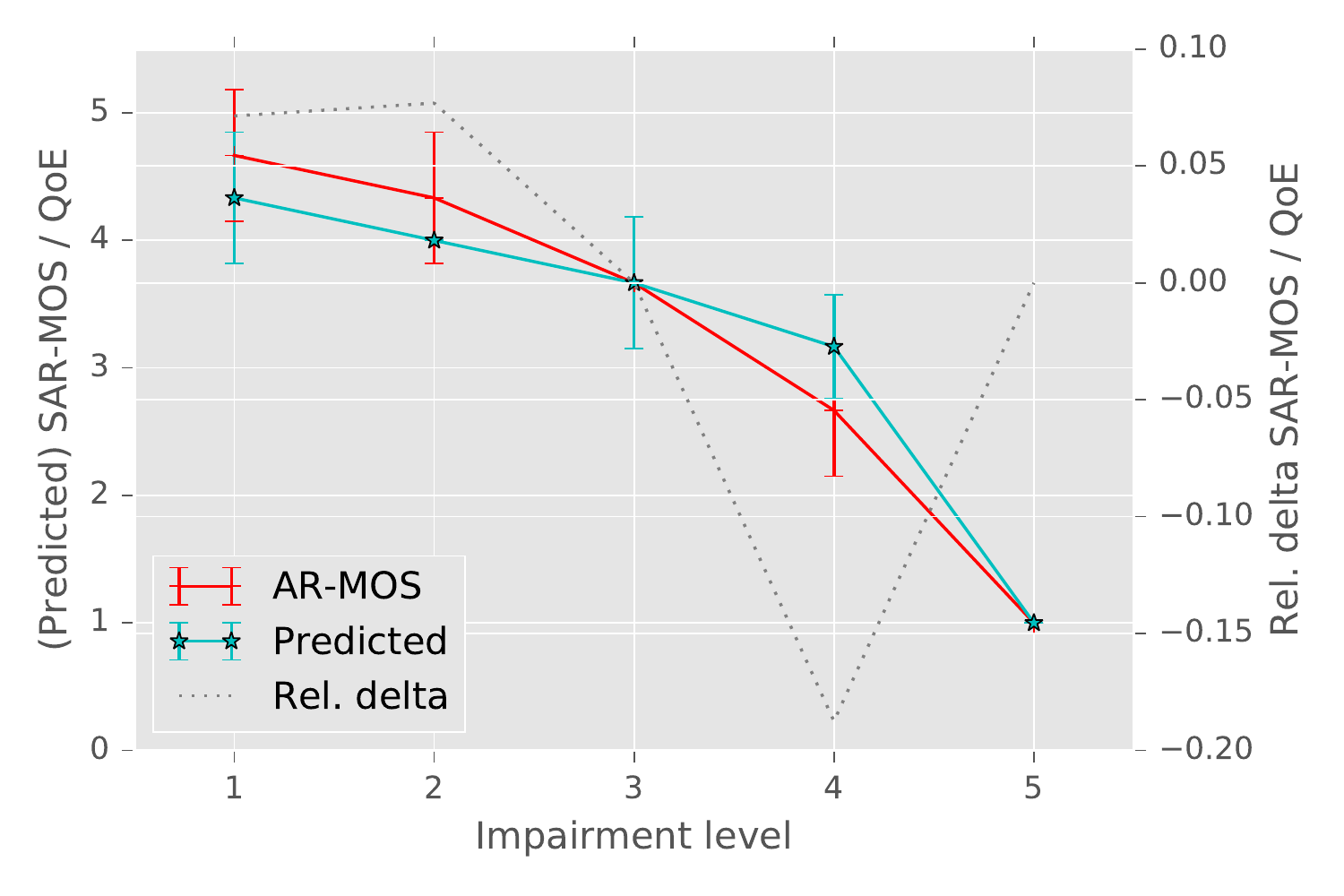}}
	\caption{Subject examples for lowest and highest attained $\hat{R}^2$ scores for linear and logistic predictions at degree $d=2$. }
	\label{fig:sar-subjects}
\end{figure*}

Here, a shortcoming of the $R^2$ score independent of other metrics comes into play, as a more detailed inspection of the actual predictions based on the different impairment levels shows that the overall subject-specific trends are nevertheless captured, as illustrated in Figure~\ref{fig:sar-subjects} for the highest and lowest $\hat{R}^2$ scores from Table~\ref{tab:sar-subjects}.
We note that on average, the individual predictions follow the overall trends for the individual users. Specifically, we observe that the linear approximation for subject 30 is only off by a maximum of 20 percent, which would equal approximately one rating level on a Likert-type scale.  For subject 19, we notice a very close following of the characteristic rating scale, but slight underestimation for low impairments, followed by an overestimation for higher impairment levels.  Overall, however, the linear approach here narrows the relative deviation between subject ratings and prediction further down.  For the logistic approach, we observe an average overestimation for subject 30, responsible for the lower $\hat{R}^2$ score attained. Interestingly, the logistic approach for subject 19 follows the same trend observed for the linear approach, but the deviations at low and high ends are more pronounced, reducing the $\hat{R}^2$ performance score in comparison.

\section{Prediction from EEG}
\label{s:predEEG}
In this section, we regard the prediction of the Quality of Experience based on EEG measurements obtained from consumer-grade off-the-shelf hardware. We initially consider ``regular'' images, i.e., those that are non-immersive, before evaluating spherical images.  Furthermore, we consider two main scenarios for the close to real-time direct prediction of the QoE, namely ($i$) with additional information for media consumption, e.g., gathered from monitored past multimedia experiences, and ($ii$) with an existing overall profile of the user under consideration.  We finally note that we employ the same prediction approach that we outlined prior for the BRISQUE metric in Section~\ref{s:predB}, except with a significantly increased parameter space.

\subsection{Traditional Image Quality Predictions}
We employ the image dataset that was previously used to evaluate the QoE (as AR-MOS) based on user ratings in~\cite{BaSe1701:Towards-Still,Se1601:Visual} for consistency and as grounded truth reference.  The images were part of the TID2013 reference data set with JPEG image compression impairments only, see~\cite{TID2013} for details concerning this reference data set.

\subsubsection{Prediction with Direct Media EEG} 
\begin{figure*}
	\centering
	\subfloat[EEG, QoE ${q}^u_{i,l}$]{\includegraphics[height=0.25\textheight]{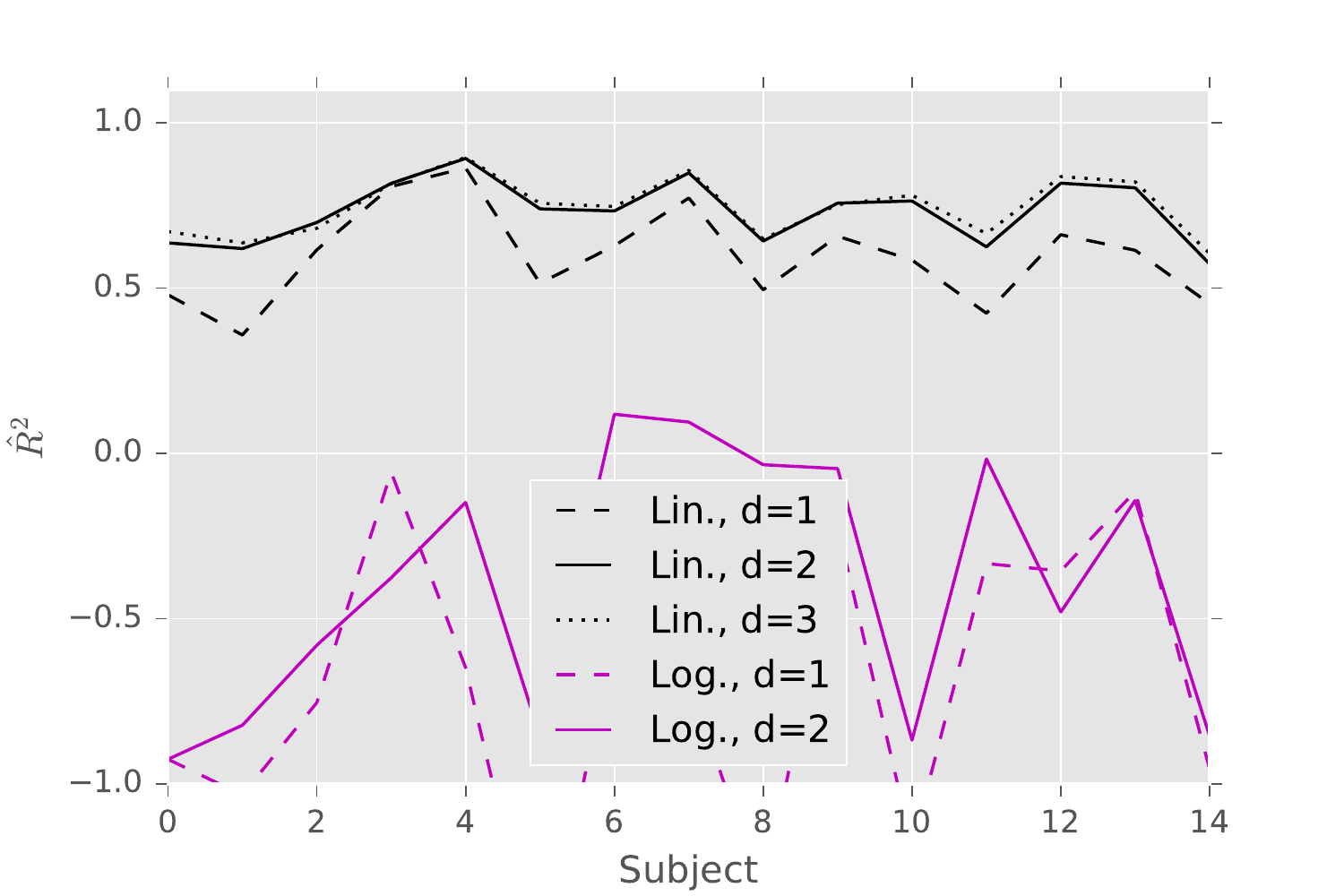}}
	\\
	\subfloat[EEG, QoS level ${l}$]{\includegraphics[height=0.25\textheight]{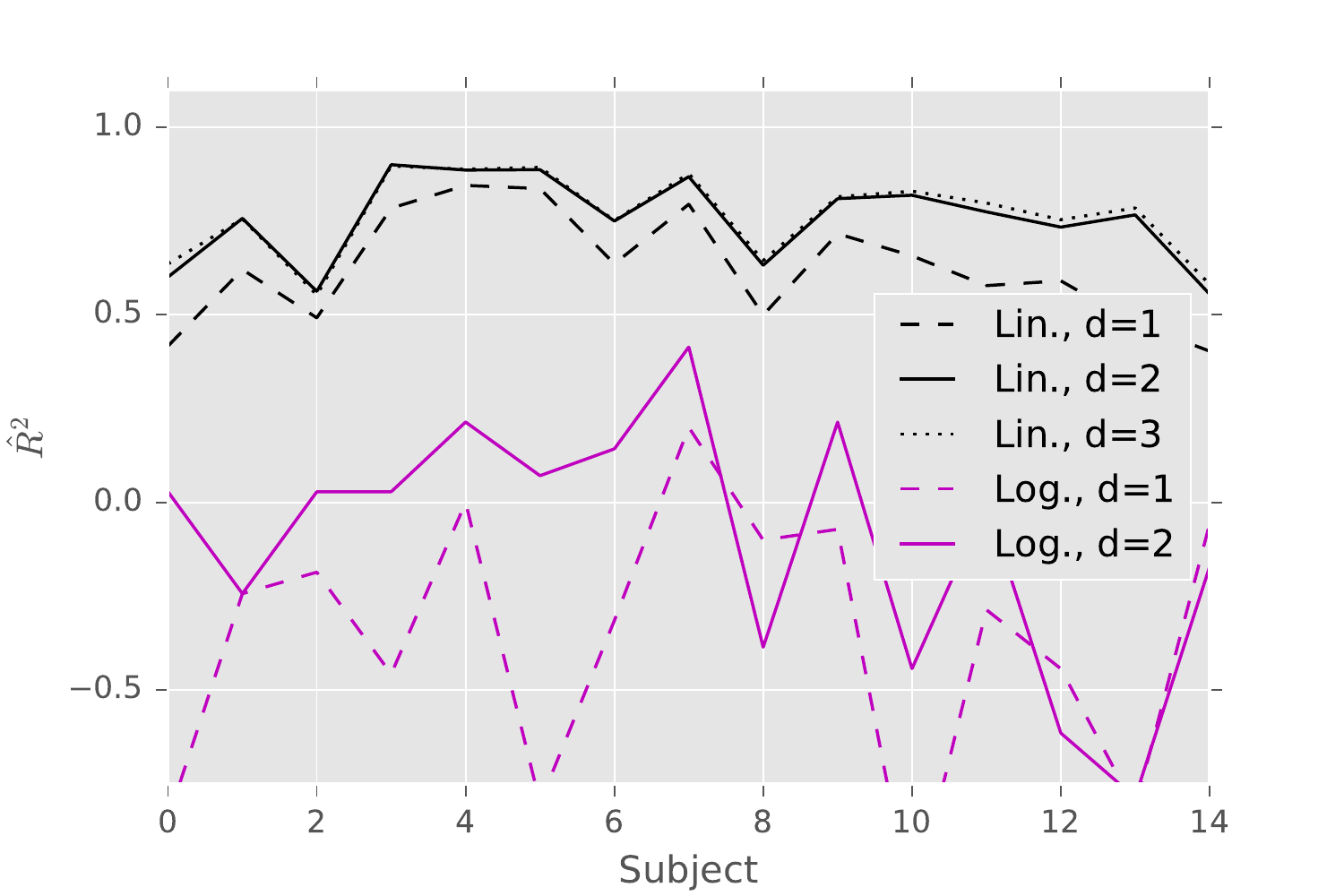}}
	\\
	\subfloat[EEG, BRISQUE value $B_{i,l}$]{\includegraphics[height=0.25\textheight]{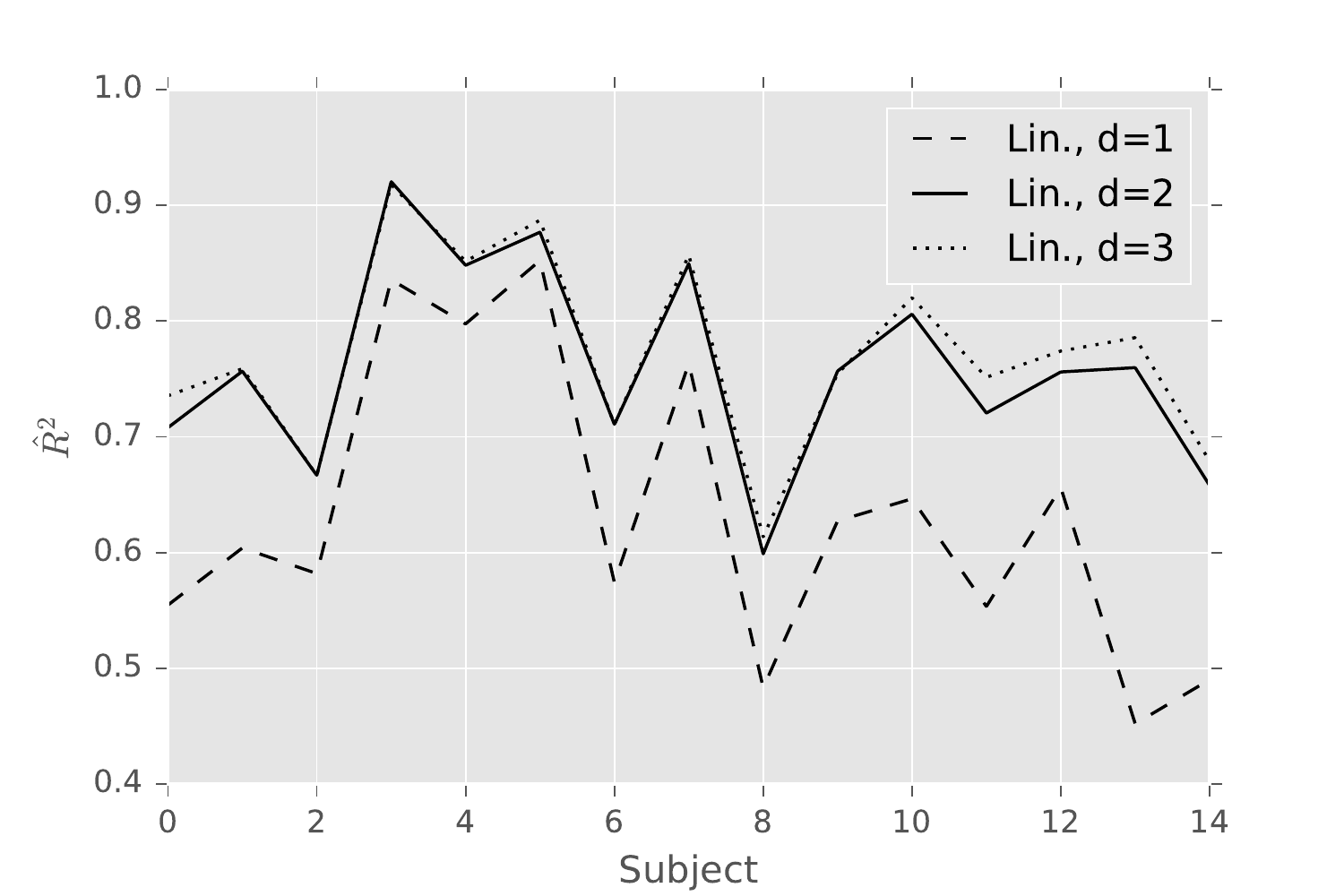}}
	\caption{Augmented Reality image quality prediction performance results as $\hat{R}^2$ for subject ratings ${q}^u_{i,l}$, image compression (QoS) levels ${l}$, and BRISQUE metric values ${B}_{i,l}$. Results are based on absolute EEG channel measurements.}
	\label{fig:ARPEEG}
\end{figure*}

The initial direct evaluation approach does only consider the power levels of the EEG signals for the different forehead positions and channels described in Section~\ref{s:approach}, whereby we split the pairs of EEG values and metric under consideration into a training set and a testing set which is used to calculate the prediction performance during the parameter approximation phase.  We illustrate the results from predictions in Figure~\ref{fig:ARPEEG}.
We initially observe that the linear prediction performance for the user-rated image quality (QoE) ranges from $\hat{R}^2=0.573$ to $\hat{R}^2=0.892$ if we consider the 2nd degree polynomial for the regression.  If only the linear regression without higher degree is employed, this range drops to $\hat{R}^2=0.358$ to $\hat{R}^2=0.863$, a noticeable decrease.  As similarly visible in ~\ref{fig:ARPEEG}, an increase in the prediction degree returns mixed results by subject, i.e., for some subjects the results improve, whereas for others the results decrease.  In turn, the second degree can be reasonably regarded as a sensible solution in terms of trade-off between prediction performance and complexity.  This relationship becomes more pronounced when considering the logistic regression as described in~\cite{YuHuLi11:Dual-coordinate}.  The regression and subsequent prediction would yield $\hat{R}^2$ scores from $\hat{R}^2=-1.605$ to $\hat{R}^2=-0.057$ for the first degree and $\hat{R}^2=-0.925$ to $\hat{R}^2=0.118$ for the second degree.  Clearly, the logistic regression, despite its general fit for ordinal scale prediction, does not capture the underlying relationship between averaged direct EEG band signal strengths and resulting subject votes.  However, in order to perform the prediction a significant amount of individually performed evaluations is required (growing exponentially with the degree), quickly increasing complexity and required time.

We now shift the view to the set image quality, i.e., the level determined oftentimes through objective image quality metrics (QoS).
For the illustrated $\hat{R}^2$ performance results in Figure~\ref{fig:ARPEEG}, we again note a considerable difference between the two regression/prediction approaches.  Compared to the QoE scenario, the logistic one performs somewhat better than in the rating prediction scenario, but still significantly lower than the linear one.  For the latter, we observe the previously identified improvement in $\hat{R}^2$ scores with an increase in the degree.  We again note that an increase from the second to higher degrees can have an adverse effect in some cases, while only slightly increasing others.  Paired with the significant increase in the computational complexity, these results indicate that it is generally not advantageous to consider degrees above the second one as viable for close to real-time application scenarios.
Comparing set and rated prediction performances, we observe that the $\hat{R}^2$ scores for the set predictions commonly outperform those for the rated ones. 

\begin{figure*}
	\centering
	\subfloat[Linear: QoE ${q}^u_{i,l}$]{\includegraphics[height=0.25\textheight]{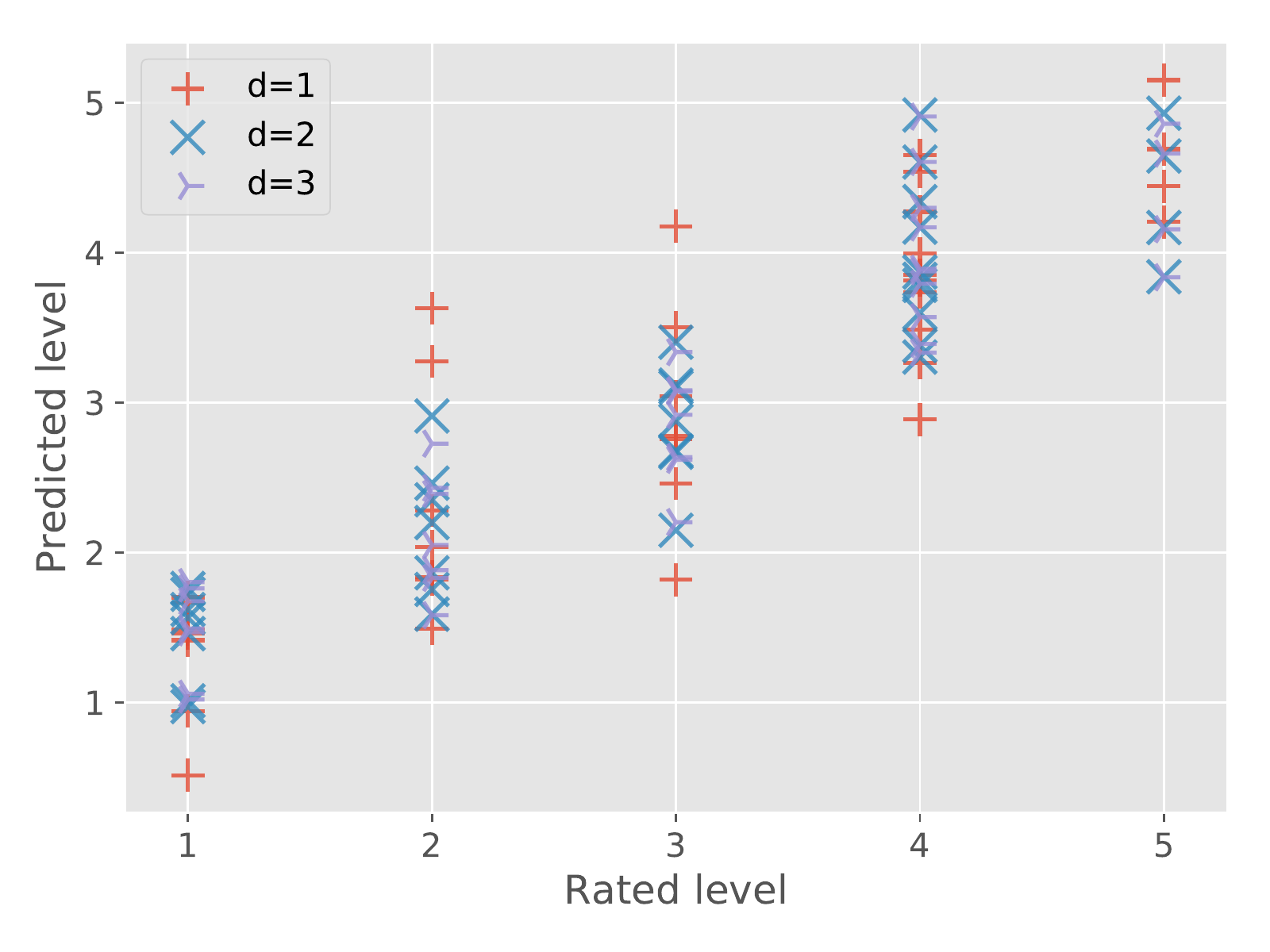}} 
	\qquad
	\subfloat[Logistic: QoE ${q}^u_{i,l}$]{\includegraphics[height=0.25\textheight]{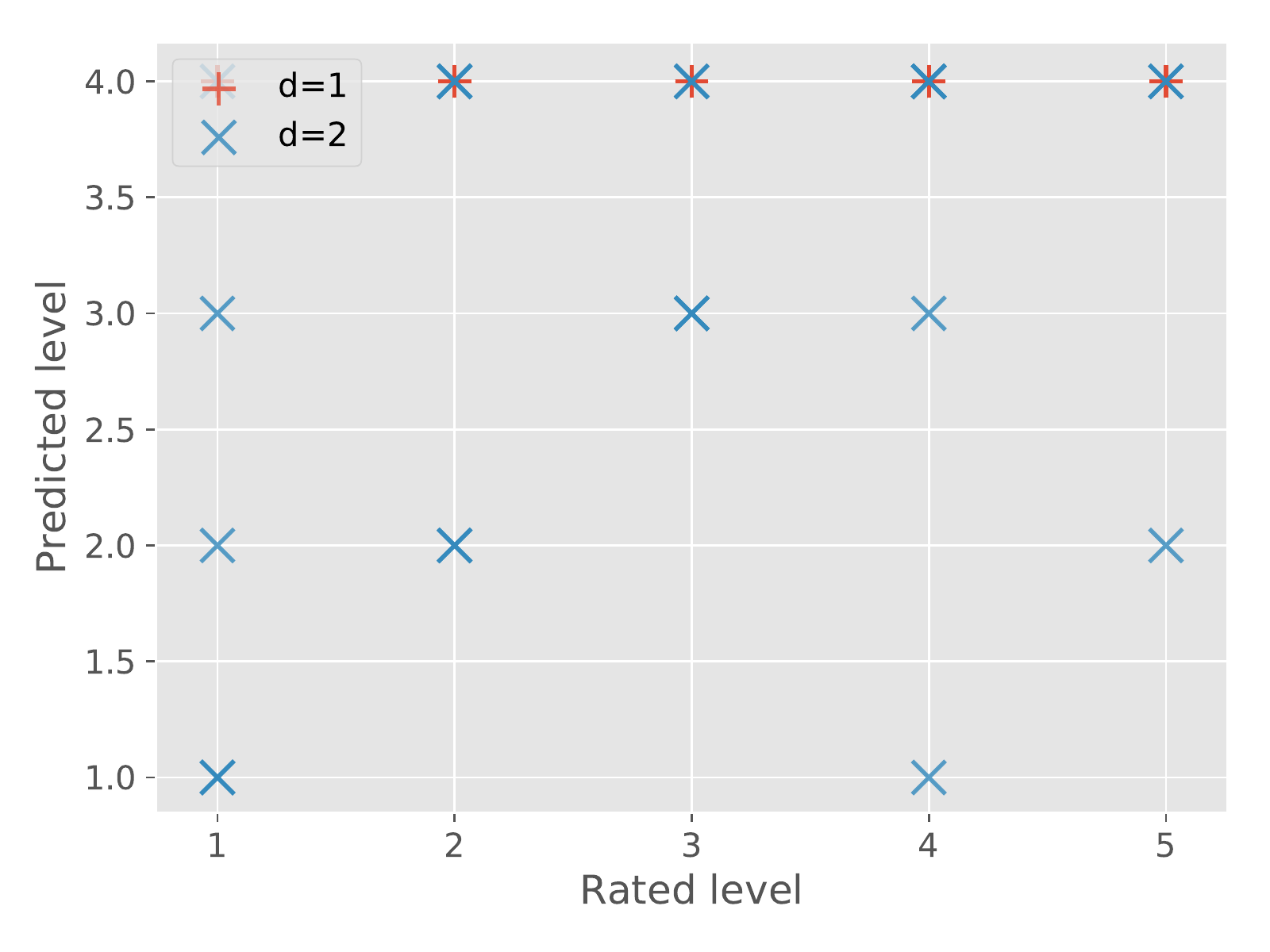}} 
	\\
	\subfloat[Linear: QoS level $l$]{\includegraphics[height=0.25\textheight]{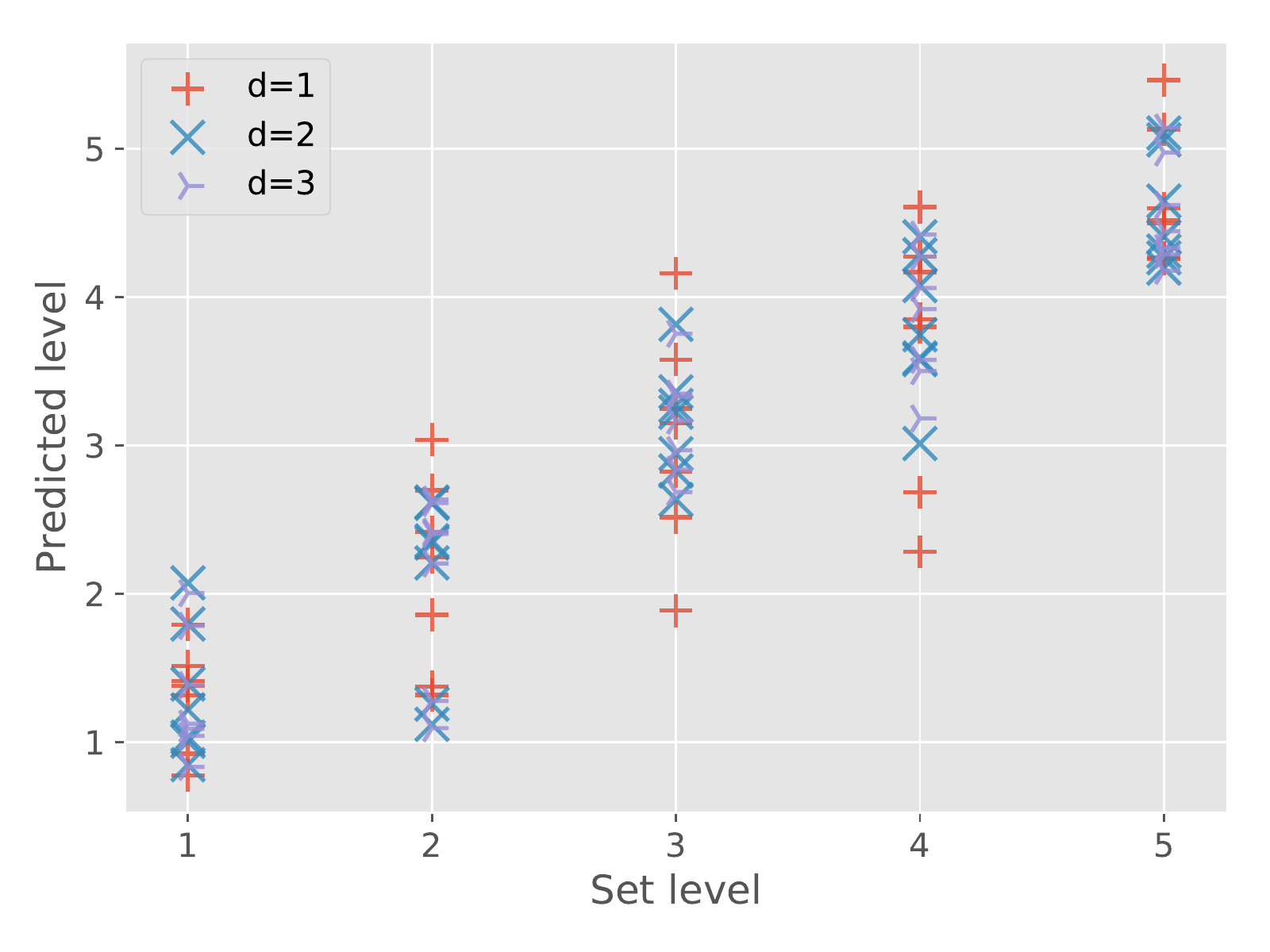}} 
	\qquad	
	\subfloat[Logistic: QoS level $l$]{\includegraphics[height=0.25\textheight]{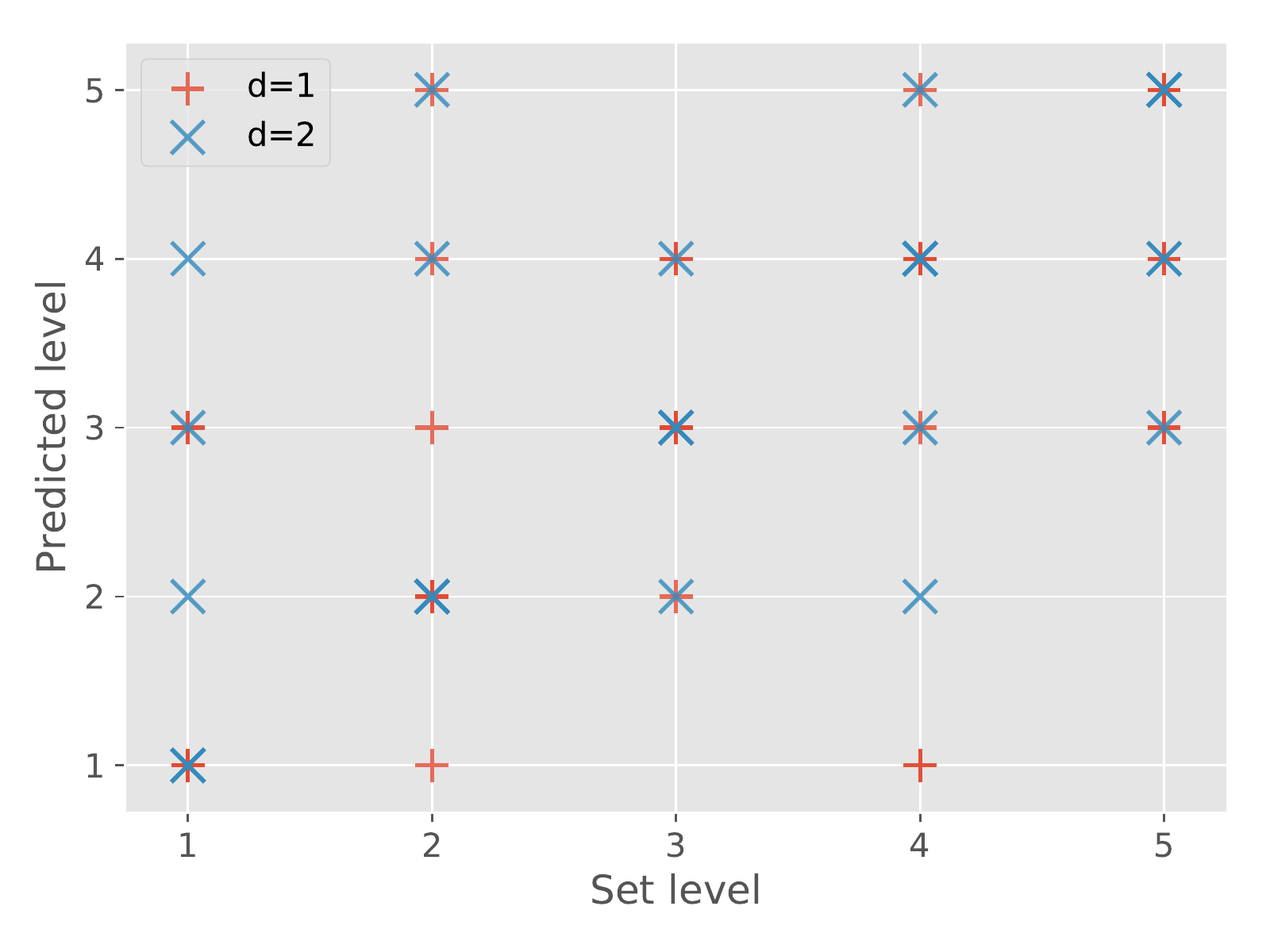}} 
	\\
	\subfloat[Linear: BRISQUE value $B_{i,l}$]{\includegraphics[height=0.25\textheight]{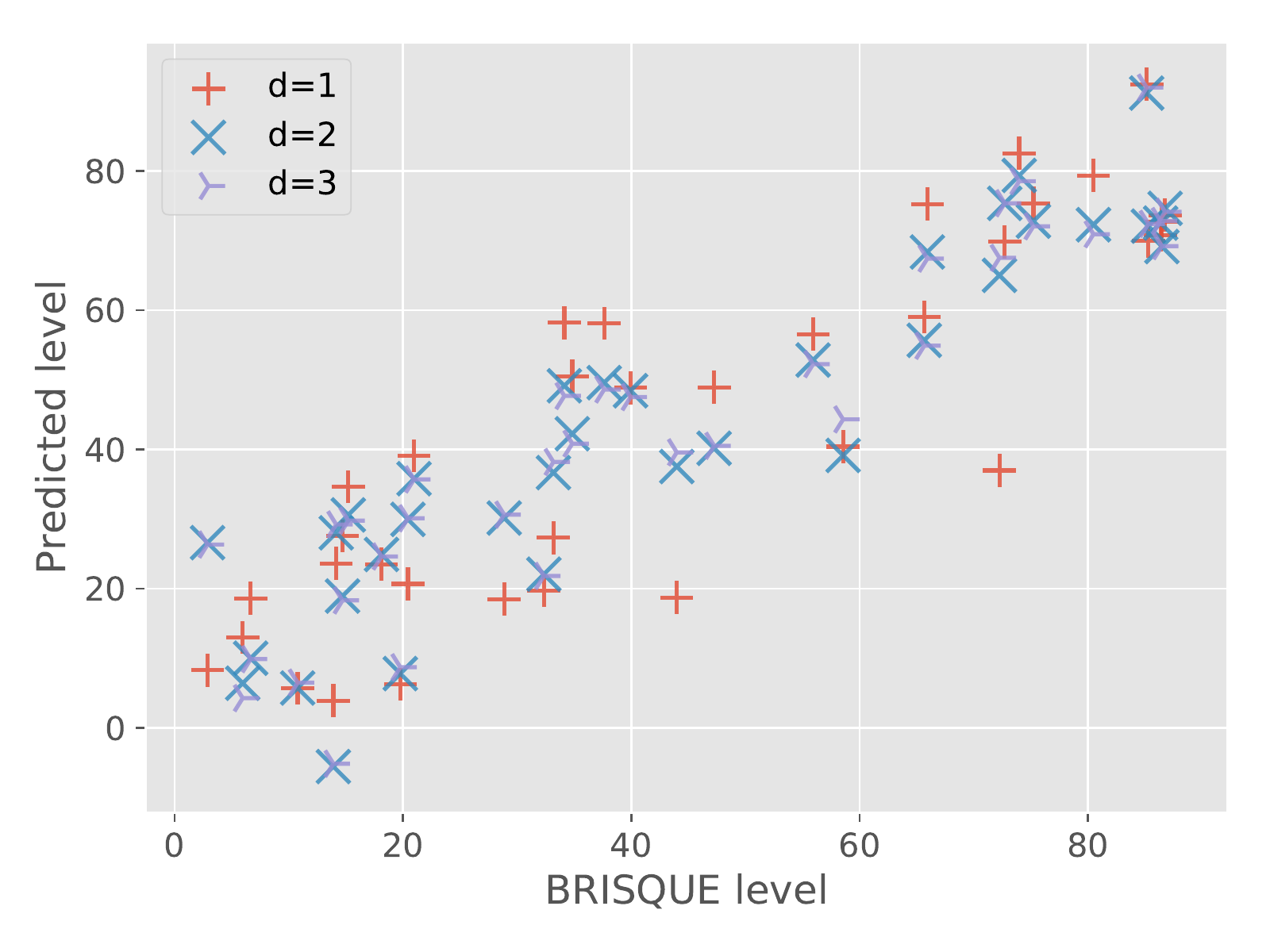}} 
	\caption{Augmented Reality image quality prediction results employing linear and logistic regression approaches for subject ratings ${q}^u_{i,l}$, image compression (QoS) levels $l$, and BRISQUE metric values $B_{i,l}$. Results are based on absolute EEG channel measurements for test subject 7.}
	\label{fig:AREEG-S7}
\end{figure*}

To shed further light on this, we additionally evaluate the linear prediction of the BRISQUE non-referential image impairment metric in Figure~\ref{fig:ARPEEG} as well.  We observe that the overall level is comparable to those observed for QoE and QoS values.  Some of the in-between subject differences, however, exhibit higher degrees of changes and could contribute to the overall performance, due to the QoS-QoE relationship we established earlier in Section~\ref{s:eval}.
An additional underlying reason for this behavior could be that users on the high and low ends of the image quality spectrum show little differentiation in their ratings (i.e., 4/5 and 1/2 ratings). To investigate the underlying mechanisms more closely, we illustrate the prediction details for an individual subject in Figure~\ref{fig:AREEG-S7}.

We initially observe for both, QoE and QoS linear predictions a fairly large spread around the targeted level. Typically, the predicted values tend to slightly overshoot their actual counterparts, leading to the overall deviations.  With higher degrees, on the other hand, the spread decreases, i.e., the values begin to more closely align towards a line.  Shifting the view to the logistic prediction, we note the discrete spread of estimated values.  For the initial degree, we furthermore notice a steady prediction outcome of four for the QoE, which minimized the underlying residuals and limits performance here.  An increase in the degree does result in a widespread estimation of QoE values, but does not yield the desired overall fit.  For the QoS levels, we notice that both degrees result in a wide spread from the correct prediction, rendering the overall accuracy low.  Lastly, we note that the prediction of BRISQUE values results in an initial wider spread around a linear trend, with higher degrees resulting in a convergence of the estimated values closer to the real ones.

\subsubsection{Prediction with User EEG Profiles}
Assuming now that we have a user-specific profile of EEG patterns available, we normalize the EEG values from beginning of the session recording to the end employ the averaged z-scored $\epsilon^z_{i,l}$ as in Equation~\ref{e:p500z} for each placement and channel as outlined in Section~\ref{s:approach} in greater detail.  To differentiate these pre-processed results from those previously employed, we utilize the term EEGz.

\begin{figure*}
	\centering
	\subfloat[EEGz, QoE $q^u_{i,l}$]{\includegraphics[height=0.275\textheight]{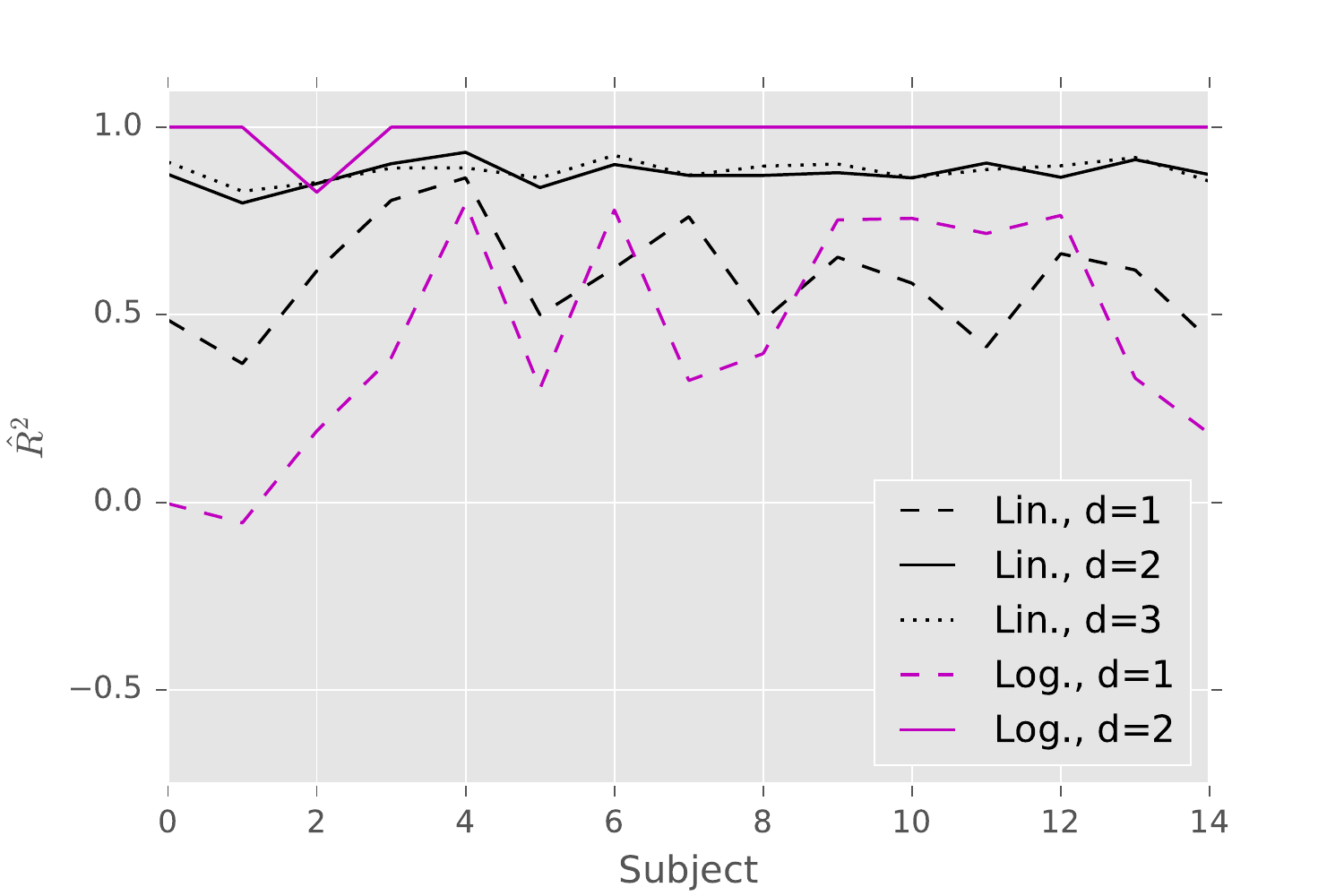}} 
	\\
	\subfloat[EEGz, QoS level $l$]{\includegraphics[height=0.275\textheight]{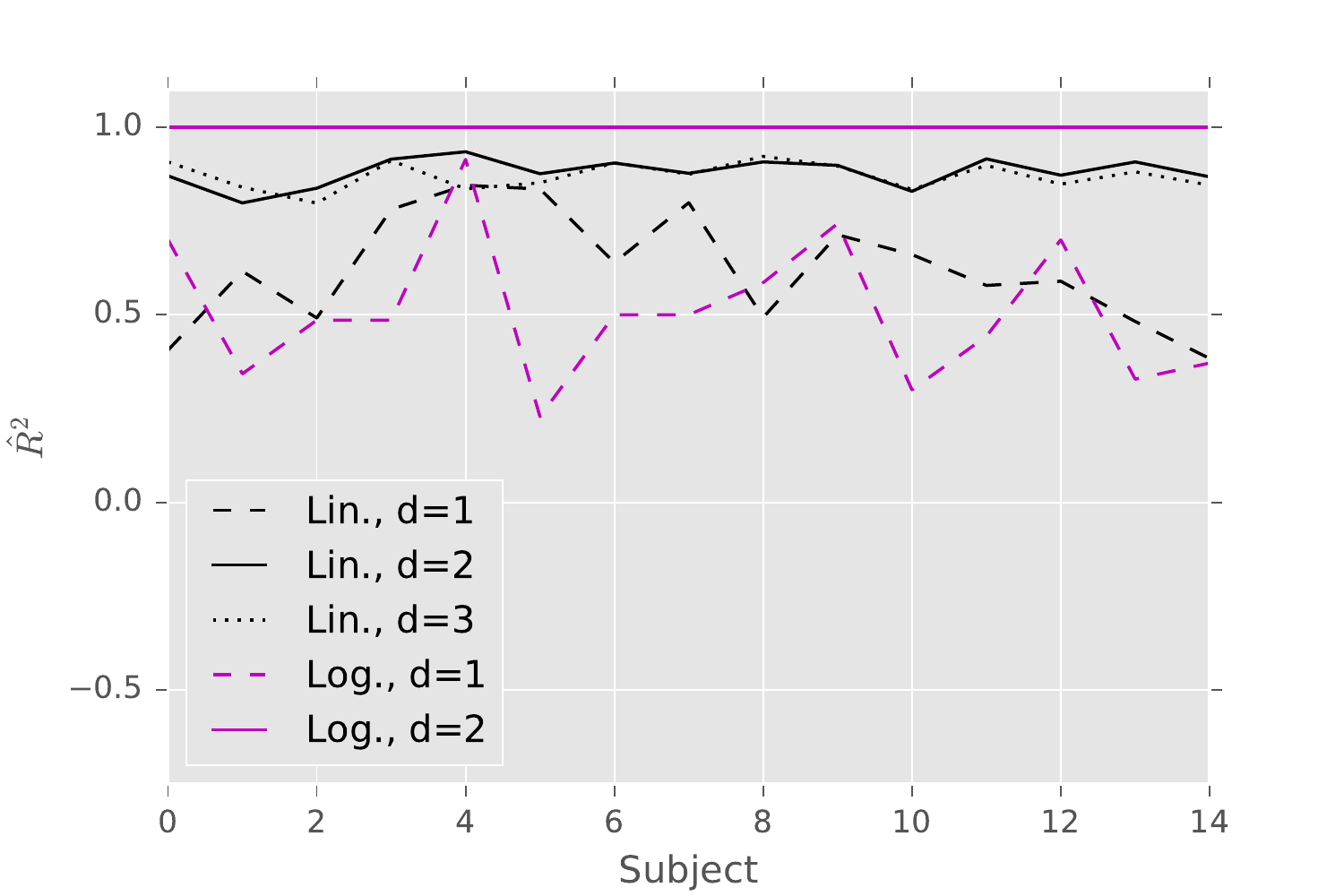}}	
	\\
	\subfloat[EEGz, BRISQUE value $B_{i,l}$]{\includegraphics[height=0.275\textheight]{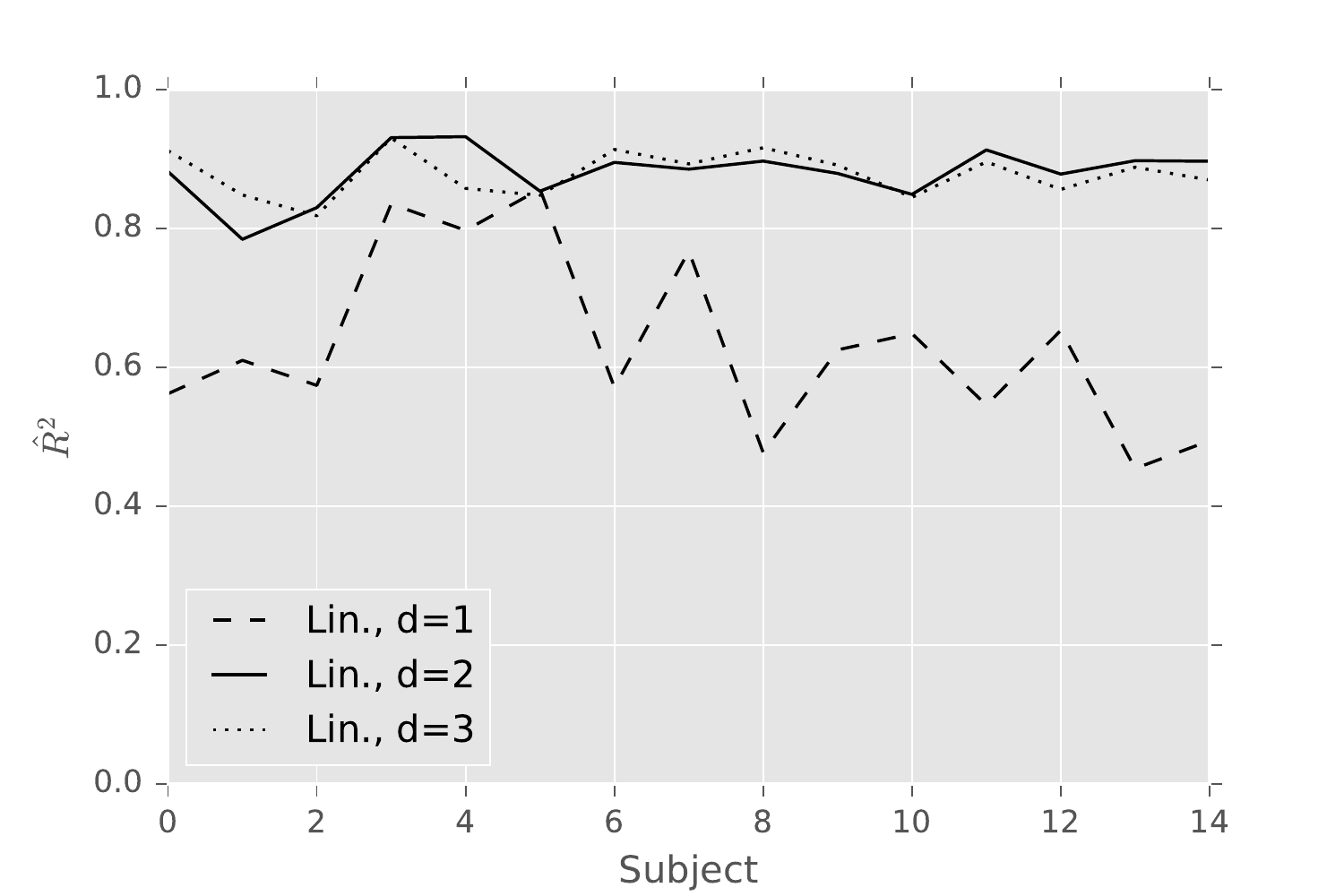}}
	\caption{Augmented Reality (AR) image quality prediction results for subject ratings (QoE) and image compression levels (QoS) based on z-score normalized EEG channel values (EEGz).}
	\label{fig:ARPEEGz}
\end{figure*}

We note that for each user, the amount of additional information at the beginning and end of the recording sessions might vary. In turn, the additional amount of noise introduced by these variations can be seen as a real-world practical scenario approach. We illustrate the $\hat{R}^2$ performance in Figure~\ref{fig:ARPEEGz}, which corresponds to the alternative approach in Figure~\ref{fig:ARPEEG}.

As for the raw EEG data, we initially consider the linear regression approach. 
For the first degree, we note a fairly variable result ranging from $\hat{R}^2=0.37$ to $\hat{R}^2=0.865$, depending on the subject
For increased degrees of two and three, this band narrows and increases to a range from $\hat{R}=0.798$ to $\hat{R}=0.933$ for degree $d=2$ and $\hat{R}=0.7829$ to $\hat{R}=0.925$ for $d=3$.
These EEGz values represent a significant increase over the raw EEG data and are high enough to approach a fairly close prediction of how user ratings are performed based on their EEGz values.
Next, we shift our view to the logistic regression approach.
For the first degree regression, we observe a significant increase over the raw EEG data scenario for the $\hat{R}^2$ scores attained.
However, the $\hat{R}^2$ scores for the first degree logistic regression and subsequent prediction still remain below those that could be achieved employing linear approaches with less computational effort.
For the second degree, however, we note that the $\hat{R}^2$ score is one for all but participating user 2.
In other words, if we employ our approach jointly with an existing EEG portfolio for a given subject, we are enabled to almost perfectly predict their rating of the image quality (i.e., within approx. 500 ms of image display, the subjective quality rating can be obtained without any user interference).
This is a significant achievement, as most wearable displays would allow for dry EEG electrode connections on the forehead and thus enable direct feedback loops of the AR-QoE.

We now shift our view to the prediction of the set image quality levels $l$. 
We again note that the linear approach yields significant increases in $\hat{R}^2$ values by increasing the degree, but has a mixed return on complexity when increasing the degree further to three.
In line with prior observations for the results for subjective ratings, we note that the EEGz-based approach yields overall better results (as manifested in the attained higher $\hat{R}^2$ scores) than raw EEG values would permit.
For the logistic regression-based prediction, we observe a similar behavior of better performance with higher degrees.
Specifically, the second order expansion captures the entire underlying variability as indicated in a perfect $\hat{R}^2$ score of one, indicating perfect prediction.
In other words, based on logistic regression, we could with a very high accuracy predict how EEGz and objective image quality levels would correlate.

\begin{figure*}
	\centering
	\subfloat[Linear: QoE $q^u_{i,l}$]{\includegraphics[height=0.25\textheight]{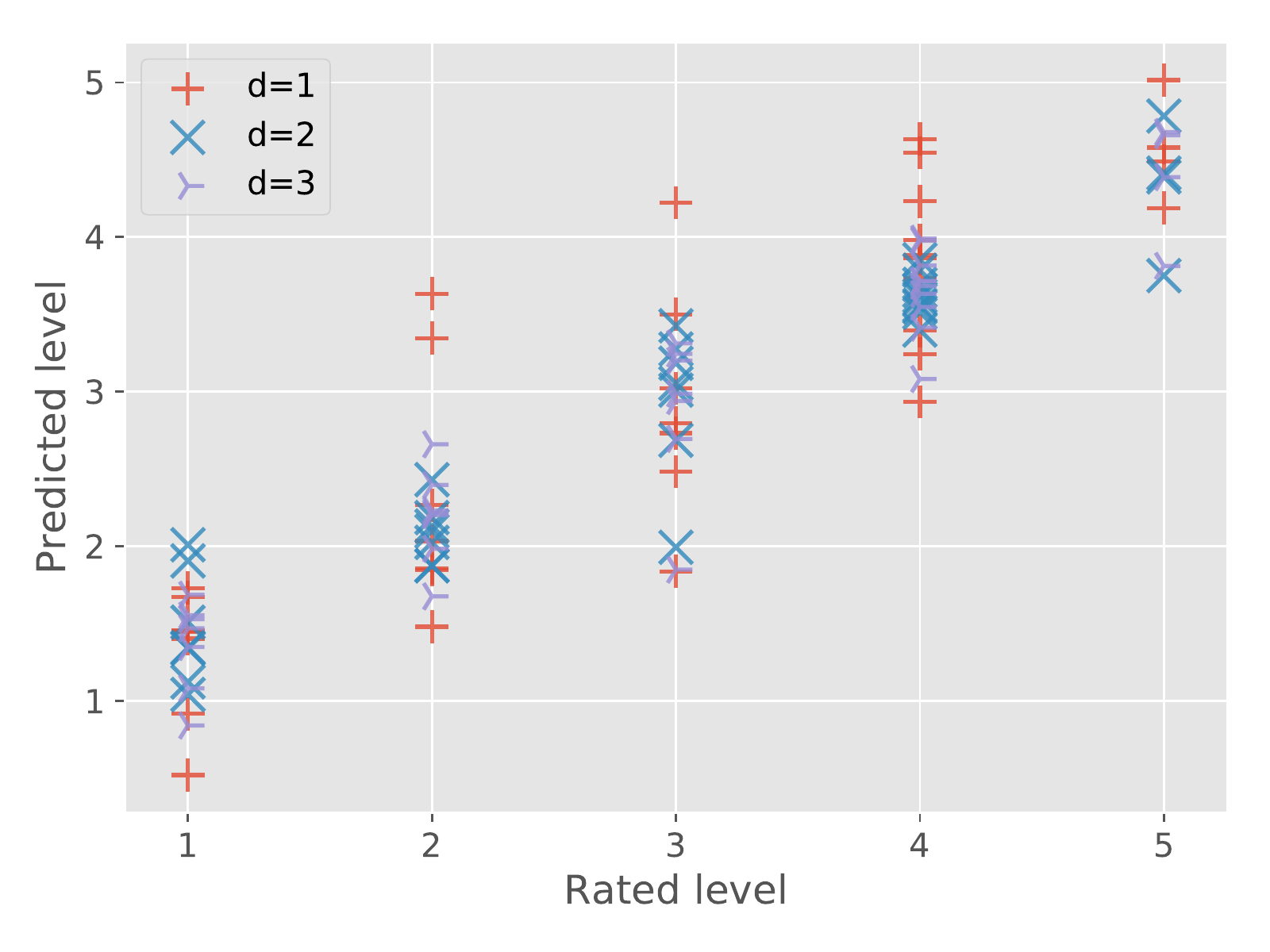}} 
	\qquad
	\subfloat[Logistic: QoE $q^u_{i,l}$]{\includegraphics[height=0.25\textheight]{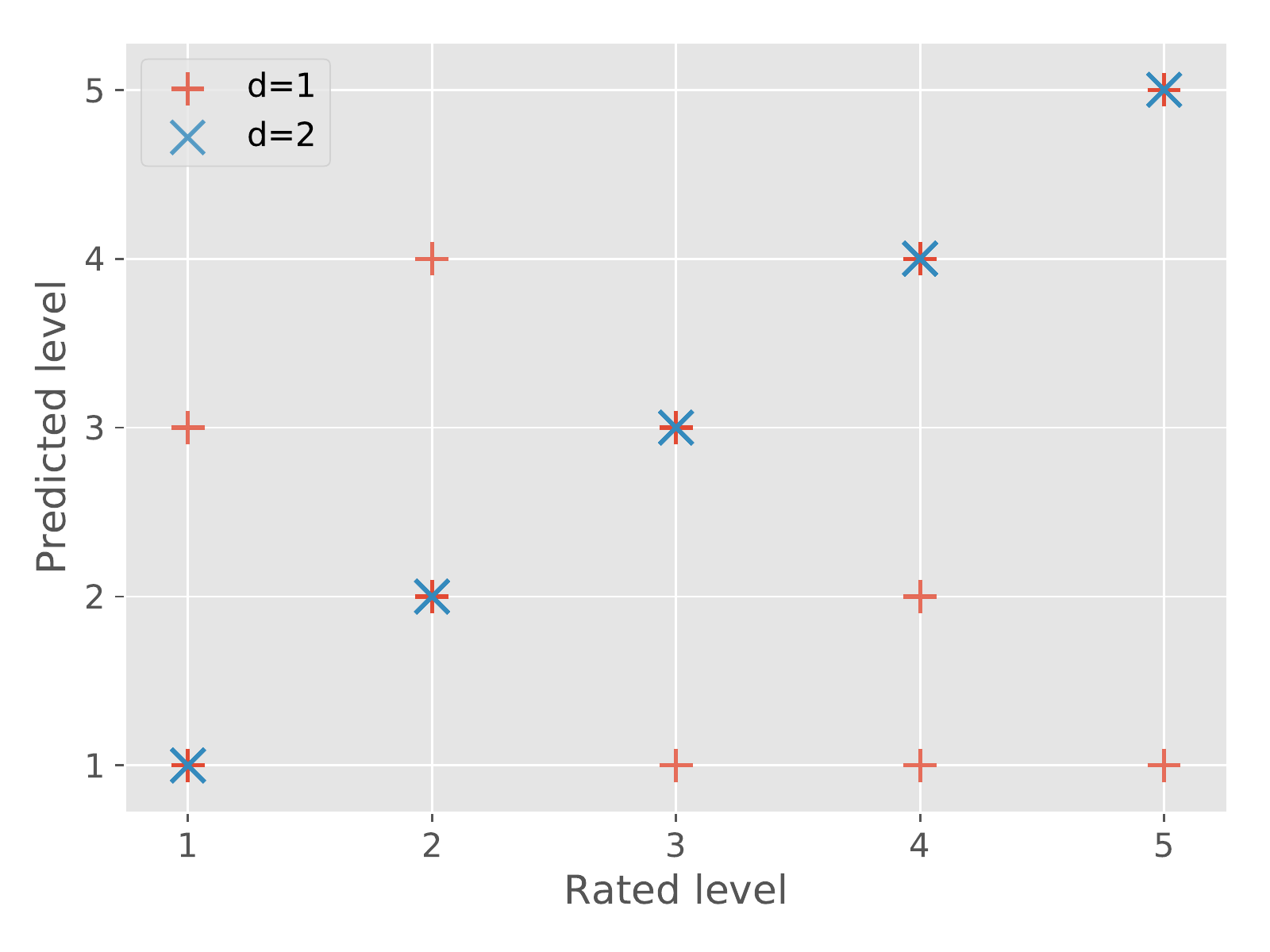}} 
	\\
	\subfloat[Linear: QoS level $l$]{\includegraphics[height=0.25\textheight]{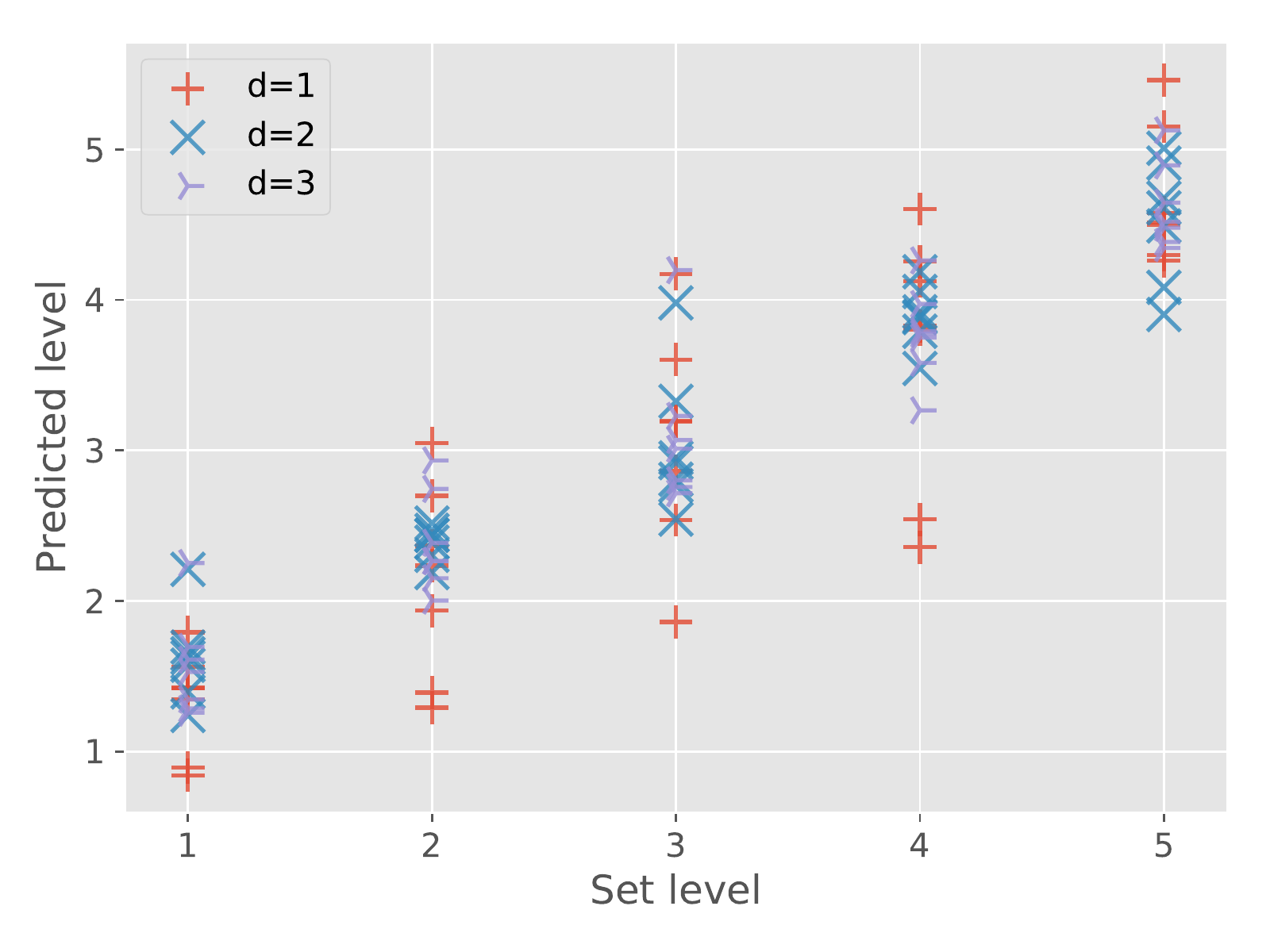}} 
	\qquad	
	\subfloat[Logistic: QoS level $l$]{\includegraphics[height=0.25\textheight]{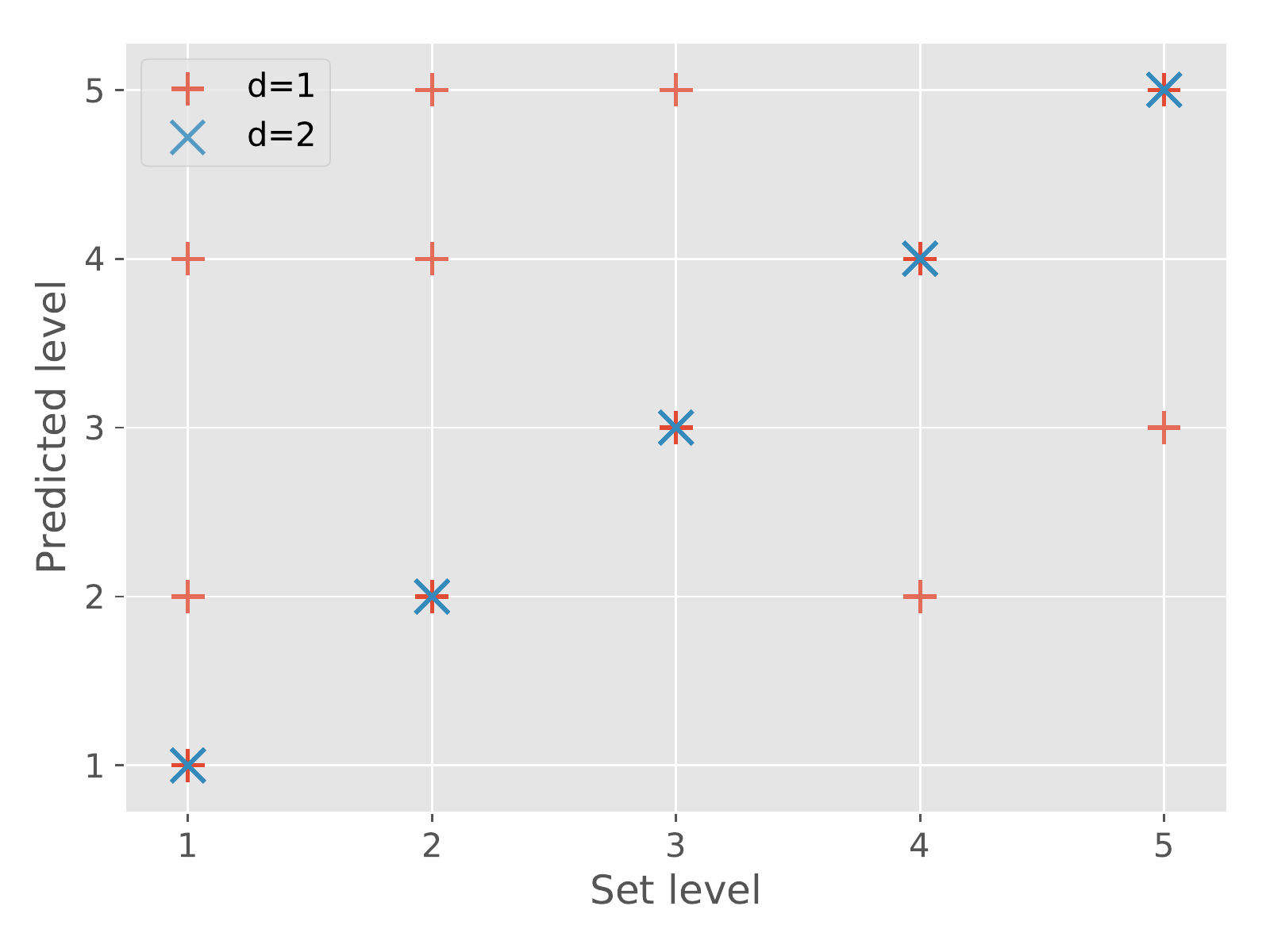}} 
	\\
	\subfloat[Linear: BRISQUE value $B_{i,l}$]{\includegraphics[height=0.25\textheight]{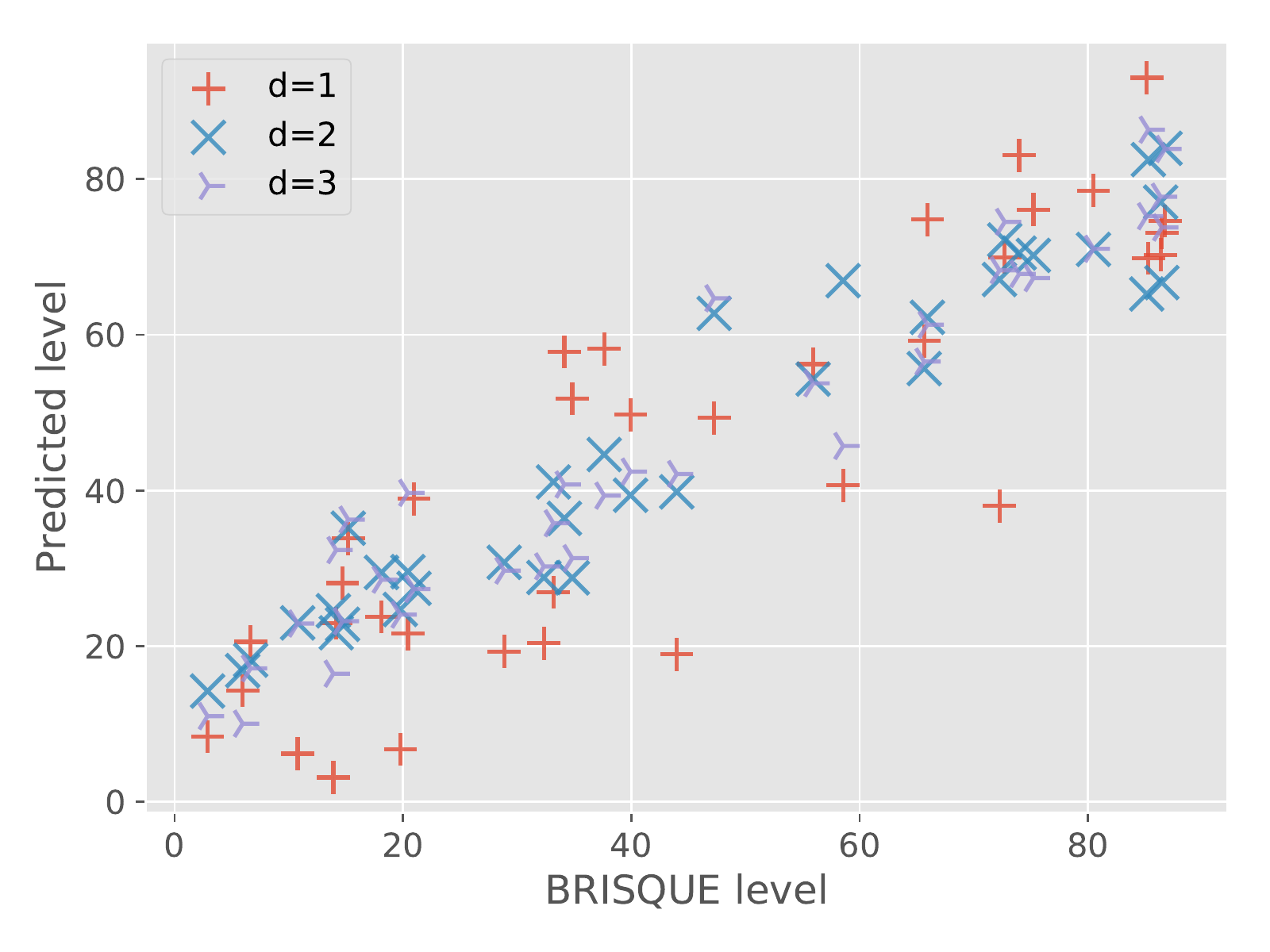}} 
	\caption{Augmented Reality (AR) image quality prediction results for subject ratings $q^u_{i,l}$, image compression (QoS) levels $l$, and BRISQUE metric values $B_{i,l}$.  Results are based on z-score normalized EEG channel measurements (EEGz)  for test subject 7.}
	\label{fig:AREEGz-S7}
\end{figure*}

We showcase the impact this approach has on individual users again for the same subject's EEGz values in Figure~\ref{fig:AREEGz-S7}.
Similar to our prior observations for QoS and QoE predictions employing linear approaches, we observe a declining spread of value deviations from the perfect prediction as degrees increase for both scenarios.  Likewise, the overall prediction of the BRISQUE metric from EEGz values exhibits the same trend (exhibiting a smaller spreading factor as well), which increases performance.  The logistic approach for the QoE prediction yields $\hat{R}^2=0.325$ in the first degree with small error.  With the second degree, however, the prediction of the QoE can be fully performed.  We note that the same underlying degree dependency is exhibited for the QoS predictions with $\hat{R}^2=0.5$ and $\hat{R}^2=1.0$ for first and second degrees, respectively.

\subsection{Spherical Image Quality Predictions}
Motivated by the findings for the traditional image display in the general field of view, we now extend the prediction to the spherical image QoE, denoted as SAR-QoE.  The main difference to the regular image configuration is that images were immersive as described in Section~\ref{s:approach} and participating subjects were free and encouraged to rotate their heads in all directions, as to determine the overall image quality experience. 

\subsubsection{Prediction with Direct Media EEG} 
\begin{figure*}[b!]
	\centering
	\subfloat[EEG, QoE $q^u_{i,l}$]{\includegraphics[width=0.475\linewidth]{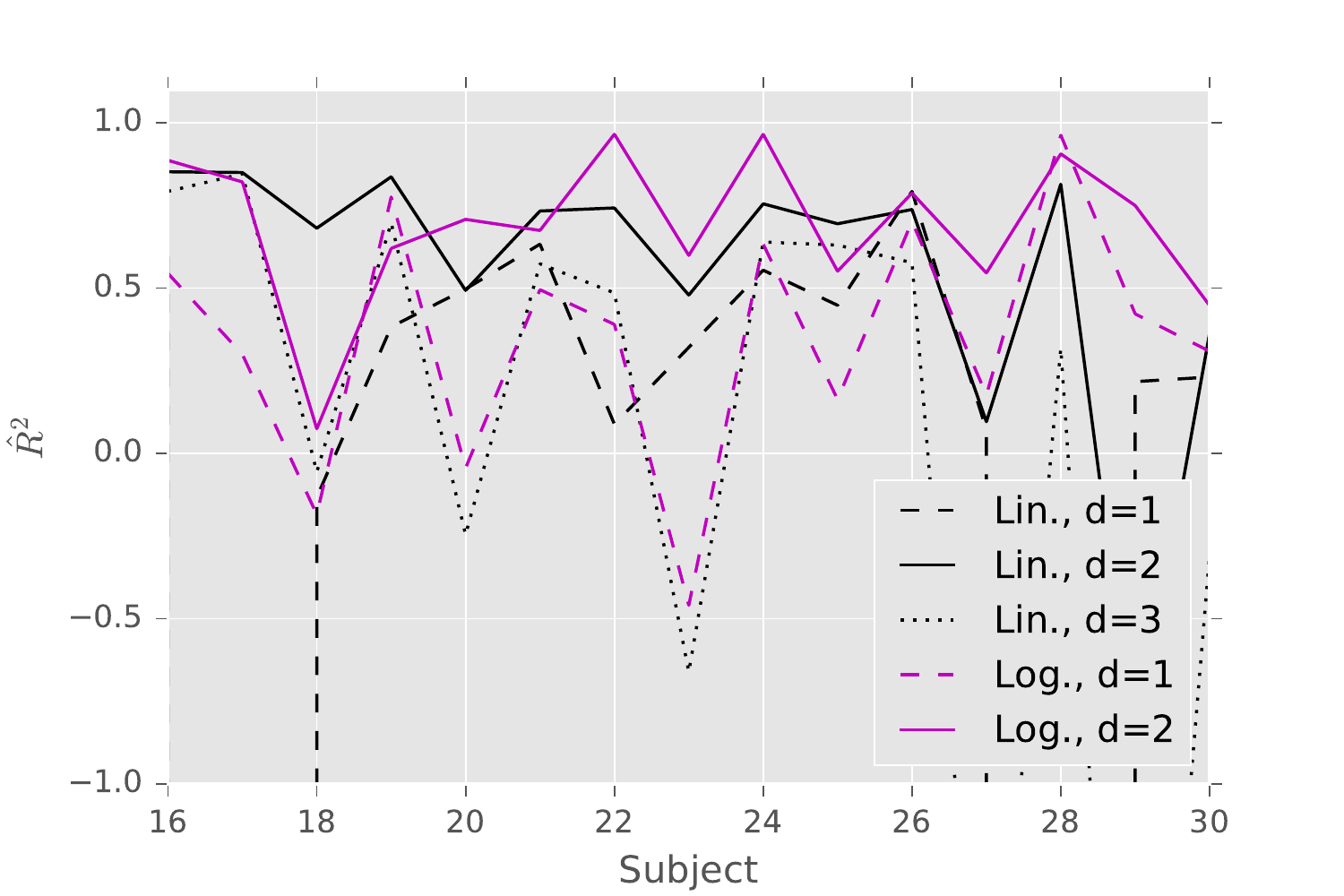}}
	\\
	\subfloat[EEG, QoS level $l$ ]{\includegraphics[width=0.475\linewidth]{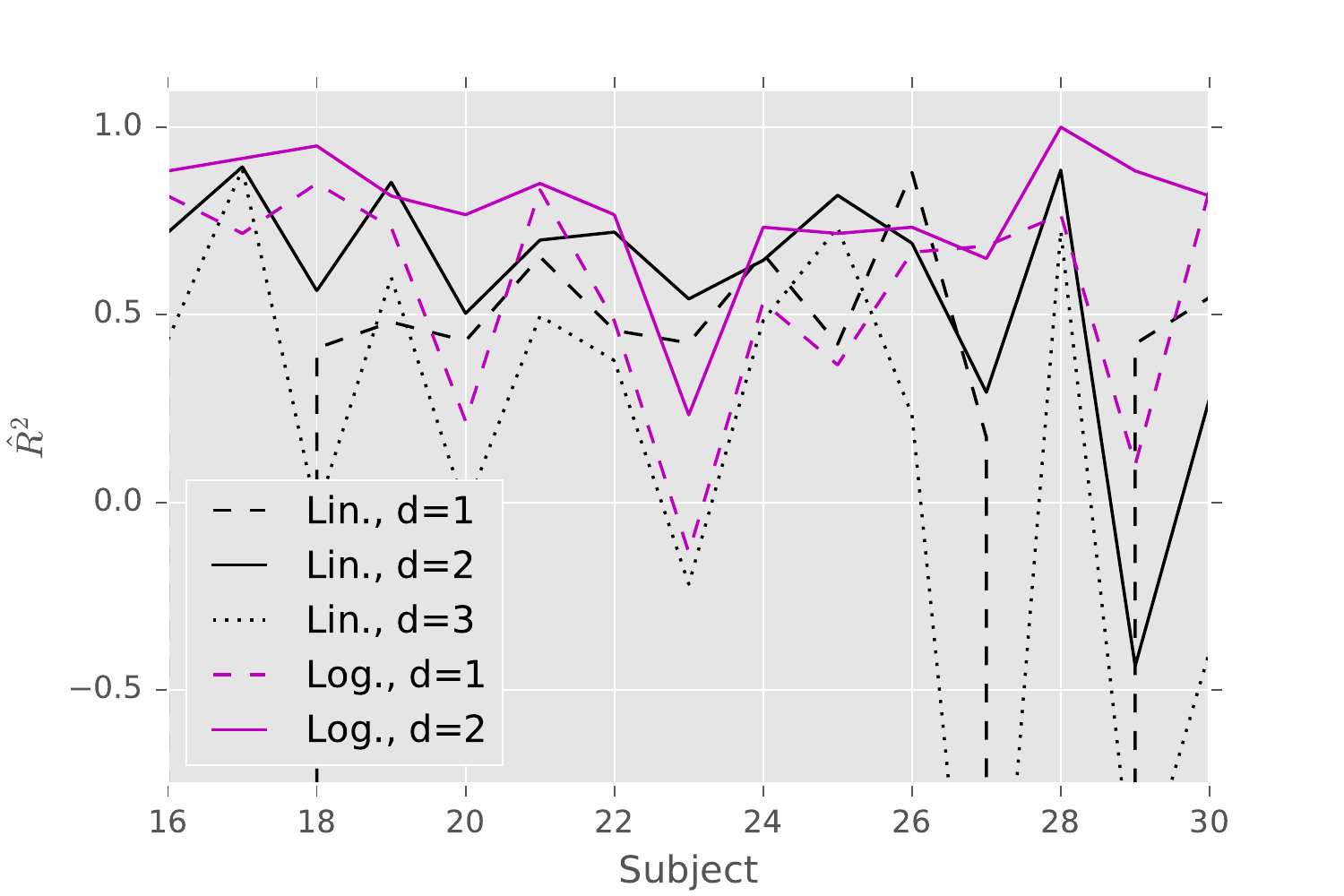}}
	\\    
	\subfloat[EEG, BRISQUE value $B_{i,l}$]{\includegraphics[width=0.475\linewidth]{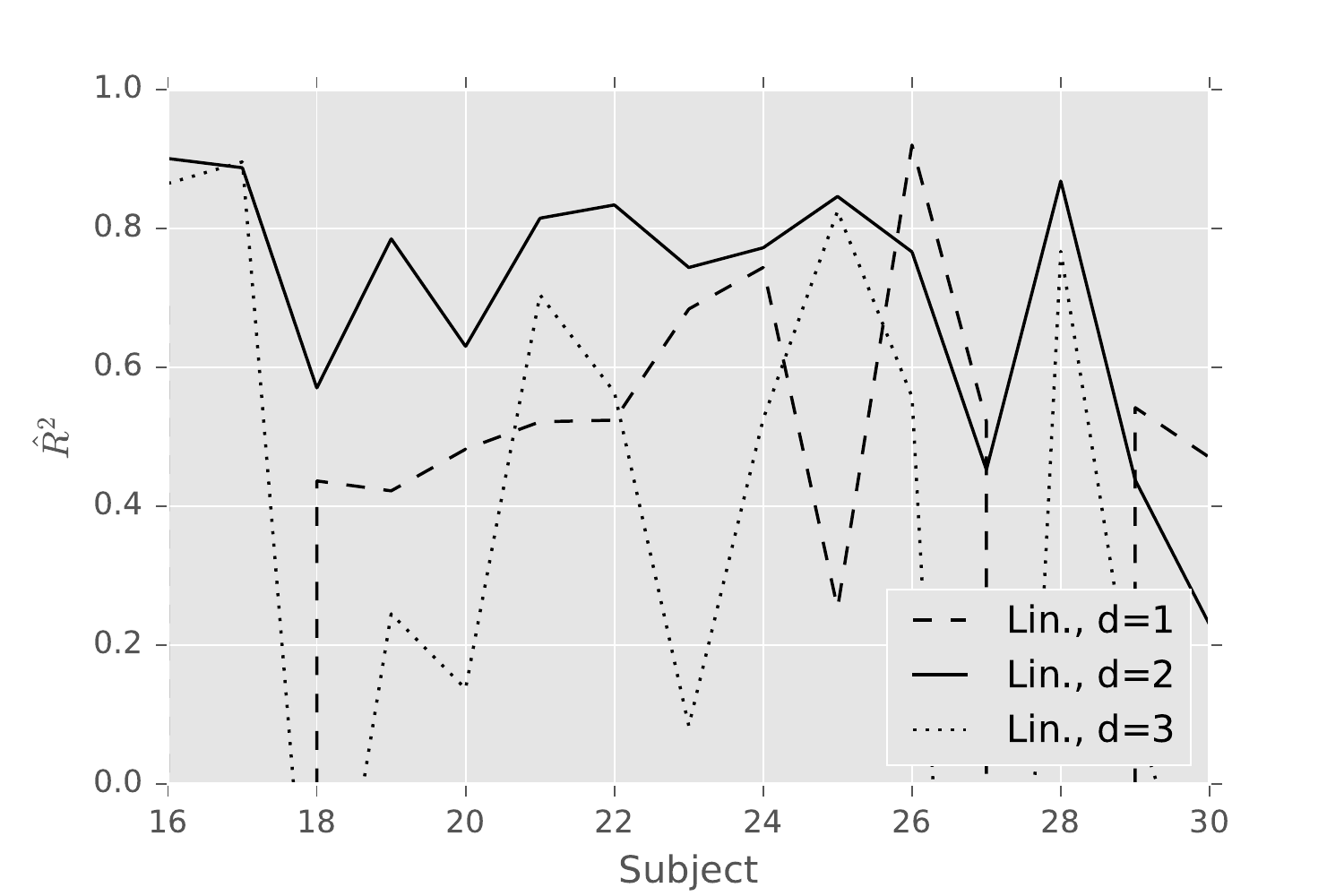}}
	\caption{Spherical AR (SAR) image quality prediction results for subject ratings $q^u_{i,l}$, image compression (QoS) levels $l$ (cmp. Figs. \ref{fig:SAR-all}, \ref{fig:SAR-all-linreg}), and BRISQUE metric values $B_{i,l}$. Results are based on absolute EEG channel measurements (EEG).}
	\label{fig:SARPEEG}
\end{figure*}

Based on the data gathered, we perform the same prediction approaches previously employed for regular images, taking the single measurement EEG into account. The resulting outcomes for the SAR values are illustrated in Figure~\ref{fig:SARPEEG}.
We initially observe for the raw EEG values that the linear regression approach to user ratings (QoE) results in potentially negative coefficients of determination for some users.  It is highly variable when considering the first degree, with the best prediction yielding $\hat{R}^2=0.792$.  Increasing the degree yields better prediction outcomes up to $\hat{R}^2=0.852$, which is below a level observed for the simple image display described earlier.  Interestingly, we again observe a reversal effect, whereby higher than quadratic polynomial extension to $d=3$ of the original data results in a lower $\hat{R}^2=0.845$ value. The overall well-performing second degree results again in the best attainable performance given the linear prediction. For the logistic approach, we note that in the majority of cases, the degree $d=1$ performs slightly worse than the linear prediction counterpart.  Increasing the degree to $d=2$, this relationship is inversed and the majority of subject-dependent predictions based on the logistic approach outperform those based on linear approaches.

\begin{figure*}
	\centering
	\subfloat[Linear: QoE $q^u_{i,l}$]{\includegraphics[height=0.25\textheight]{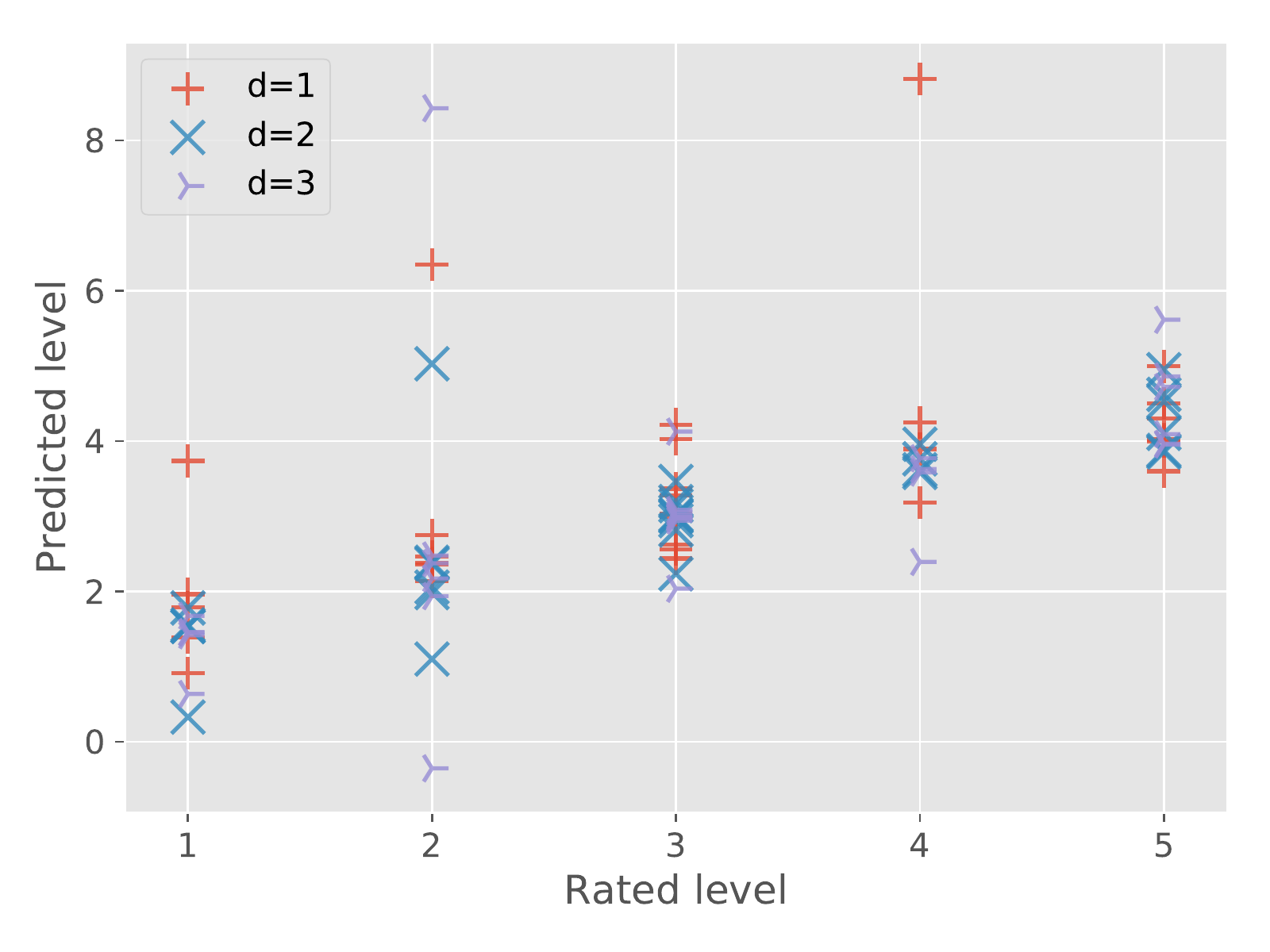}} 
	\qquad
	\subfloat[Logistic: QoE $q^u_{i,l}$]{\includegraphics[height=0.25\textheight]{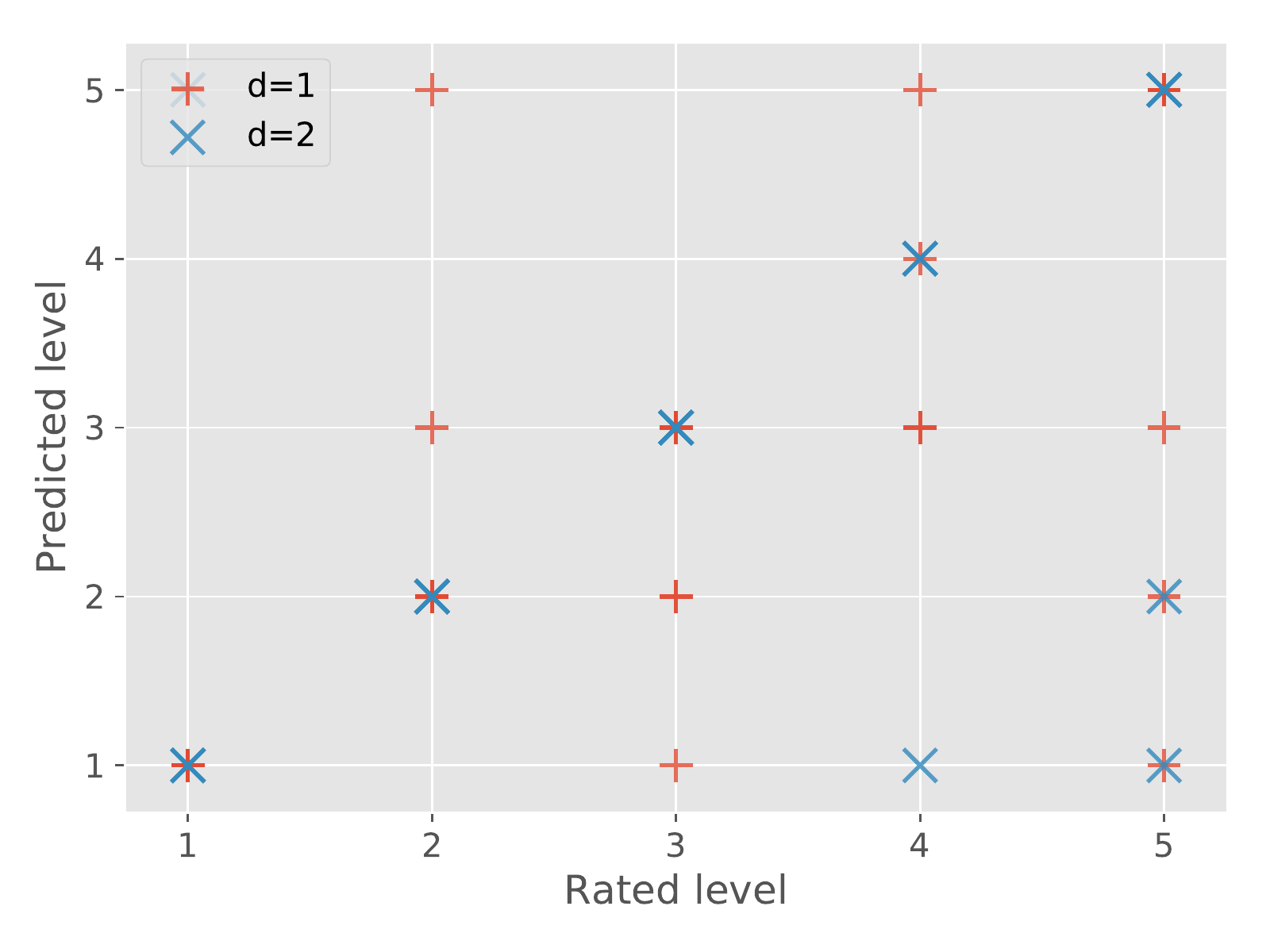}} 
	\\
	\subfloat[Linear: QoS level $l$]{\includegraphics[height=0.25\textheight]{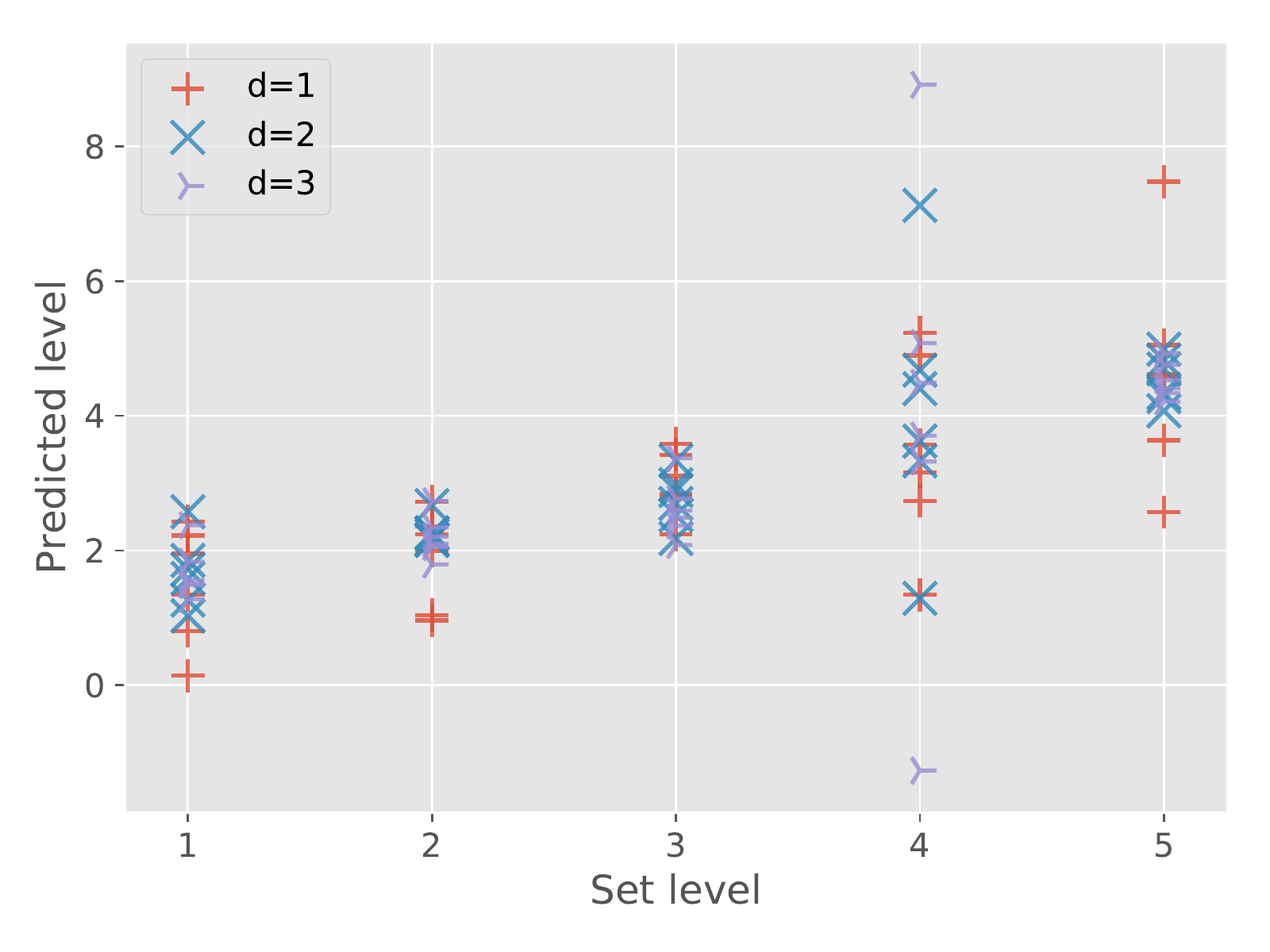}} 
	\qquad	
	\subfloat[Logistic: QoS level $l$]{\includegraphics[height=0.25\textheight]{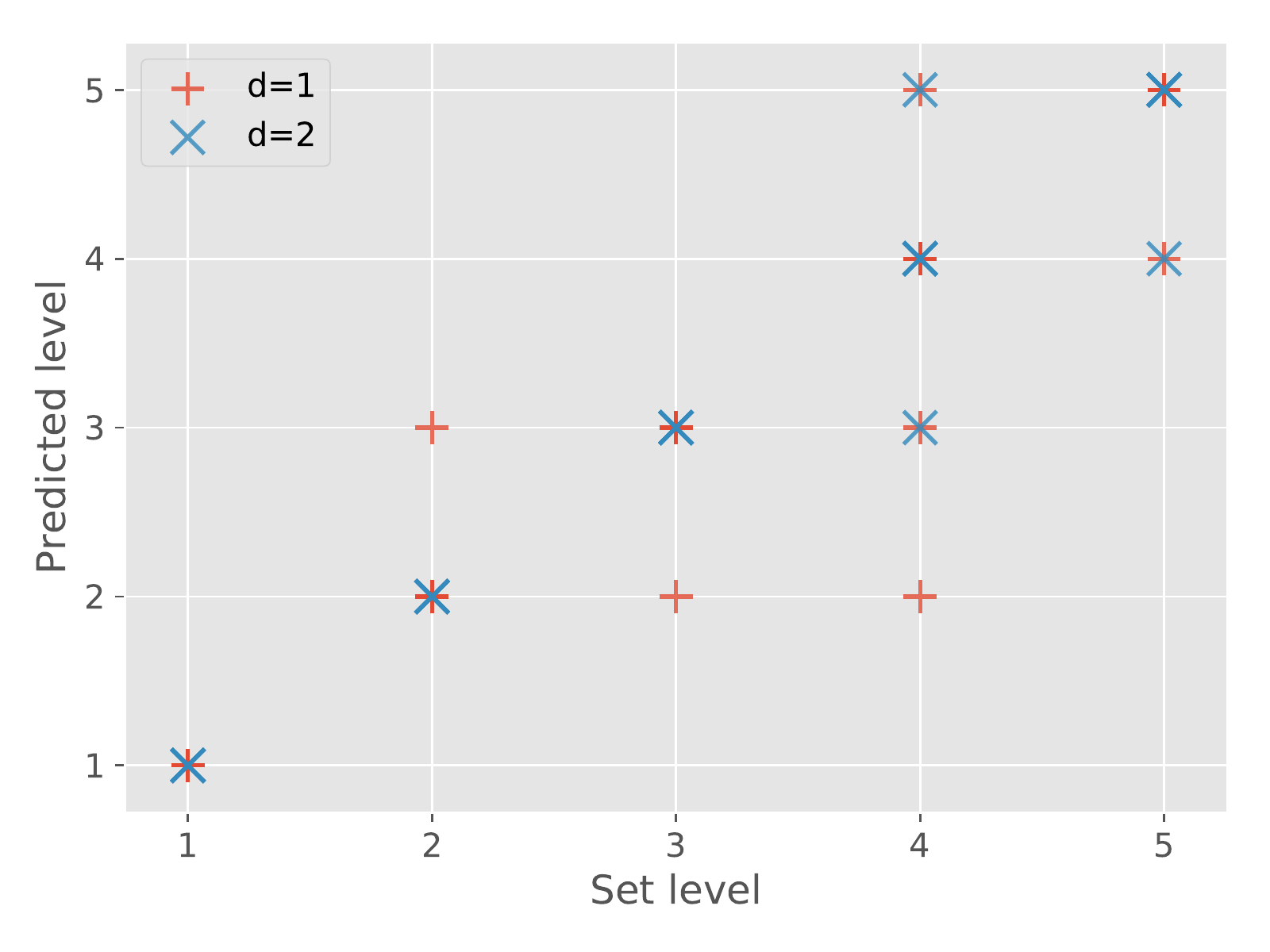}} 
	\\
	\subfloat[Linear: BRISQUE value $B_{i,l}$]{\includegraphics[height=0.25\textheight]{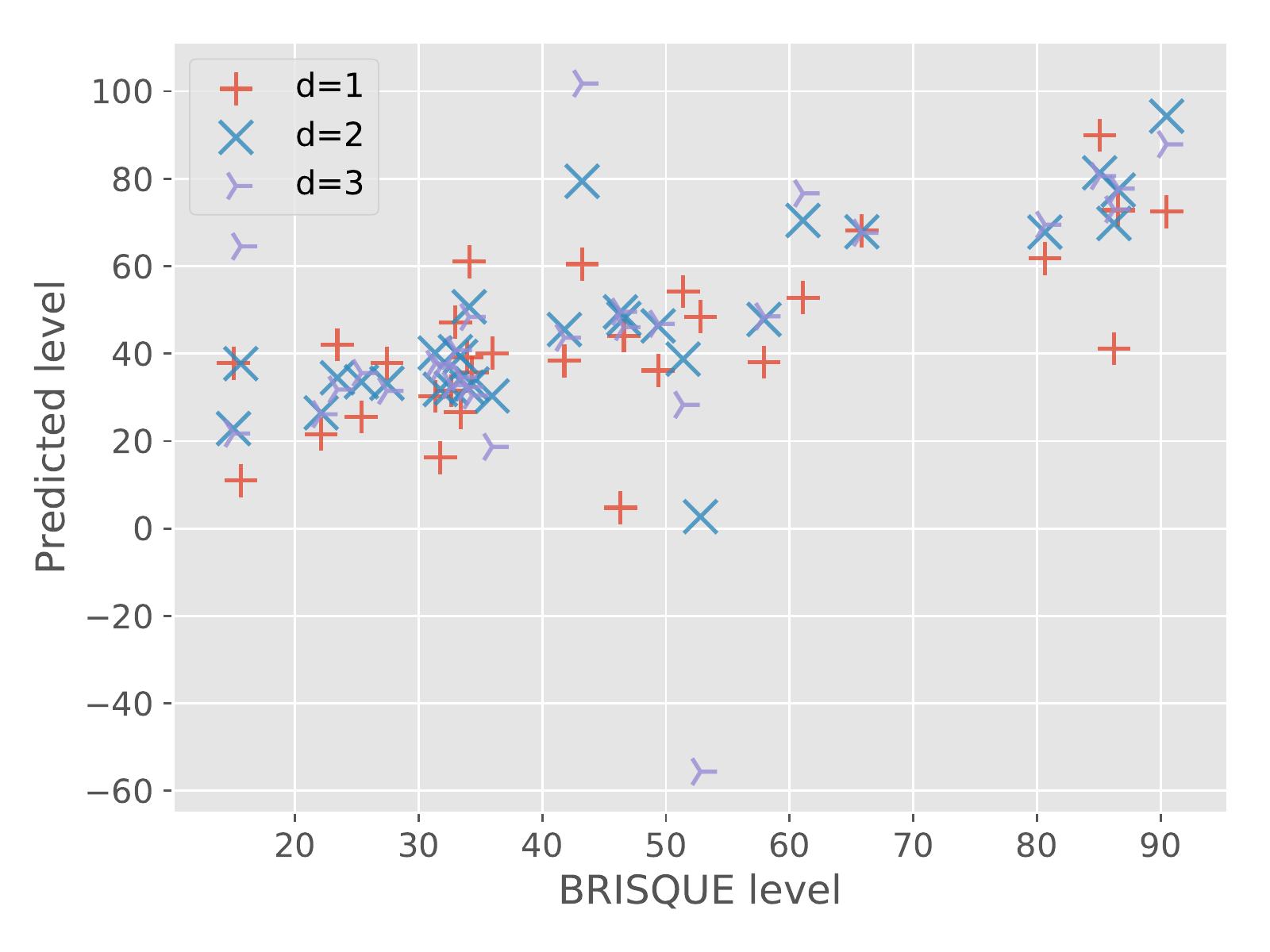}} 
	\caption{Spherical AR (SAR) image quality prediction results for subject ratings $q^u_{i,l}$, image compression (QoS) levels $l$, and BRISQUE metric values $B_{i,l}$.  Results are based on absolute EEG channel measurements (EEG) employing test subject 18 as example.}
	\label{fig:SARPEEG-S18}
\end{figure*}

Shifting the view to the prediction of the set quality levels in Figure~\ref{fig:SARPEEG}, we note that both linear and logistic predictions of a linear degree maintain the commonly lower performance under $\hat{R}^2=0.5$. Again, the second degree yields the best prediction results with the majority of subjects being predicted slightly better by employing the logistic approach.  For the BRISQUE prediction, we observe again that the second degree provides the best prediction results.  We also note a more pronounced reversal effect when considering higher than cubic polynomial extension degrees for the forecasting of BRISQUE values.
In comparison to the non-immersive images, we notice that the overall performance levels for all three evaluated subject-level predictions are generally lower when considering the immersive images.
We provide a more detailed view on the impacts for individual users in Figure~\ref{fig:SARPEEG-S18}, exemplary employing subject 18.
We note, as for regular images, that the spread for the predicted QoE levels narrows with the overall degree and increases after $d=2$. For the logistic approach, the reduction in the deviation from the actual ratings becomes even more visible, and only a handful of significant outliers for the higher levels are responsible for the overall deviation from the actual values.  For the QoS predictions with linear and logistic approaches, we observe a similar behavior. Employing the linear prediction on the BRISQUE metric directly, we notice a significant amount of large outliers in all degrees, albeit at $d=3$, we observe the highest deviation from actual values.


\subsubsection{Prediction with User EEG Profiles}
\begin{figure*}[b!]
	\centering
	\subfloat[EEGz, QoE $q^u_{i,l}$]{\includegraphics[width=0.475\linewidth]{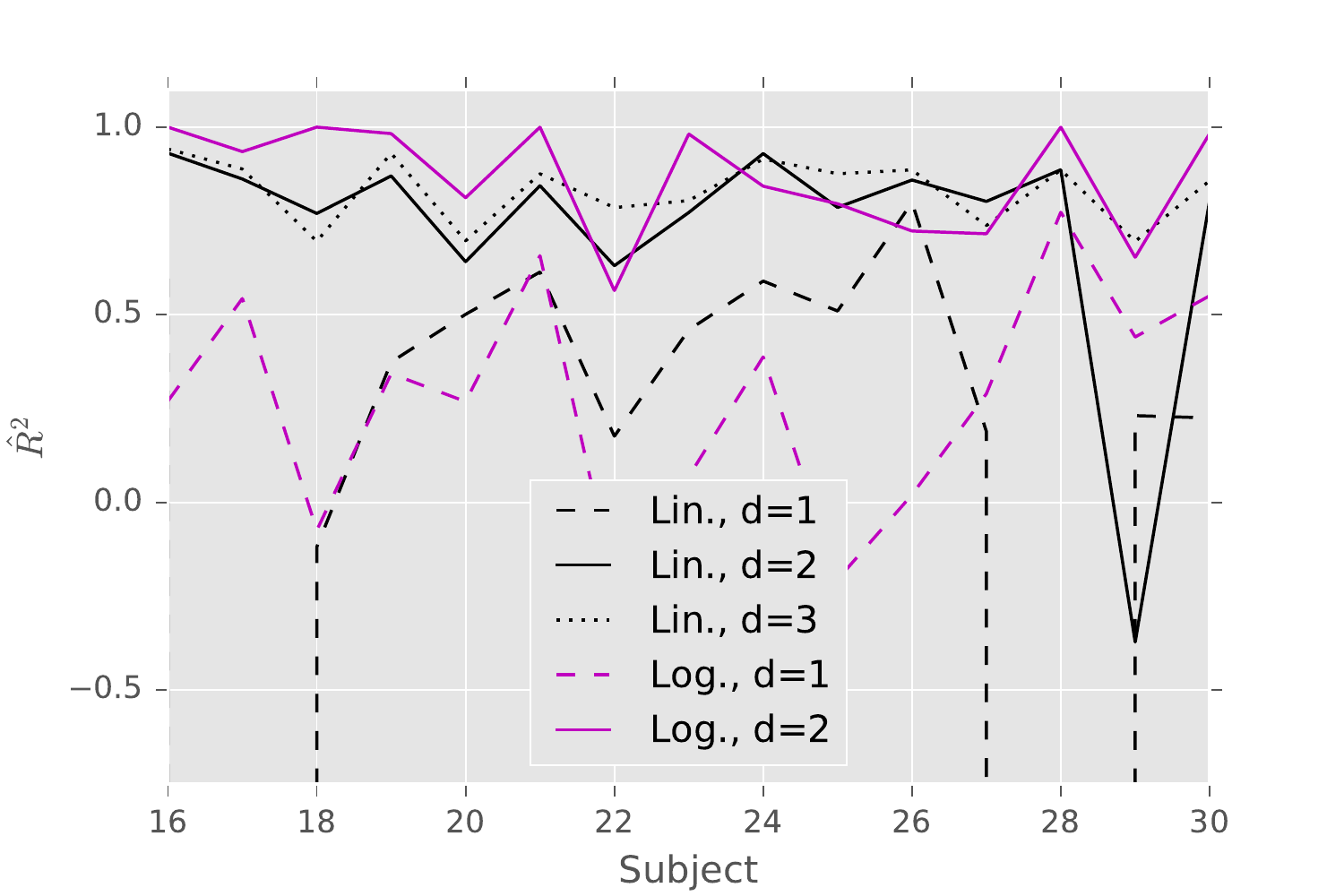}}
	\\
	\subfloat[EEGz, QoS level $l$]{\includegraphics[width=0.475\linewidth]{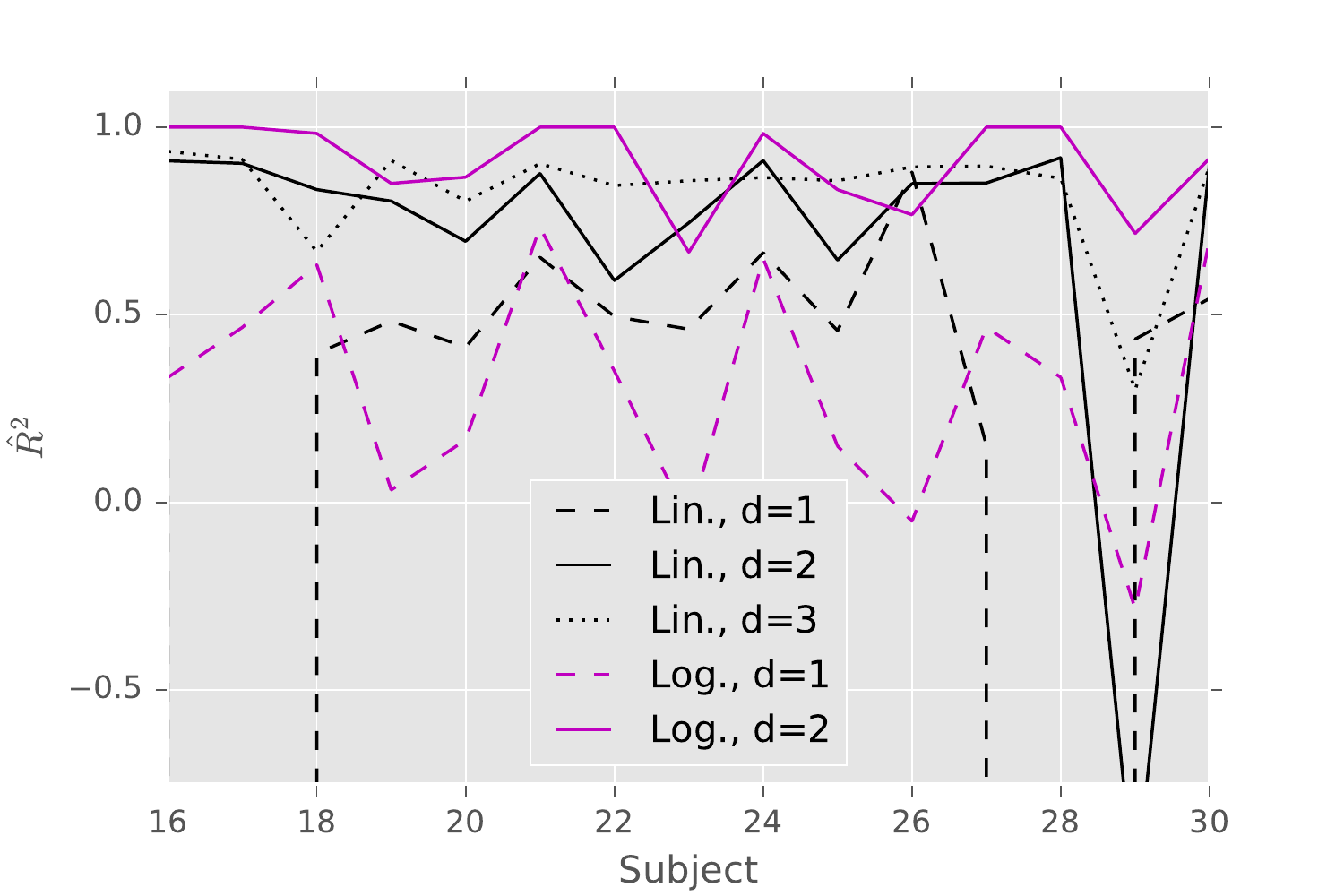}}
	\\    
	\subfloat[EEGz, BRISQUE value $B_{i,l}$]{\includegraphics[width=0.475\linewidth]{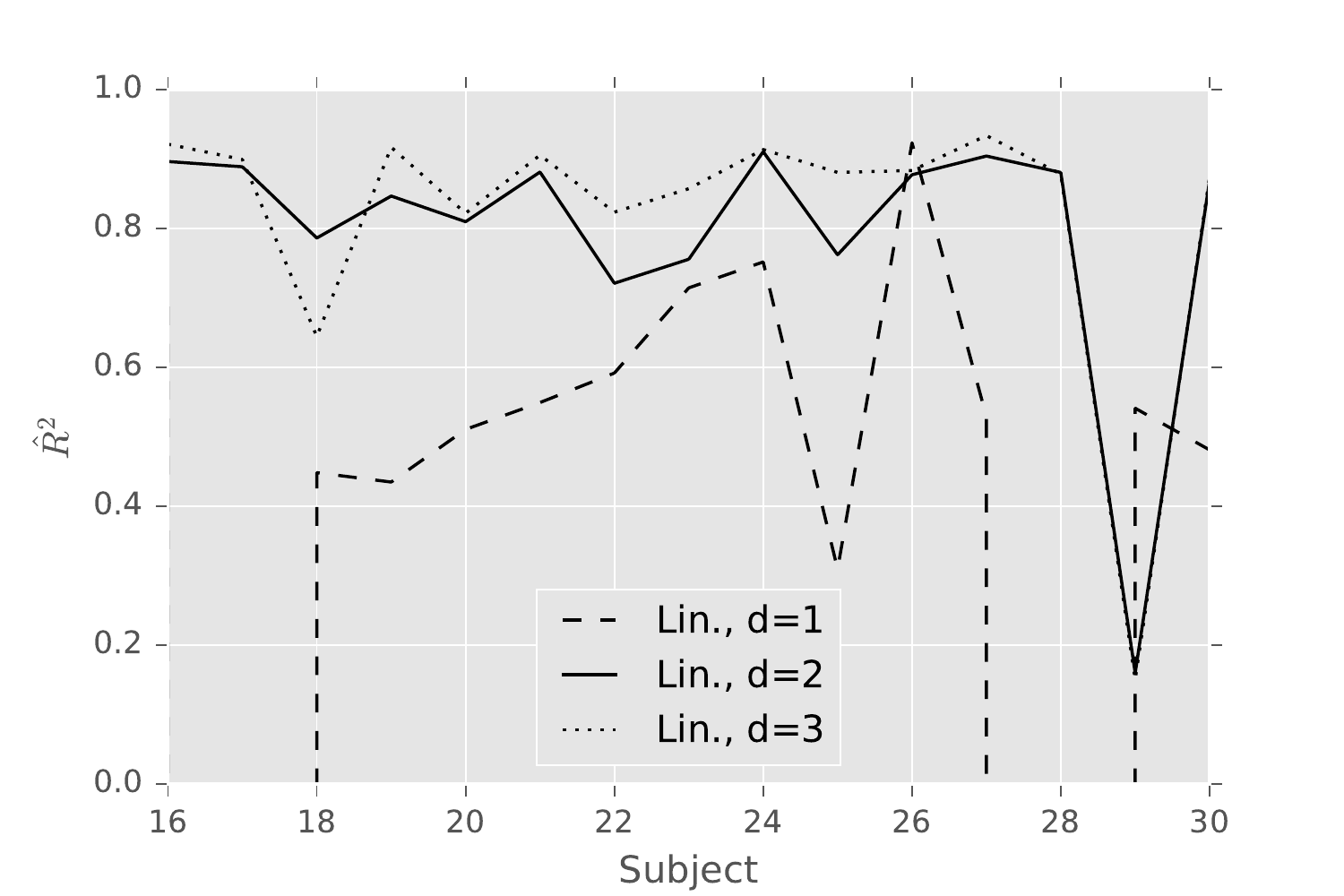}}
	\caption{Spherical AR (SAR) image quality prediction results for subject ratings $q^u_{i,l}$, and image compression (QoS) levels $l$ (cmp. Figs. \ref{fig:SAR-all}, \ref{fig:SAR-all-linreg}). Results are based on z-score normalized EEG channel values (EEGz).}
	\label{fig:SARPEEGz}
\end{figure*}

We continue our prediction evaluations with the assumption that an overall profile for the user would be available to a content provider and EEG signals could be z-score normalized. This EEGz-based approach is identical to the one outlined for regular image display with results illustrated in Figure~\ref{fig:SARPEEGz}.
We initially observe that the general observations made concerning approach and degrees apply here for the prediction of the QoE, but slightly different.  In particular, we observe that the outlier-free average for the linear approach yields average prediction performances of $\hat{R}^2=0.396$, $\hat{R}^2=0.734$, and $\hat{R}^2=0.832$ for degrees one to three, respectively.  Overall, the increase in degrees here results directly in an increase in the performance, which was not the case for the corresponding EEG evaluations.  For the logistic approach to the prediction of the subject-experienced QoE, we derive outlier-cleaned averages of $\hat{R}^2=0.275$ and $\hat{R}^2=0.866$ for degrees $d=1$ and $d=2$, respectively.  While lower than for regular image counterparts, the logistic prediction approach here visibly outperforms the linear approach for the majority of the subjects at the second degree with several perfect $\hat{R}^2=1$ scores. 

For the set (QoS) levels, we notice a similar result of applied prediction models at varying degrees.  The linear prediction again exhibits directly improving prediction performances with degree increases from $\hat{R}^2=0.504$ over $\hat{R}^2=0.692$ to $\hat{R}^2=0.827$ as degrees increase to three.  As for the QoE prediction, the logistic approach yields an overall better performance at the second degree, with an average of $\hat{R}^2=0.906$ -- again with a large number of perfect prediction performance scores.  Lastly, we evaluate the BRISQUE metric prediction, which results in degree-dependent average performance values of $\hat{R}^2=0.575$, $\hat{R}^2=0.797$, and $\hat{R}^2=0.820$.  While not attaining a perfect prediction performance score, the overall values are fairly high, indicating an generally good prediction achievement.  

Compared to the EEG-only prediction, we observe an increase in the overall performance with the logistic approach yielding better prediction results for QoE and QoS predictions.  Due to the increased complexity and suitability for employment near real-time, higher degrees for logistic prediction become non-preferential and are not considered here.  Similarly, the spherical immersive image prediction performance is in general below that observed for static images.

\begin{figure*}
	\centering
	\subfloat[Linear: QoE $q^u_{i,l}$]{\includegraphics[height=0.25\textheight]{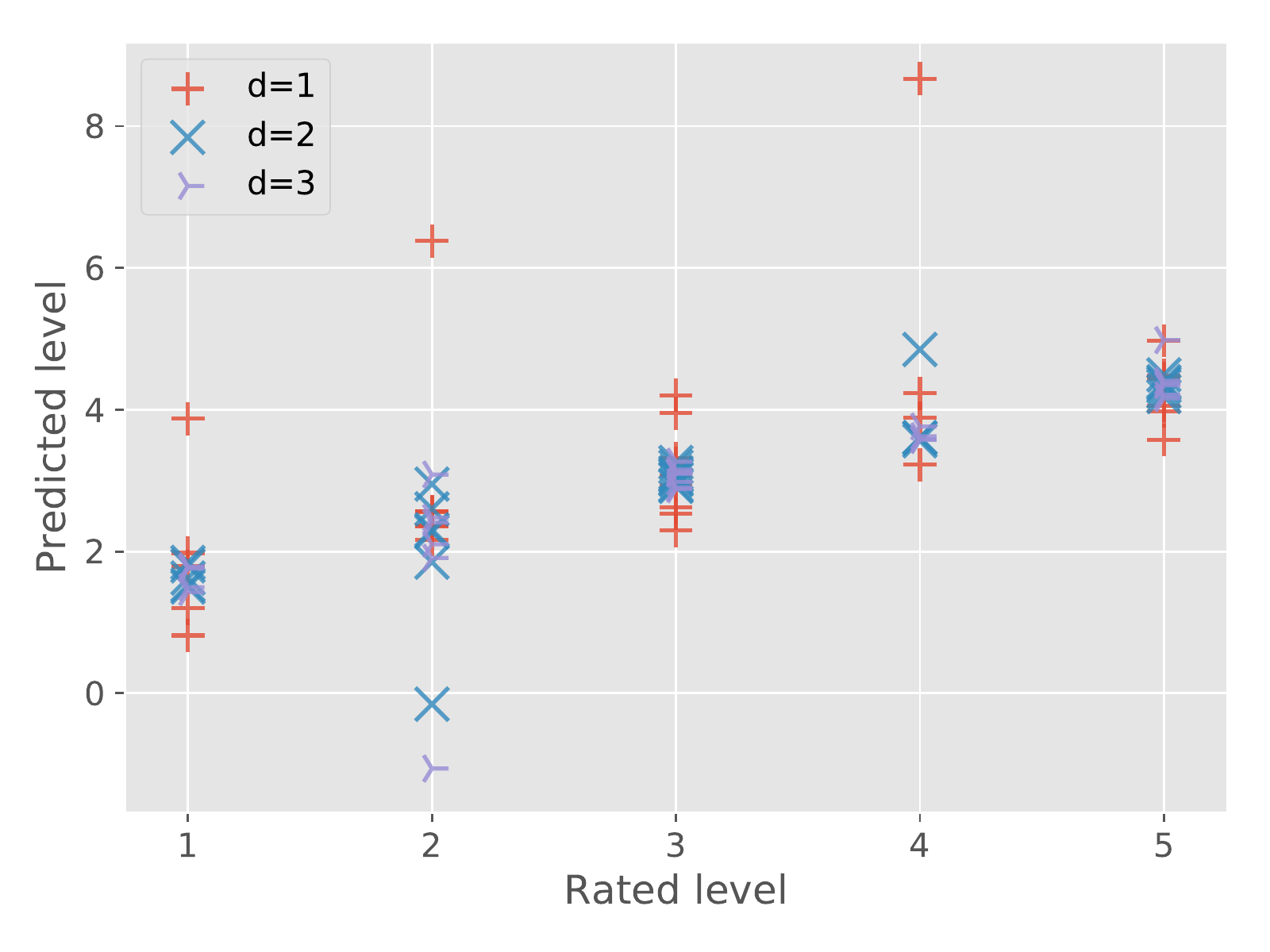}} 
	\qquad
	\subfloat[Logistic: QoE, $q^u_{i,l}$]{\includegraphics[height=0.25\textheight]{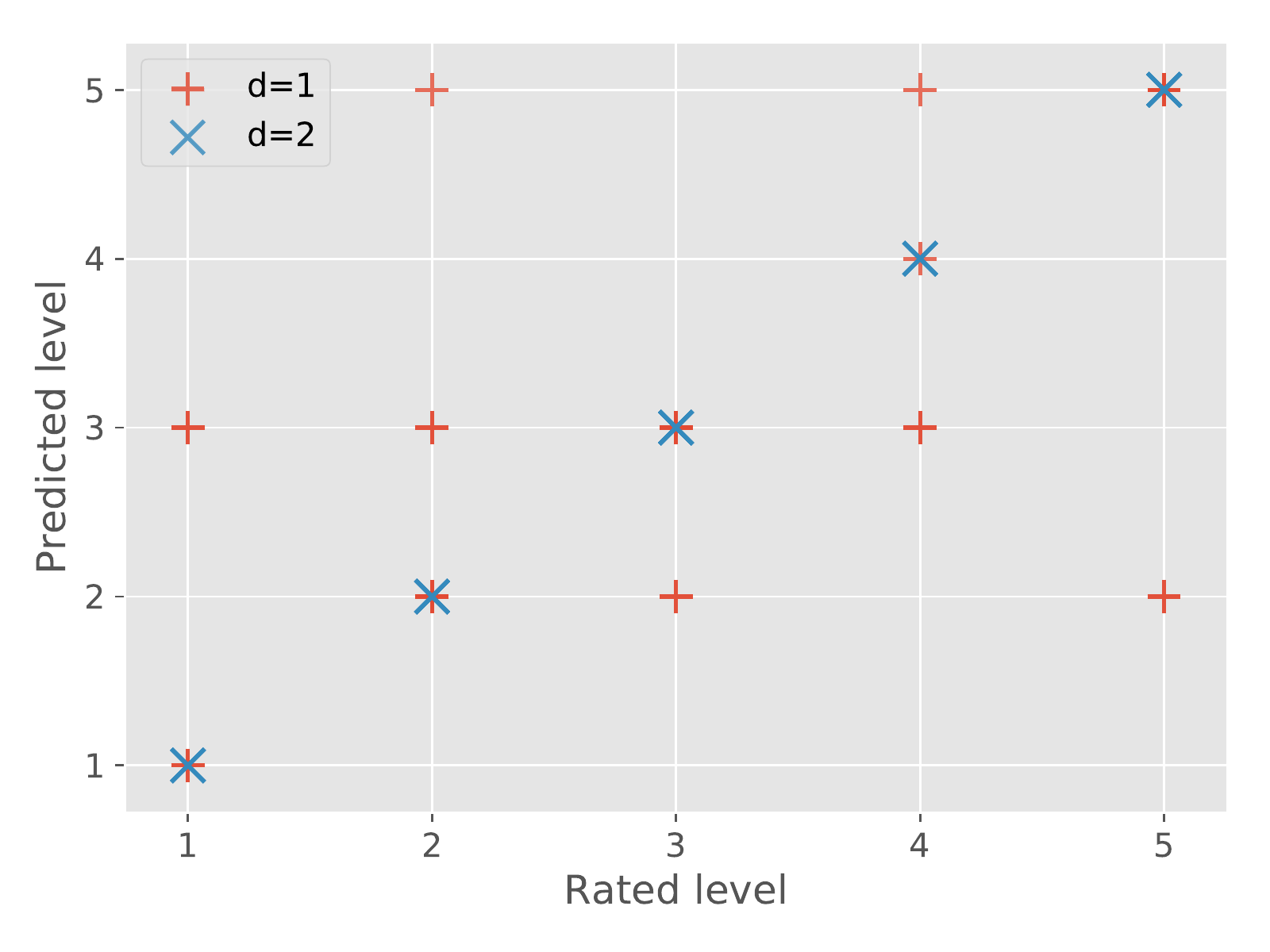}} 
	\\
	\subfloat[Linear: QoS level $l$]{\includegraphics[height=0.25\textheight]{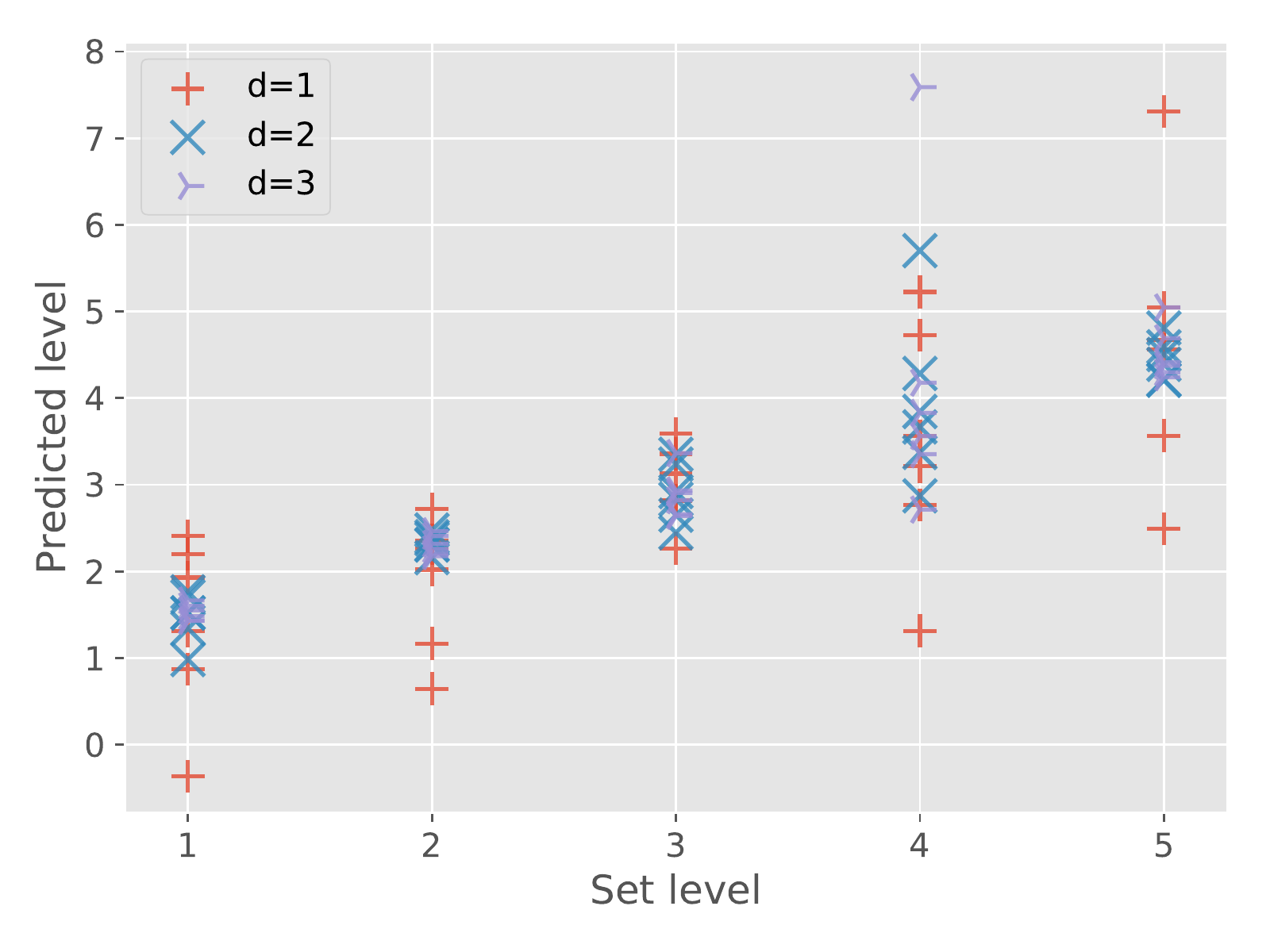}} 
	\qquad	
	\subfloat[Logistic: QoS level $l$]{\includegraphics[height=0.25\textheight]{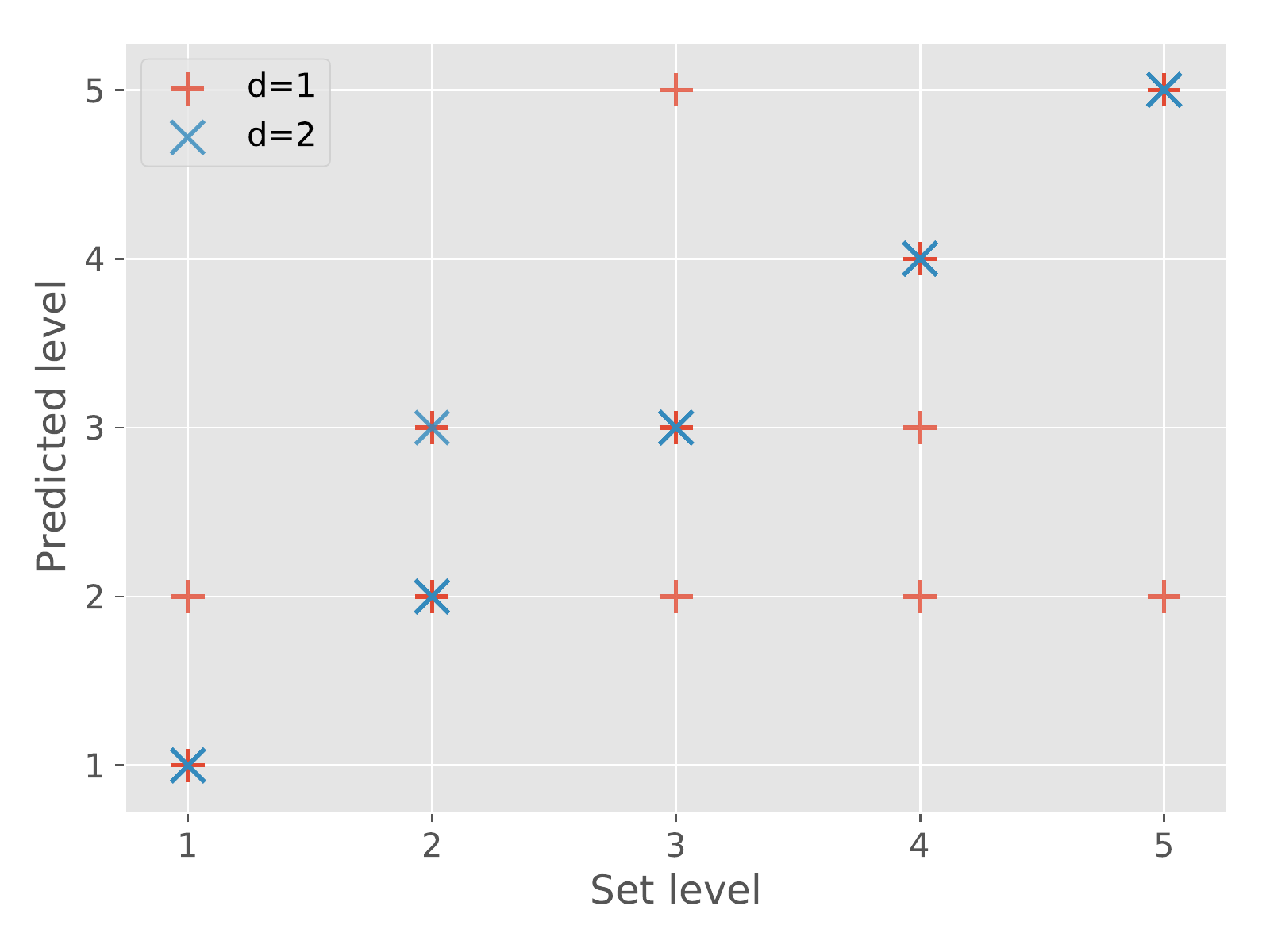}} 
	\\
	\subfloat[Linear: BRISQUE value $B_{i,l}$]{\includegraphics[height=0.25\textheight]{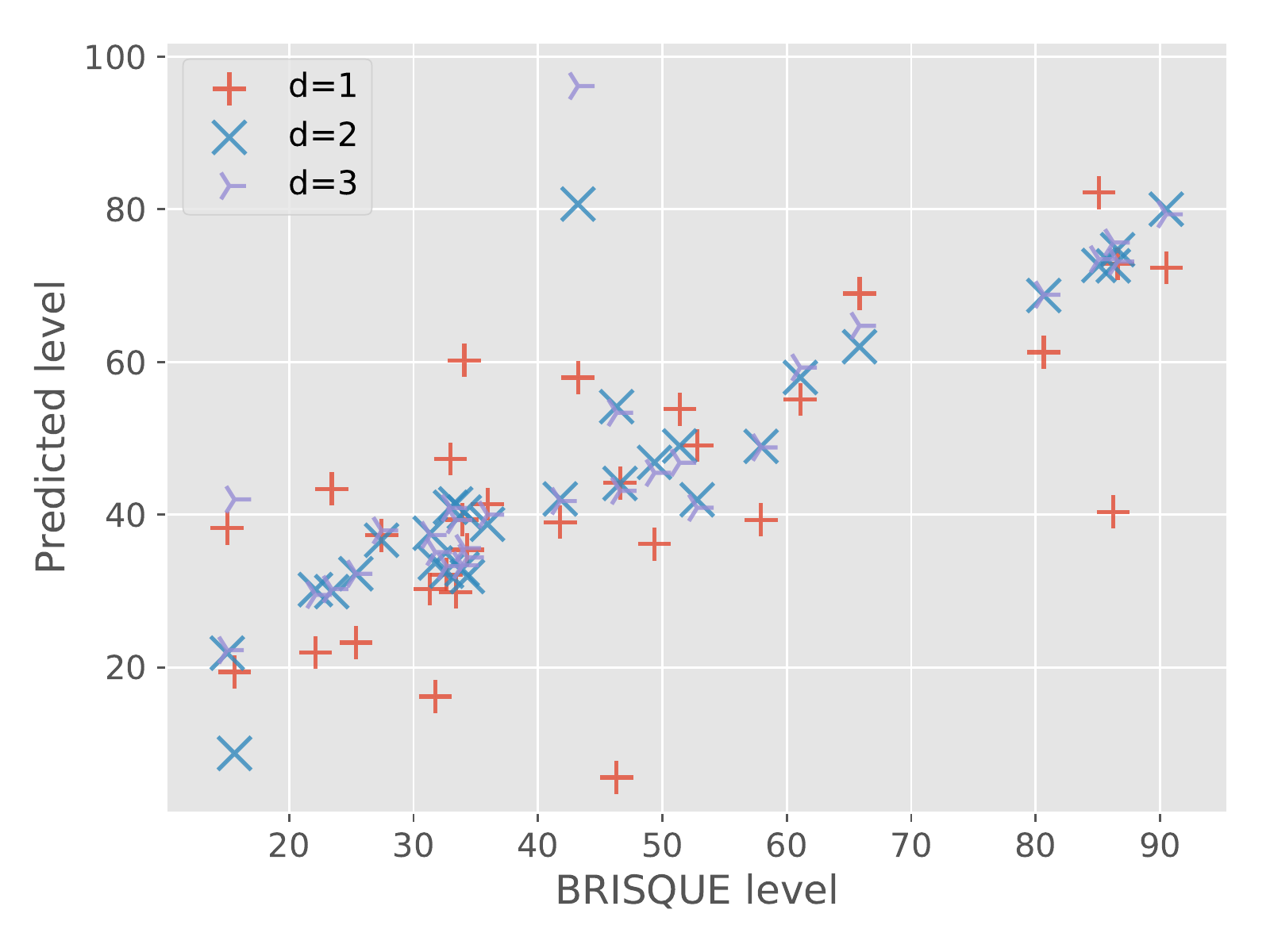}} 
	\caption{Spherical AR (SAR) image quality prediction results for subject ratings $q^u_{i,l}$, image compression (QoS) levels $l$, and BRISQUE metric values $B_{i,l}$.  Results are based on z-score normalized EEG channel measurements (EEGz) employing test subject 18 as example.}
	\label{fig:SARPEEGz-S18}
\end{figure*}

Next, we illustrate the corresponding individual user examples in Figure~\ref{fig:SARPEEGz-S18}.
For the QoE predictions based on a linear approach, we note a quick reduction of the overall spread of values with an increase of $d$ towards a linear trend.  We furthermore note that the number of outliers decreases as well.  Compared to the EEG-based approach in Figure~\ref{fig:SARPEEG-S18}, we additionally observe that the spread here is smaller and more concise, which results in the better prediction performance as measured by the $\hat{R}^2$ score.  
The logistic prediction based on EEGz-values exhibits a wide spread of predicted values for $d=1$, which after an increase to $d=2$ disappears for a perfect prediction of the QoE as rated by subject 18.  The logistic prediction here is capable of capturing all user ratings based on the EEGz.

The relationship between QoE and QoS results in a similar behavior observable for the predictions of the QoS levels in Figure~\ref{fig:SARPEEGz-S18} as well.  Specifically, we notice a slightly higher level of dispersion of predicted values, which results in a worse prediction performance for the QoS level.  The reversal effect for higher degrees of the underlying polynomial additionally renders the prediction above $d=2$ non-favorable.  For the logistic approach, we observe the remaining of a prediction error for QoS level two, which results in a less than perfect, albeit close to, prediction performance at degree $d=2$. 

Lastly, for the BRISQUE metric predictions, we notice the reduction of a predicted value spread as the degree increases from $d=1$ to $d=2$. However, the reversal effect we noticed throughout reintroduces significant outliers as the degree increases further.  Compared to the EEG-based prediction of the BRISQUE metric, we observe a significant increase in the prediction performance resulting in $\hat{R}^2=0.786$ compared to $\hat{R}^2=0.570$ when no subject profile were available.

\section{Discussion}
\label{s:disc}
Throughout this article, we described the QoE properties for immersive, i.e., spherical, images when viewed through an augmented reality heads-up display.  We find that the characteristics for this particular type of image are in line with those we found in our prior works for regular images, see~\cite{}.
Specifically, we note that the BRISQUE non-referential image quality metric (employed as QoS metric) exhibits excellent properties that result in a high correlation with the subjective QoE ratings (as SAR-MOS and SAR-DMOS) by users.  

The result is an overall attainable linear relationship between BRISQUE values and QoE.  Overall, the BRISQUE image quality metric shows very good potential to be employed in the prediction of the overall QoE of spherical images that are displayed in an AR setting. As a non-referential image quality metric, BRISQUE can be employed independently at different stages of image delivery and display in augmented reality settings and be used to adapt content in a dynamic fashion.
We validate this approach by employing prediction based on linear, bound linear (i.e., quantized to the 5-point Likert scale), and logistic regression and subsequent prediction based on machine learning approaches.  Given an increase in complexity and resulting computational time required, we notice that the second degree polynomial extension for the BRISQUE values yields a sensible trade-off, as for some content, a reversal of performance was found with higher degrees $d$.  In summary, we note that prediction of the image QoE (indicated on a 5-point Likert scale MOS) based on the BRISQUE metrics is achievable, whereby some images might pose additional challenges and result in reduced prediction accuracies.  Cross-evaluating the prediction approach based on subjects, rather than content alone, we notice that despite $\hat{R}^2$ values above 0.5 common at $d=2$, the subjective nature of the QoE results in potentially sub-optimal prediction outcomes.  (We note, however, that the deviation here typically is within 20\% on average, which might be tolerable for content and network service providers.)

Due to these potential problems, we additionally evaluated the possibility of predicting the QoE from a user's EEG signals.  For this approach, common devices that employ dry electrodes are already available and could be embodied into heads-up displays in the future.  Jointly with current advances in ultra-low latency networking referred to as the ``tactile internet,'' such an approach would similarly enable providers to adjust content display based on user feedback, but without user interventions.  
We focus our EEG-based predictions on two different scenarios, namely ($i$) utilizing only user EEG measurements relating to the image quality measurement time point and ($ii$) assuming an overall EEG user profile normalization, for which we employ all captured EEG signals.  Additionally, we consider both, regular and immersive images in our prediction performance evaluation.  In alignment with our initial observations, we find that the second degree logistic prediction outperforms the other approaches and yields very close predictions of the user-experienced QoE, especially if a user profile were available. In this case for regular images, the prediction in most cases is perfect, for spherical images in almost half.

For application scenarios, this substances the possibility of employing measurement and feedback channels to adjust the displayed quality based on biofeedback rather than QoS metrics.  The main potential benefit is the possibility to dynamically adjust content quality levels in very short time spans within original display, which could prove highly beneficial for cognitive strain of operators of AR devices in the field that rely on the overlaid information.  Given that we are able to perform this prediction with already available COTS hardware and within a networked environment, our ongoing efforts are targeting future refinements within the EEG data acquisition and prediction pipeline, as well as subsequent networked media delivery optimization approaches.

\section{Conclusion}
\label{s:conc}
In this article, we considered a general approach to predictions of the QoE and QoS for augmented reality image presentations, typically in form of binocular vision augmentation.  Both, the non-referential image quality metric BRISQUE as well consumer-grade EEG headband determined user EEG stimuli were found to enable high levels of QoE prediction performance. Prediction complexity limitations are joined by an uncovered prediction performance reversal, resulting in the second degree polynomial extensions yielding the best performance for both linear and logistic predictions.
The results we obtained are highly motivating for automatic user-dependent content adaptation scenarios without further intervention, which is an avenue for future research.  For content and network providers alike, this opens the possibilities for well-grounded content adaptation before the last, typically rate-restricted wireless link. 


\end{document}